\documentclass {aastex63}
\usepackage{epsf}
\usepackage{graphicx}
\usepackage{amsmath}
\usepackage{amssymb}
\usepackage{gensymb}
\usepackage{chngcntr}
\usepackage{longtable}
\usepackage{latexsym}

\newcommand {\nth}{N$_2$H$^+$}
\newcommand{\ntd}{N$_2$D$^+$}

    {\pagebreak[4]\global\pdfpageattr\expandafter{\the\pdfpageattr/Rotate 90}}%
    {\pagebreak[4]\global\pdfpageattr\expandafter{\the\pdfpageattr/Rotate 0}}%
    
\usepackage{natbib,graphicx,morefloats,multirow,pbox,booktabs, makecell}
\usepackage{hyperref}
\usepackage{soul}

\begin{document}

\title{Nobeyama Survey of Inward Motions toward Cores in Orion Identified by SCUBA-2}

\correspondingauthor{Ken'ichi Tatematsu}
\email{k.tatematsu@nao.ac.jp}

\author[0000-0002-8149-8546]{Ken'ichi Tatematsu}
\affil{Nobeyama Radio Observatory, National Astronomical Observatory of Japan,
National Institutes of Natural Sciences,
Nobeyama, Minamimaki, Minamisaku, Nagano 384-1305, Japan}
\affiliation{Department of Astronomical Science,
The Graduate University for Advanced Studies, SOKENDAI,
2-21-1 Osawa, Mitaka, Tokyo 181-8588, Japan}

\author[0000-0001-9304-7884]{You-Ting Yeh}
\affiliation{Academia Sinica Institute of Astronomy and Astrophysics, 11F of Astronomy-Mathematics Building, AS/NTU. No.1, Sec. 4, Roosevelt Rd, Taipei 10617, Taiwan, R.O.C.}

\author[0000-0001-9304-7884]{Naomi Hirano}
\affiliation{Academia Sinica Institute of Astronomy and Astrophysics, 11F of Astronomy-Mathematics Building, AS/NTU. No.1, Sec. 4, Roosevelt Rd, Taipei 10617, Taiwan, R.O.C.}

\author[0000-0003-4603-7119]{Sheng-Yuan Liu}
\affiliation{Academia Sinica Institute of Astronomy and Astrophysics, 11F of Astronomy-Mathematics Building, AS/NTU. No.1, Sec. 4, Roosevelt Rd, Taipei 10617, Taiwan, R.O.C.}

\author[0000-0002-5286-2564]{Tie Liu}
\affiliation{Shanghai Astronomical Observatory, Chinese Academy of Sciences, 80 Nandan Road, Shanghai 200030, P. R. China}
\affiliation{Korea Astronomy and Space Science Institute,
776 Daedeok-daero, Yuseong-gu, Daejeon 34055, South Korea}
\affiliation{East Asian Observatory, 660 N. A'ohoku Place, Hilo, HI 96720, USA}

\author[0000-0002-2338-4583]{Somnath Dutta}
\affiliation{Academia Sinica Institute of Astronomy and Astrophysics, 11F of Astronomy-Mathematics Building, AS/NTU. No.1, Sec. 4, Roosevelt Rd, Taipei 10617, Taiwan, R.O.C.}

\author[0000-0002-4393-3463]{Dipen Sahu}
\affiliation{Academia Sinica Institute of Astronomy and Astrophysics, 11F of Astronomy-Mathematics Building, AS/NTU. No.1, Sec. 4, Roosevelt Rd, Taipei 10617, Taiwan, R.O.C.}

\author[0000-0001-5175-1777]{Neal J. Evans II}
\affiliation{Department of Astronomy, The University of Texas at Austin, 2515 Speedway, Stop C1400, Austin, TX 78712$-$1205, USA}

\author[0000-0002-5809-4834]{Mika Juvela}
\affiliation{Department of Physics, P.O. Box 64, FI-00014, University of Helsinki, Finland}

\author[0000-0003-0537-5461]{Hee-Weon Yi}
\affiliation{Korea Astronomy and Space Science Institute,
776 Daedeok-daero, Yuseong-gu, Daejeon 34055, Republic of Korea}
\affiliation{School of Space Research, Kyung Hee University, Seocheon-Dong, Giheung-Gu, Yongin-Si, Gyeonggi-Do, 446-701, Republic of Korea}

\author[0000-0003-3119-2087]{Jeong-Eun Lee}
\affiliation{School of Space Research, Kyung Hee University, Seocheon-Dong, Giheung-Gu, Yongin-Si, Gyeonggi-Do, 446-701, Republic of Korea}

\author[0000-0002-7125-7685]{Patricio Sanhueza}
\affiliation{National Astronomical Observatory of Japan,
National Institutes of Natural Sciences,
2-21-1 Osawa, Mitaka, Tokyo 181-8588, Japan}
\affiliation{Department of Astronomical Science,
The Graduate University for Advanced Studies, SOKENDAI,
2-21-1 Osawa, Mitaka, Tokyo 181-8588, Japan}

\author[0000-0003-1275-5251]{Shanghuo Li}
\affiliation{Korea Astronomy and Space Science Institute,
776 Daedeok-daero, Yuseong-gu, Daejeon 34055, South Korea}
\affiliation{Shanghai Astronomical Observatory, Chinese Academy of Sciences, 80 Nandan Road, Shanghai 200030, P. R. China}
\affiliation{Center for Astrophysics | Harvard \& Smithsonian, 60 Garden Street, Cambridge, MA 02138, USA}
\affiliation{University of Chinese Academy of Sciences, 19A Yuquanlu, Beijing 100049, P. R. China}

\author[0000-0002-5881-3229]{David Eden}
\affiliation{Astrophysics Research Institute, Liverpool John Moores University, IC2, 
Liverpool Science Park, 146 Brownlow Hill, Liverpool L3 5RF, UK}

\author[0000-0003-2011-8172]{Gwanjeong Kim}
\affil{Nobeyama Radio Observatory, National Astronomical Observatory of Japan,
National Institutes of Natural Sciences,
Nobeyama, Minamimaki, Minamisaku, Nagano 384-1305, Japan}

\author[0000-0002-3024-5864]{Chin-Fei Lee}
\affiliation{Academia Sinica Institute of Astronomy and Astrophysics, 11F of Astronomy-Mathematics Building, AS/NTU. No.1, Sec. 4, Roosevelt Rd, Taipei 10617, Taiwan, R.O.C.}

\author{Yuefang Wu}
\affiliation{Department of Astronomy, Peking University, 100871, Beijing, China}

\author[0000-0003-2412-7092]{Kee-Tae Kim}
\affiliation{Korea Astronomy and Space Science Institute,
776 Daedeok-daero, Yuseong-gu, Daejeon 34055, Republic of Korea}
\affiliation{University of Science and Technology, Korea (UST), 217 Gajeong-ro, Yuseong-gu, Daejeon 34113, Republic of Korea}

\author[0000-0002-5310-4212]{L. Viktor T\'{o}th}
\affiliation{Department of Astronomy, E\"{o}tv\"os Lor\'{a}nd University, P\'{a}zm\'{a}ny P\'{e}ter s\'{e}t\'{a}ny 1/A, H-1117 Budapest, Hungary}

\author{Minho Choi}
\affiliation{Korea Astronomy and Space Science Institute,
776 Daedeok-daero, Yuseong-gu, Daejeon 34055, South
Korea}

\author[0000-0002-5016-050X]{Miju Kang}
\affiliation{Korea Astronomy and Space Science Institute,
776 Daedeok-daero, Yuseong-gu, Daejeon 34055, South
Korea}

\author[0000-0001-5392-909X]{Mark A. Thompson}
\affiliation{School of Physics \& Astronomy, University of Leeds, Leeds, LS2 9JT, UK}
\affiliation{Centre for Astrophysics Research, Science \& Technology Research Institute, University of Hertfordshire, Hatfield, AL10 9AB, UK}

\author[0000-0001-8509-1818]{Gary A. Fuller}
\affiliation{Jodrell Bank Centre for Astrophysics, School of Physics and Astronomy, University of Manchester, Oxford Road, Manchester, M13 9PL, UK}

\author[0000-0003-3010-7661]{Di Li}
\affiliation{National Astronomical Observatories, Chinese Academy of Sciences, Beijing, 100012, China}

\author[0000-0002-7237-3856]{Ke Wang}
\affiliation{The Kavli Institute for Astronomy and Astrophysics, Peking University,
5 Yiheyuan Road, Haidian District, Beijing 100871, P. R. China}
\affiliation{European Southern Observatory, Karl-Schwarzschild-Str. 2 D-85748 Garching bei M\"{u}nchen, Germany}

\author[0000-0003-4521-7492]{Takeshi Sakai}
\affiliation{Graduate School of Informatics and Engineering, The University of Electro-Communications, Chofu, Tokyo 182-8585, Japan}

\author[0000-0003-2610-6367]{Ryo Kandori}
\affiliation{Astrobiology Center of NINS,
2-21-1 Osawa, Mitaka, Tokyo 181-8588, Japan}

\author[0000-0002-1369-1563]{Shih-Ying Hsu}
\affiliation{National Taiwan University, No. 1, Sec. 4, Roosevelt Rd., Taipei 10617, Taiwan, R.O.C.}
\affiliation{Academia Sinica Institute of Astronomy and Astrophysics, 11F of Astronomy-Mathematics Building, AS/NTU. No.1, Sec. 4, Roosevelt Rd, Taipei 10617, Taiwan, R.O.C.}

\author[0000-0001-9304-7884]{Chau-Ching Chiong}
\affiliation{Academia Sinica Institute of Astronomy and Astrophysics, 11F of Astronomy-Mathematics Building, AS/NTU. No.1, Sec. 4, Roosevelt Rd, Taipei 10617, Taiwan, R.O.C.}

\author{JCMT Large Program ``SCOPE'' collaboration}
\author{``ALMASOP'' collaboration}


\begin{abstract}
In this study, 36 cores (30 starless and 6 protostellar)
identified in Orion
were surveyed to search 
for inward motions.
We used the Nobeyama 45 m radio telescope, and mapped
the cores in the $J = 1\rightarrow0$ transitions of HCO$^+$, H$^{13}$CO$^+$, N$_2$H$^+$, HNC, and HN$^{13}$C.
The asymmetry parameter $\delta V$, which was the ratio of 
the difference between 
the HCO$^+$ and H$^{13}$CO$^+$ peak velocities 
to the H$^{13}$CO$^+$ line width,
was biased toward negative values, suggesting that
inward motions were more dominant than outward motions.
Three starless cores (10\% of all starless cores surveyed)  were identified as cores with blue-skewed line profiles
(asymmetric profiles with more intense blue-shifted emission),
and another 
two starless cores (7\%)
were identified as candidate blue-skewed line profiles.
The peak velocity difference between HCO$^+$ and H$^{13}$CO$^+$ of them was
up to 0.9~km~s$^{-1}$, suggesting that some inward motions exceeded
the speed of sound for the quiescent gas  ($\sim10-17$ K).
The mean of $\delta V$ of the 
five
aforementioned starless cores 
was derived to be 
$-$0.5$\pm$0.3.
One core, G211.16$-$19.33North3, observed using the ALMA ACA in DCO$^+$ $J = 3\rightarrow2$
exhibited blue-skewed features.
Velocity offset in the blue-skewed line profile with a dip 
in the DCO$^+$ $J = 3\rightarrow2$ line was larger ($\sim 0.5$~km~s$^{-1}$)
than that in HCO$^+$ $J = 1\rightarrow0$ ($\sim 0.2$~km~s$^{-1}$),
which may represent gravitational acceleration of inward motions.
It seems that this core is at the
last stage  in the starless phase,
judging from the chemical evolution factor version 2.0 (CEF2.0).
\end{abstract}

\section{INTRODUCTION}
The formation of stars has been a topic of considerable research attention.
It has been suggested that denser cores have shorter lifetimes in units of free-fall time;
the lifetime of a core with an H$_2$ density of
10$^4$ cm$^{-3}$ is approximately 10$^6$ yr, which is longer than 
the free-fall time ($t_{ff} = 3\times10^5$ yr), whereas
that of a core with an H$_2$ density of
10$^6$ cm$^{-3}$ seems close to the free-fall time ($t_{ff} = 3\times10^4$ yr)
\citep{2000MNRAS.311...63J, 2014prpl.conf...27A, 2015A&A...584A..91K, 2020ApJ...899...10T}.
Here, the lifetime is estimated from the fraction of
cores that contain protostars or young stellar objects (YSOs).
Starless cores with an H$_2$ density of
10$^{4-5}$ cm$^{-3}$ do not appear to be collapsing dynamically, and may exhibit dynamically stable equilibrium states. 
If such stable cores finally form stars,
a mechanism may arise to make them unstable.
Mass inflow into cores is one such mechanism \citep{2007ApJ...669.1042G}.
Hence, it is a matter of interest to determine when starless cores initiate inward motions.
For terminology,
we adopt that of 
\cite{2020ApJ...900...82P}, i.e., 
inward motions on scales of 1$-$10 pc (filaments to cores), 0.01$-$0.1 pc (inside cores), and 100$-$1000 au (disks to stars)
are called inflow, infall, and accretion, respectively.
Infall within the core may form a protostar or protostars.
B335 and
L1544 provided the first observational examples of infall in the protostellar and starless phases,
respectively 
\citep{1993ApJ...404..232Z,1998ApJ...504..900T}.
Time evolution of infall motions in isothermal spheres differs with different initial radial density distributions
\citep{1969MNRAS.145..271L,1977ApJ...214..488S,1977ApJ...218..834H,1992ApJ...394..204Z,1993ApJ...416..303F}.
For example, \cite{2021ApJS..256...25T} suggested that the SCUBA-2 core in Orion studied by \cite{2018ApJS..236...51Y} had radial density profiles 
with a power-lar index considerably shallower than $-$2. 
Our aim was to detect inward motions of either inflow into cores or infall in cores by
investigating kinematics on scales of 0.1 pc,
mainly for starless cores.

This required the estimation of the evolutionary stage of a core.
\cite{2017ApJS..228...12T} and \cite{2020ApJS..249...33K}
developed a chemical evolution factor (CEF), 
a measure of the evolution of the starless core based on the logarithmic deuterium fraction of {\ntd} and DNC.
It is known that the deuterium fractions of molecules formed in the gas phase (e.g., DNC/HNC and {\ntd}/{\nth}) 
increase monotonically 
with core evolution during the starless core phase
\citep{2005ApJ...619..379C, 2006ApJ...646..258H, 2009A&A...496..731E, 2019ApJ...883..202F}.

Another tracer for starless core evolution is 
the distribution of
density. 
The ACA (Atacama Compact Array; a.k.a. the Morita Array) of the ALMA (Atacama Large Millimeter/submillimeter Array)  
is sensitive to the gas concentration near the core center.
\cite{2020ApJ...899...10T} used detectability in the dust continuum emission using the ALMA ACA;
detection of strong continuum with a 6$\farcs$5 (980 au at $\sim$150 pc distance) 
suggested the existence of 
structures with densities of $\ga 8 \times 10^5$~cm$^{-3}$,
and is considered to represent later stages in the starless core phase.
\cite{2021ApJ...907L..15S} reported
the detection of five centrally concentrated dense structures 
with sizes of $\sim$ 2000 au using the
ALMA 12 m Array ($\sim$300 au resolution),
which most likely represent the last stages of starless cores.
Inflow of matter into cores along and/or across filaments was also studied
\citep{2011A&A...533A..34H,2013MNRAS.436.1513,2013A&A...550A..38P,2018ApJ...855....9L,2020ApJ...891...84C,2021ApJ...915L..10S}.

There have been surveys in various star-forming regions for inward motions, which include
\cite{1995ApJ...454..217W}, \cite{1997ApJ...489..719M}, \cite{1999ApJ...526..788L}, \cite{2000ApJ...533..440G}, \cite{2000ApJ...538..260G}, \cite{2005A&A...442..949F},
\cite{2008ApJ...688L..87V},
\cite{2011ApJ...740...40R},
\cite{2018ApJ...861...14C},
\cite{2019ApJ...870....5J},
and
\cite{2021ApJS..254...14Y}.
A popular method is to search for a blue-skewed line profile with a dip in optically thick lines, 
such as HCO$^+$.
Blue-skewed profiles are asymmetric profiles with more intense blue-shifted emission in an optically thicker line with respect to those
in an optically thinner line.
\cite{2000ApJ...533..440G}, who conducted an inward-motion survey
toward Class 0 and Class I protostars using HCO$^+$ $J = 3\rightarrow2$, concluded 
that it was difficult to make unambiguous claims regarding infall,
because that infall velocities were roughly the same speed or less than
turbulent and outflow motions, and also because
the beam size was not high enough to separate outflow from infall.
Clear cases of the signature of infall in terms of redshifted absorption include
observations using the ALMA by
\cite{2015ApJ...814...22E}, \cite{2020ApJ...891...61Y}, and \cite{2021ApJ...909..199O}.
There is a need to investigate whether the observed profiles truly represent inward motions from various aspects.
Contrarily, it is true that a gas with inward motions may not disclose its signature in the observations due to limitation of sensitivity 
and so on.

We selected cores in the Orion molecular clouds 
that have been
previously observed using the ALMA in the
ALMA Survey of Orion Planck Galactic Cold Clumps (ALMASOP) collaboration \citep{2020ApJS..251...20D}
and/or using the Nobeyama 45 m telescope \citep{2020ApJS..249...33K,2021ApJS..256...25T}.
These cores were originally taken from the catalog of cores
identified
by the Submillimetre Common-User Bolometer Array 2 (SCUBA-2) 
inside Planck Galactic cold clumps (PGCCs) in Orion \citep{2018ApJS..236...51Y}.
All the SCUBA-2 cores identified in PGCCs in the SCOPE 
(SCUBA-2 Continuum Observations of Pre-protostellar Evolution) collaboration
\citep{2015PKAS...30...79L,2018ApJS..234...28L} were cataloged by \cite{2019MNRAS.485.2895E}.
Temperatures of PGCCs run from 10 to 17 K.
Please refer to \cite{2011AA...536A..23P} and \cite{2016AA...594A..28P} for PGCCs.
SCUBA-2 cores in Orion have similar distances \citep[420$\pm$40 pc,][]{2017ApJ...834..142K,2019MNRAS.487.2977G}
such that we can avoid distance-dependent effects, e.g., different beam dilutions \citep{2020ApJS..249...33K}.
We mainly selected 30 starless cores at the Nobeyama resolution.
In addition, we also included six protostellar cores at the Nobeyama resolution which
were found to be starless cores at the ALMA ACA resolution (these are treated as protostellar cores in this paper).
Prior to our study, \cite{2008ApJ...688L..87V} surveyed inward motions toward starless cores in Orion;
however, in reality, the samples did not show overlapping except for one core.
They surveyed two sub-regions in the Ori A cloud, whereas
we selected the cores in wider Orion areas containing the
$\lambda$ Ori region, Ori A cloud, and Ori B cloud.
We selected cores in PGCCs, which are known to have low temperatures, suggesting earlier core evolutionary stages,
whereas \cite{2008ApJ...688L..87V} did not use such a selection.
In addition, at the time the aforementioned study was conducted,
information on YSOs  was limited.
Presently, more sensitive YSO information, such as \cite{2016ApJS..224....5F} is available.
For example, 
HOPS-10, -11, -12, -87, -88, -89, -91, -92, -178, -182, and -380
are associated with cores that were regarded as starless by \cite{2008ApJ...688L..87V}.
We found that 52\% of their starless cores are actually associated with YSOs,
and 59\% are associated with YSOs or YSO candidates.
If we exclude YSOs and YSO candidates,
eleven out of their 27 starless cores (41\%) are still starless on the basis of the current YSO information.
Then, overlapping is actually small.
Using the KVN (Korean VLBI Network) 21 m telescope (30$\arcsec$ beam),
\cite{2021ApJS..254...14Y} surveyed inward motions in HCO$^+$ ($J$ = 1$\rightarrow$0) toward
80 cores (45 starless and 35 protostellar) out of the 119 SCUBA-2 cores in Orion \citep{2018ApJS..236...51Y}. 
The current observations demonstrate 1.8 times better spatial resolution.
Observations by \cite{2021ApJS..254...14Y} were made in single pointing toward SCUBA-2 cores, whereas this study mapped cores.
We will examine their work later in this paper.

\section{OBSERVATIONS\label{sec:obs}}

We carried out mapping observations of 33 fields containing 36 SCUBA-2 cores 
using the 45 m radio telescope
of the Nobeyama Radio Observatory\footnote{Nobeyama Radio Observatory
is a branch of the National Astronomical Observatory of Japan,
National Institutes of Natural Sciences} (program ID = CG201001; PI = Gwanjeong Kim). 
Table \ref{tbl:sample} summarizes the numbers of clumps and cores in the related studies.
The observed lines and their rest frequencies 
are listed in Table \ref{tbl:line}. 
The critical and effective densities are adopted from
\cite{2015PASP..127..299S}.
Observations were conducted from 2020 December to 2021 February. 
For the receiver frontend, the ``FOREST'' \citep[FOur-beam REceiver System on the 45 m Telescope, ][]{2016SPIE.9914E..1ZM} 
was employed for simultaneous observations of up to
four lines.
HCO$^+$, H$^{13}$CO$^+$, HNC, and HN$^{13}$C were observed simultaneously.
The half-power beam width (HPBW) and main-beam efficiency $\eta_{mb}$ at 86 GHz were 
18$\farcs2~\pm~0\farcs$3 (0.037~pc at 420~pc distance) and 50.4{\%}~$\pm$~3.2{\%}, respectively. 
For the receiver backend, the ``SAM45'' \citep[Spectral Analysis Machine for the 45 m telescope, ][]{2012PASJ...64...29K} 
was employed with a spectral resolution of 30.52~kHz, which corresponds to $\sim$0.1~km~s$^{-1}$ at 82~GHz.
Table \ref{tbl:line} provides the precise values of the velocity resolution.

On-the-fly (OTF) mapping observations \citep{2008PASJ...60..445S} of 33 fields,
were performed in the R.A. and decl. directions
in most cases to minimize striping effects.  
However, 30\% of the OTF observations were made only in one direction.
The map size was 3{\arcmin}~$\times~$3{\arcmin}.
Three neighboring cores 
(cores 6, 12, and 21) were mapped simultaneously with cores 5, 11, and 20, respectively.
The JCMT SCUBA-2 core names and their coordinates 
were taken from \cite{2018ApJS..236...51Y}; they
are summarized in Table \ref{tbl:coord}.
The association with YSOs
was obtained from \cite{2020ApJS..249...33K}, who made observations using the Nobeyama 45 m telescope.
30 cores out of the 36 observed cores were found to be starless, whereas 
the remaining six cores were protostellar cores.
However, identification of either starless or protostellar cores depends on observational spatial resolutions.
These six protostellar cores were classified as starless with offset YSOs by \cite{2020ApJS..251...20D} 
using the ALMA ACA at a spatial resolution of
$7\farcs6 \times 4\farcs1$.
For the current study, we adopted classification by \cite{2020ApJS..249...33K},
who used the same telescope.
The position switching mode was employed, and
the reference positions were ($\Delta$R.A., $\Delta$decl.) = $(-30\arcmin, 0\arcmin), (20\arcmin, 20\arcmin), (-20\arcmin, -20\arcmin)$,
and $(20\arcmin, 20\arcmin)$ from the reference center of the map, for cores 1$-$12, 13$-$16, 17$-$18, and 19$-$36, respectively.
It took approximately 3 h to achieve an rms noise level of 0.1~K for a 3{\arcmin}~$\times~$3{\arcmin} map. The typical system temperature was 200 K. 
The telescope pointing calibration was performed at 1.0$-$1.5 h intervals toward the SiO maser source, Orion KL, which resulted in a pointing accuracy of $\lesssim$5$\arcsec$. 

Linear baselines were subtracted from the spectral data, and the data were stacked on a grid of 6$\arcsec$ with the Bessel$-$Gauss function in the NOSTAR program
of the Nobeyama Radio Observatory \citep{2008PASJ...60..445S}. 
The R.A. and decl. scan data were basket-weaved using the method proposed by \cite{1988A&A...190..353E}.
The line intensity was expressed in terms of the antenna temperature $T_{\rm A}^{\ast}$ corrected for atmospheric extinction using standard chopper wheel calibration.

We also used the N$_2$H$^+$ data from our previous observations reported in \cite{2021ApJS..256...25T} in this analysis to avoid redundant N$_2$H$^+$ observations. 
In addition, we adopted the SCUBA-2 core properties
such as the HWHM (half-width-at-half-maximum) radius, mass, and density from \cite{2018ApJS..236...51Y} (Table \ref{tbl:coord}). 
We converted 
the FWHM (full-width-at-half-maximum) diameter 
of \cite{2018ApJS..236...51Y} to  HWHM radius $R$ by dividing by two.
\cite{2018ApJS..236...51Y} adopted distances of
380 pc and 420 pc for the $\lambda$ Ori region and the Ori A and B regions taken from
\cite{1997A&A...323L..49P} and a combination of \cite{2007MNRAS.376.1109J} and \cite{2007ApJ...667.1161S}, respectively;
we adopted their physical parameters based on these assumed distances.

\floattable
\begin{deluxetable}{lrrl}
\tablecaption{Clump and Core Samples \label{tbl:sample}}
\tablecolumns{4}
\tablenum{1}
\tablewidth{0pt}
\tablehead{
\colhead{Category} &
\colhead{Clump Number} &
\colhead{Core Number} &
\colhead{Reference} 
}
\startdata
PGCC                                               &  13188  &   . . .    &  \cite{2016AA...594A..28P}   \\
SCUBA-2 core in PGCC  (SCOPE)          &  558         &   3528      &  \cite{2019MNRAS.485.2895E}   \\
SCUBA-2 core in Orion                       &   40       &   119    &  \cite{2018ApJS..236...51Y}  \\
SCUBA-2 core in Orion studied here   &   . . .            &  36    &   This work \\
\enddata
\end{deluxetable}

\floattable
\begin{deluxetable}{lcccccc}
\rotate
\tablecaption{Observed Lines \label{tbl:line}}
\tablecolumns{7}
\tablenum{2}
\tablewidth{0pt}
\tablehead{
\colhead{Line} &
\colhead{Frequency} &
\colhead{Frequency Reference}&
\colhead{Velocity Resolution}&
\colhead{Upper Energy Level $E_u$}&
\colhead{Critical Density at 10 K}&
\colhead{Effective Density at 10 K}\\
\colhead{} &
\colhead{GHz} &
\colhead{}&
\colhead{km~s$^{-1}$}&
\colhead{K}&
\colhead{cm$^{-3}$}&
\colhead{cm$^{-3}$}
}
\startdata
HCO$^+$ $J$ = 1$\rightarrow$0            &89.1885230 &\citet{1998JQSRT..60..883P} & 0.103 & 4.3    & 6.8$\times10^4$  &9.5$\times10^2$    \\
H$^{13}$CO$^+$ $J$ = 1$\rightarrow$0   &86.7542884 &\citet{1998JQSRT..60..883P} & 0.105 & 4.2   &  6.2$\times10^4$ & 3.9$\times10^4$  \\
HNC                 $J$ = 1$\rightarrow$0   &90.6635930 &\citet{1998JQSRT..60..883P} & 0.101 & 4.4   &  1.4$\times10^5$ &1.0$\times10^4$   \\
HN$^{13}$C        $J$ = 1$\rightarrow$0   &87.0908500 &\citet{1998JQSRT..60..883P} & 0.105 & 4.2   &  . . .                   & . . .                   \\
N$_2$H$^+$       $J$ = 1$\rightarrow$0   &93.1737767 &\citet{1995ApJ...455L..77C}  &0.098 & 4.5    &  6.1$\times10^4$ &6.8$\times10^4$   \\
\enddata
\end{deluxetable}

\floattable
\begin{deluxetable}{cllllcrrr}
\rotate
\tablecaption{Name, Coordinates, SCUBA-2 Radius, and Mass of the SCUBA-2 Core \label{tbl:coord}}
\tablecolumns{9}
\tablenum{3}
\tablewidth{0pt}
\tablehead{
\colhead{Core Number} &
\colhead{JCMT Core Name} &
\colhead{R.A. (J2000)}&
\colhead{Decl. (J2000)} &
\colhead{YSO Ass.}&
\colhead{ALMA Source Name} &
\colhead{$R$} &
\colhead{$M$} &
\colhead{$n{\rm (H_2)}$} \\
\colhead{} &
\colhead{} &
\colhead{h~m~s}&
\colhead{$\arcdeg~\arcmin~\arcsec$}&
\colhead{} &
\colhead{} &
\colhead{pc} &
\colhead{$M_{\odot}$} &
\colhead{10$^5$ cm$^{-3}$}
}
\startdata
1  &  G191.90$-$11.21North  &  5  31  28.99   &  12  58  55.0   &  starless  &  G191.90$-$11.21N    &  0.050   &  0.34  $\pm$  0.07  &  0.8   $\pm$  0.2  \\
2  &  G198.69$-$09.12North1  &  5  52  29.61   &  8  15  37.0   &  starless  &  G198.69$-$09.12N1  &  0.045   &  0.43  $\pm$  0.07  &  2.0   $\pm$  0.4  \\
3  &  G198.69$-$09.12North2  &  5  52  25.30   &  8  15  8.8   &  starless  &  G198.69$-$09.12N2  &  0.030   &  0.21  $\pm$  0.02  &  2.1   $\pm$  0.4  \\
4  &  G201.72$-$11.22  &  5  50  54.53   &  4  37  42.6   &  starless  &  . . .  &  1.125   &  1.83  $\pm$  0.21  &  1.3   $\pm$  0.4  \\
5  &  G203.21$-$11.20East2  &  5  53  47.90   &  3  23  8.9   &  starless  &  G203.21$-$11.20E2  &  0.060   &  2.65   $\pm$  1.73   &  10.3   $\pm$  0.7  \\
6  &  G203.21$-$11.20East1  &  5  53  51.11   &  3  23  4.9   &  starless  &  G203.21$-$11.20E1  &  0.060   &  2.50   $\pm$  0.43   &  9.8   $\pm$  0.6  \\
7  &  G205.46$-$14.56North1  &  5  46  5.49   &  0  9  32.4   &  starless  &  G205.46$-$14.56M3  &  0.015   &  0.39   $\pm$  0.06   &  17.0   $\pm$  1.9  \\
8  &  G206.21$-$16.17North  &  5  41  39.28   &  -1  35  52.9   &  starless  &  G206.21$-$16.17N  &  0.105   &  4.97   $\pm$  0.05   &  4.7   $\pm$  0.8  \\
9  &  G206.21$-$16.17South  &  5  41  34.23   &  -1  37  28.8   &  starless  &  G206.21$-$16.17S  &  0.055   &  1.05   $\pm$  0.46   &  3.6   $\pm$  0.8  \\
10  &  G206.93$-$16.61East1  &  5  41  40.54   &  -2  17  4.3   &  starless  &  G206.93$-$16.61E1  &  0.060   &  5.37   $\pm$  0.54   &  17.1   $\pm$  2.9  \\
11  &  G206.93$-$16.61West4  &  5  41  25.84   &  -2  19  28.4   &  starless  &  . . .  &  0.040   &  1.25   $\pm$  0.67   &  9.0   $\pm$  0.2  \\
12  &  G206.93$-$16.61West5  &  5  41  28.77   &  -2  20  4.3   &  starless  &  G206.93$-$16.61W4  &  0.055   &  7.10   $\pm$  0.72   &  23.5   $\pm$  1.6  \\
13  &  G207.36$-$19.82North2  &  5  30  50.67   &  -4  10  15.6   &  protostellar  &  G207.36$-$19.82N2  &  0.020   &  0.44   $\pm$  0.14   &  4.9   $\pm$  0.9  \\
14  &  G207.36$-$19.82North4  &  5  30  44.81   &  -4  10  27.6   &  starless  &  G207.36$-$19.82N4  &  0.015   &  0.15   $\pm$  0.05   &  4.2   $\pm$  0.8  \\
15  &  G207.36$-$19.82South  &  5  30  46.81   &  -4  12  29.4   &  starless  &  G207.36$-$19.82S  &  0.095   &  2.01   $\pm$  0.91   &  1.1   $\pm$  0.2  \\
16  &  G208.68$-$19.20North2  &  5  35  20.45   &  -5  0  53.0   &  protostellar  &  G208.68$-$19.20N2  &  0.025   &  2.22   $\pm$  1.15   &  18.9   $\pm$  3.6  \\
17  &  G209.05$-$19.73North  &  5  34  3.96   &  -5  32  42.5   &  starless  &  . . .  &  0.115   &  2.83   $\pm$  0.15   &  0.9   $\pm$  0.6  \\
18  &  G209.05$-$19.73South  &  5  34  3.12   &  -5  34  11.0   &  starless  &  . . .  &  0.070   &  1.65   $\pm$  0.29   &  1.6   $\pm$  0.4  \\
19  &  G209.29$-$19.65North1  &  5  35  0.25   &  -5  40  2.4   &  starless  &  G209.29$-$19.65N1  &  0.035   &  0.70   $\pm$  0.10   &  2.4   $\pm$  0.8  \\
20  &  G209.29$-$19.65South1  &  5  34  55.99   &  -5  46  3.2   &  starless  &  G209.29$-$19.65S1  &  0.035   &  1.49   $\pm$  0.26   &  5.1   $\pm$  0.4  \\
21  &  G209.29$-$19.65South2  &  5  34  53.81   &  -5  46  12.8   &  starless  &  G209.29$-$19.65S2  &  0.040   &  2.31   $\pm$  0.97   &  6.2   $\pm$  0.5  \\
22  &  G209.55$-$19.68North2  &  5  35  7.01   &  -5  56  38.4   &  starless  &  G209.55$-$19.68N2  &  0.025   &  0.28   $\pm$  0.07   &  3.1   $\pm$  0.6  \\
23  &  G209.77$-$19.40West  &  5  36  21.19   &  -6  1  32.7   &  starless  &  G209.77$-$19.40W  &  0.045   &  0.50   $\pm$  0.07   &  1.7   $\pm$  0.2  \\
24  &  G209.77$-$19.40East2  &  5  36  32.19   &  -6  2  4.7   &  starless  &  G209.77$-$19.40E2  &  0.035   &  1.20   $\pm$  0.56   &  6.4   $\pm$  0.9  \\
25  &  G209.77$-$19.40East3  &  5  36  35.94   &  -6  2  44.7   &  starless  &  G209.77$-$19.40E3  &  0.020   &  0.26   $\pm$  0.12   &  5.4   $\pm$  0.7  \\
26  &  G209.94$-$19.52North  &  5  36  11.55   &  -6  10  44.8   &  protostellar  &  G209.94$-$19.52N  &  0.065   &  2.81   $\pm$  0.28   &  3.1   $\pm$  0.5  \\
27  &  G209.79$-$19.80West  &  5  35  11.19   &  -6  14  0.7   &  starless  &  G209.79$-$19.80W  &  0.135   &  7.06   $\pm$  2.56   &  4.4   $\pm$  0.4  \\
28  &  G209.94$-$19.52South1  &  5  36  24.96   &  -6  14  4.7   &  starless  &  G209.94$-$19.52S1  &  0.080   &  3.52   $\pm$  0.09   &  2.5   $\pm$  0.2  \\
29  &  G210.37$-$19.53North  &  5  36  55.03   &  -6  34  33.2   &  starless  &  G210.37$-$19.53N  &  0.030   &  0.28   $\pm$  0.02   &  1.1   $\pm$  0.2  \\
30  &  G210.82$-$19.47North2  &  5  37  59.84   &  -6  57  9.9   &  starless  &  G210.82$-$19.47N2  &  0.030   &  0.15   $\pm$  0.05   &  0.7   $\pm$  0.1  \\
31  &  G211.16$-$19.33North5  &  5  38  46.00   &  -7  10  41.9   &  protostellar  &  G211.16$-$19.33N5  &  0.060   &  0.64   $\pm$  0.18   &  1.2   $\pm$  0.2  \\
32  &  G211.16$-$19.33North3  &  5  39  2.26   &  -7  11  7.9   &  starless  &  G211.16$-$19.33N3  &  0.050   &  0.41   $\pm$  0.12   &  1.0   $\pm$  0.1  \\
33  &  G211.16$-$19.33North4  &  5  38  55.67   &  -7  11  25.9   &  protostellar  &  G211.16$-$19.33N4  &  0.065   &  0.51   $\pm$  0.04   &  0.7   $\pm$  0.1  \\
34  &  G211.72$-$19.25South1  &  5  40  19.04   &  -7  34  28.8   &  starless  &  G211.72$-$19.25S1  &  0.060   &  1.15   $\pm$  0.82   &  1.3   $\pm$  0.5  \\
35  &  G212.10$-$19.15North1  &  5  41  21.56   &  -7  52  27.7   &  protostellar  &  G212.10$-$19.15N1  &  0.090   &  3.52   $\pm$  1.68   &  3.1   $\pm$  0.4  \\
36  &  G215.44$-$16.38  &  5  56  58.45   &  -9  32  42.3   &  starless  &  G215.44$-$16.38  &  0.030   &  0.19   $\pm$  0.04   &  1.0   $\pm$  0.2  \\
\enddata
\end{deluxetable}

\section{RESULTS \label{sec:res}}

\subsection{Infall Velocity and  Asymmetry Parameter}

We searched for observational evidence of inward motions.
The double-peaked profile due to absorption by foreground lower excitation-temperature gas
is called ``self-reversal''
\citep{1976ApJ...209..466L,1980ApJ...241..676B}.
The ``self-reversal'' profile contains a dip (suppression) between two emission peaks.
If the cloud or cloud core has systematic motions with respect to the core center
such as inward or outward motions, the profile of an optically thick line
will exhibit asymmetry in shape.
The profile may also have an absorption dip. 
Identification of inward or outward motions needs careful investigation
through a comparison of line profiles on the optically thick and thin lines and also through
spatial distribution of the line profile shape across the sky.
Later, we explained how we investigated these motions.
Cases of infall were explained in detail by \cite{1992ApJ...394..204Z} and \cite{1993ApJ...404..232Z}.
In such cases, we will see blue-skewed line profiles with dips, whose lower-velocity (blue)
peaks are brighter than its higher-velocity (red) peaks.
Starless cores are likely to have kinetic-temperature gradients decreasing
inward toward the core center \citep{2001ApJ...557..193E}; 
however, if the radial density gradient decreasing outward is adequately steep,
cores may have excitation-temperature gradients decreasing outward, which will cause
self-reversal or blue-skewed line profiles with dips.

Table \ref{tbl:intensity}
summarizes the peak intensities and peak velocities 
(velocities of the peak intensities)
of the molecular lines toward the SCUBA-2 core centers.
The peak was determined from the spectrum without any fitting; thus, the nominal peak velocity accuracy is equal 
to the spectral resolution (0.1~km~s$^{-1}$).
Differences between the line velocities facilitate identification of
possible inward or outward motions.
Table \ref{tbl:veldiff} summarizes
velocity differences.
The HCO$^+$ line is often optically thick, based on a comparison with the H$^{13}$CO$^+$ emission in
intensity and line shape, and is considered to
represent gas motions in the line shape such as line profile asymmetry
\citep{2000ApJ...533..440G,2000ApJ...538..260G,2005A&A...442..949F,2008ApJ...688L..87V,2011ApJ...740...40R,2019ApJ...870....5J,2021ApJS..254...14Y}.
The velocity difference between the HCO$^+$ and H$^{13}$CO$^+$ peak velocities
might reflect whether the gas may move inward or outward,
if the sign of the velocity difference is negative or positive, respectively.
Some cores may oscillate \citep{2003ApJ...586..286L,2007ApJ...665..457A}, although the spatial resolution of the current observations with respect to the core size
may be insufficient to identify it.
As an independent measure, we also used 
velocity differences between HCO$^+$ and N$_2$H$^+$ ($F_1, F$ = 0, 1$\rightarrow$1, 2)
and between HNC and HN$^{13}$C.
The N$_2$H$^+$ ($F_1, F$ = 0, 1$\rightarrow$1, 2) and HN$^{13}$C line emissions are optically thinner than the
HCO$^+$ and HNC line emissions, respectively.
The N$_2$H$^+ J = 1\rightarrow0$ emission exhibits hyperfine splitting \citep{1995ApJ...455L..77C}, and
we adopted the isolated hyperfine component 
($F_1, F$ = 0, 1$\rightarrow$1, 2) to avoid overlapping with neighboring components.
The absolute value of the velocity difference ranged up to 0.9~km~s$^{-1}$.
\cite{1999ApJ...526..788L} concluded that inward motions in low-mass starless cores were subsonic, which is not consistent with our conclusion.
\cite{2005A&A...442..949F} studied cores associated with candidate high-mass protostars and obtained large values of inward motions
of 0.1$-$1~km~s$^{-1}$.
Core properties including velocity differences may depend on the core radius and mass.
We compared the radii and masses of these samples.
The HWHM radius and mass of our Orion cores, which are located in an intermediate mass star-forming region, 
were 0.015$-$1.125~pc (with a mean of 0.083~pc and a median of 0.050 pc) 
and 0.15$-7.1~M_{\odot}$ (with a mean of 1.8~$M_{\odot}$ and a median of 1.2~$M_{\odot}$),
respectively.
\cite{1997ApJ...489..719M}, \cite{1999ApJ...526..788L}, and \cite{2000ApJ...538..260G} studied cores
in low-mass star-forming regions, and some of them were observed by \cite{2005MNRAS.360.1506K} 
with the SCUBA-2.
The HWHM radius and mass of these cores were found to be 
0.008$-$0.37 pc (with a mean of 0.020~pc and a median of 0.020~pc) 
and 0.20$-$1.6~$M_{\odot}$ (with a mean of 0.55~$M_{\odot}$ and a median of 0.35~$M_{\odot}$),
respectively.
Our Orion cores were larger and more massive than their dark cloud cores.
The cores in high-mass star-forming regions that were studied by \cite{2005A&A...442..949F}
were mostly distributed in a mass range of 0$-100 M_{\odot}$ with a distribution tail to an even higher mass, and thus,
they were more massive than our Orion cores.

It is to be noted that we need to consider various aspects to determine whether spectra mean inward or outward motions.
It is possible that all the observed velocity shifts may not represent
the velocity field of a single core.
The existence of multiple velocity components may affect the velocity difference.
Table \ref{tbl:velcomp} lists the number of velocity components observed in
H$^{13}$CO and N$_2$H$^+$ ($F_1, F$ = 0, 1$\rightarrow$1, 2).
Later, we investigated the molecular line spectra to determine the effects of multiple velocity components.

\begin{figure}
\epsscale{0.7}
\figurenum{1}
\includegraphics[bb=0 0 505 800, width=10cm]{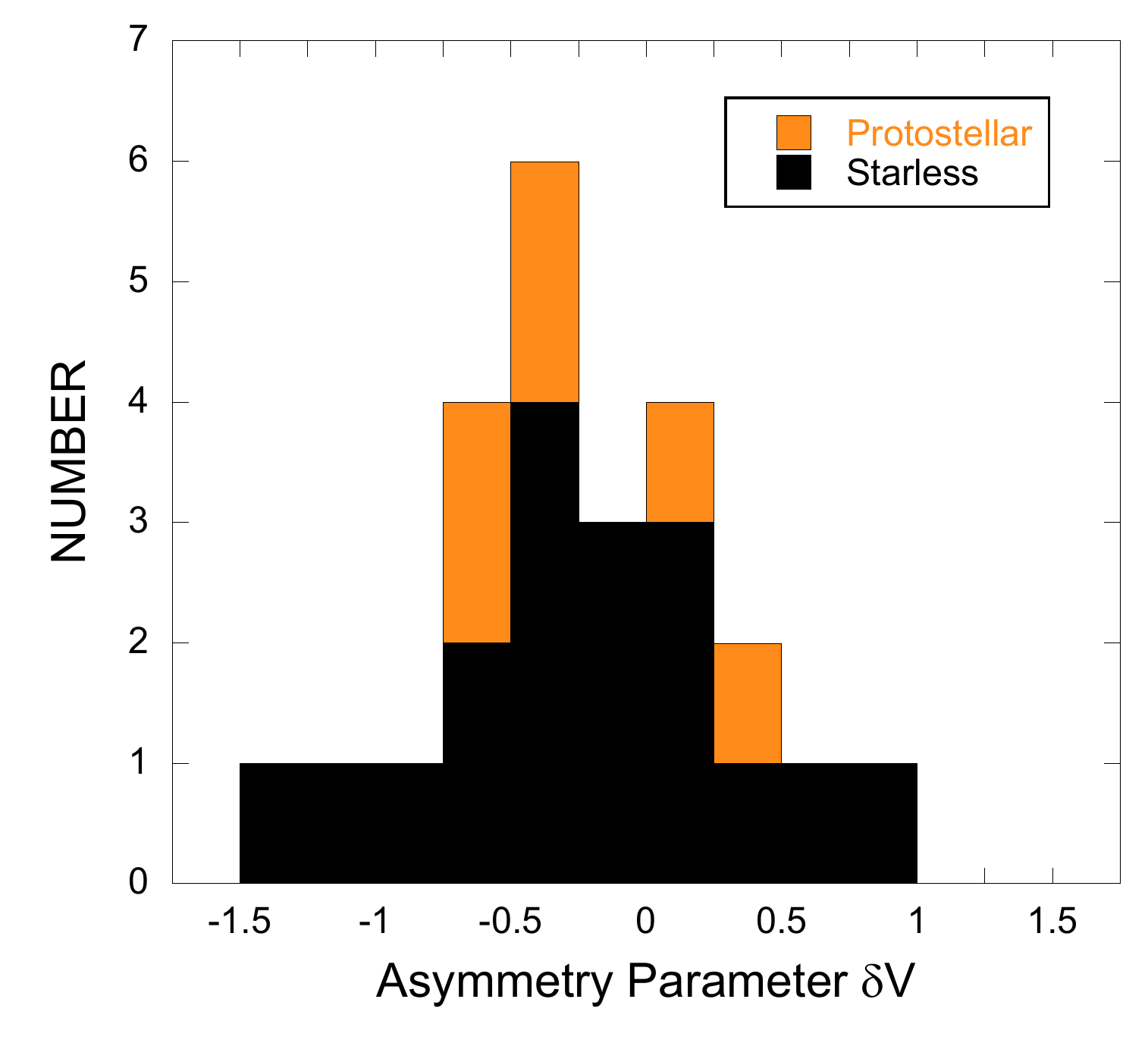}
\caption{Histogram of the asymmetry parameter (line velocity difference normalized to the H$^{13}$CO$^+$ FWHM line width) $\delta V$
between HCO$^+$ and H$^{13}$CO$^+$ for the starless and protostellar cores.
Here, we do not include cores that show multiple velocity components in H$^{13}$CO$^+$ ($\delta V$ in parentheses in Table \ref{tbl:veldiff}).
\label{fig:AsymmParmHist}}
\end{figure}

We adopted the asymmetry parameter $\delta V$ \citep{1997ApJ...489..719M},
which is the velocity difference between the optically thicker and thinner lines, normalized to the FWHM line width of the optically thinner line.
Significant detection of inward/outward motions is judged to be achieved if the absolute value of the asymmetry parameter exceeds the criterion of 0.25.
The criterion of 0.25 was chosen by them to screen
out random contributions, and has often been used by following researches based on the asymmetry parameter.
We defined our asymmetry parameter as $\delta V = [v$(HCO$^+$)$-v$(H$^{13}$CO$^+$)]/$\Delta v$(H$^{13}$CO$^+$),
and listed it in Table \ref{tbl:veldiff}.
Here, $\Delta v$ is the FWHM line width.
The number in parentheses indicates that multiple velocity components are observed in H$^{13}$CO$^+$.
The error in $\delta V$ mainly originates from the velocity difference [$v$(HCO$^+$)$-v$(H$^{13}$CO$^+$)].
The uncertainty in $\delta V$ is approximately 30\%, considering the nominal velocity resolution (0.1~km~s$^{-1}$),
the effect of the signal-to-noise (S/N) ratio on the accuracy of the velocity difference, the
accuracy in velocity of H$^{13}$CO$^+$ Gaussian fitting ($<$ 0.1~km~s$^{-1}$), and the typical H$^{13}$CO$^+$ line width.
The absolute values of our asymmetry parameters ranging up to 
1.38
were similar to 
those in \cite{1997ApJ...489..719M}, who studied low-mass molecular cloud cores containing candidate protostars 
at a convolved spatial resolution of 37$\arcsec$ (0.027 pc at a distance of 150 pc).
Their maximum absolute value was 1.16.
\cite{2000ApJ...538..260G} studied starless cores at a spatial resolution of 26$\arcsec$  (0.019 pc at 150 pc distance) and the maximum absolute value of
the asymmetry parameter was 0.68, which was smaller than 
our maximum absolute value.
Figure \ref{fig:AsymmParmHist} shows the histogram of the asymmetry parameter $\delta V$
for the starless and protostellar cores.
The distribution was more skewed to negative values, suggesting that
inward motions are more dominant than outward motions.
\cite{1999ApJ...526..788L} studied low-mass starless cores with a spatial resolution of 27$\arcsec$
 (0.020 pc at 150 pc distance),
and obtained 
a similar negative bias in the asymmetry parameter.
The asymmetry parameter runs from $-$1.3 to 0.7, which is similar to our range.
\cite{2008ApJ...688L..87V} studied 27 cores in Orion in HCO$^+$ $J$ = 3$\rightarrow$2
and reported inward motions in nine cores and outward motions in ten cores
at a spatial resolution of 26$\arcsec$ (0.053 pc at 420 pc distance).
Their $\delta V$ ranged from $-$1.25 to 0.86 with a bimodal distribution of blue and red.
They regarded all the observed cores as starless; however, protostellar cores could be detected if we used the updated YSO information as noted previously.
There is only one core overlapping between our and their Orion cores.
Their ORI1\_9 and core 16 are the same protostellar cores, and they consistently show red asymmetry suggesting expansion.
The $\delta V$ values obtained for them are 0.35 and 0.40, respectively, which are judged to be consistent.
\cite{2019ApJ...870....5J} studied high-mass, dense molecular clumps
and reported
an overall blue asymmetry 
with a large sample of
1093 sources at a spatial resolution of 38$\arcsec$.

We started our research with starless cores. We attempted to identify evidence of inward motion in the starless cores. 
First, 
we investigated inward motions by comparing the peak velocities of the optically thick and thin lines.
We
focused on cores whose $v$(HCO$^+$)$-v$(H$^{13}$CO$^+$) and $v$(HCO$^+$)$-v$(N$_2$H$^+$) 
are both $\leqq-0.2$~km s$^{-1}$.
We assume that negative shifts
more than twice the velocity resolution (0.1 km s$^{-1}$) are significant.
Starless cores 5, 10, 11, 14, 17, 18, 20, 21, 23, 27, 28, and 32
(a total of 12 cores out of 30, or 40\%)
satisfy this condition.
Measurements of peak velocities were rather straightforward.
The velocity differences between 
$v$(HCO$^+$)$-v$(H$^{13}$CO$^+$), 
$v$(HCO$^+$)$-v$(N$_2$H$^+$), and 
$v$(HNC)$-v$(HN$^{13}$C) in the Nobeyama observations (Table \ref{tbl:veldiff})
were rather consistent.
Then, we judged that the S/N ratio would not significantly affect the measurements of the velocity differences.
We judged that the velocity difference was measured with precision corresponding to the instrumental velocity resolution.

\begin{figure}
\epsscale{0.7}
\figurenum{2}
\includegraphics[bb=0 0 600 400,width=23cm,angle=90]{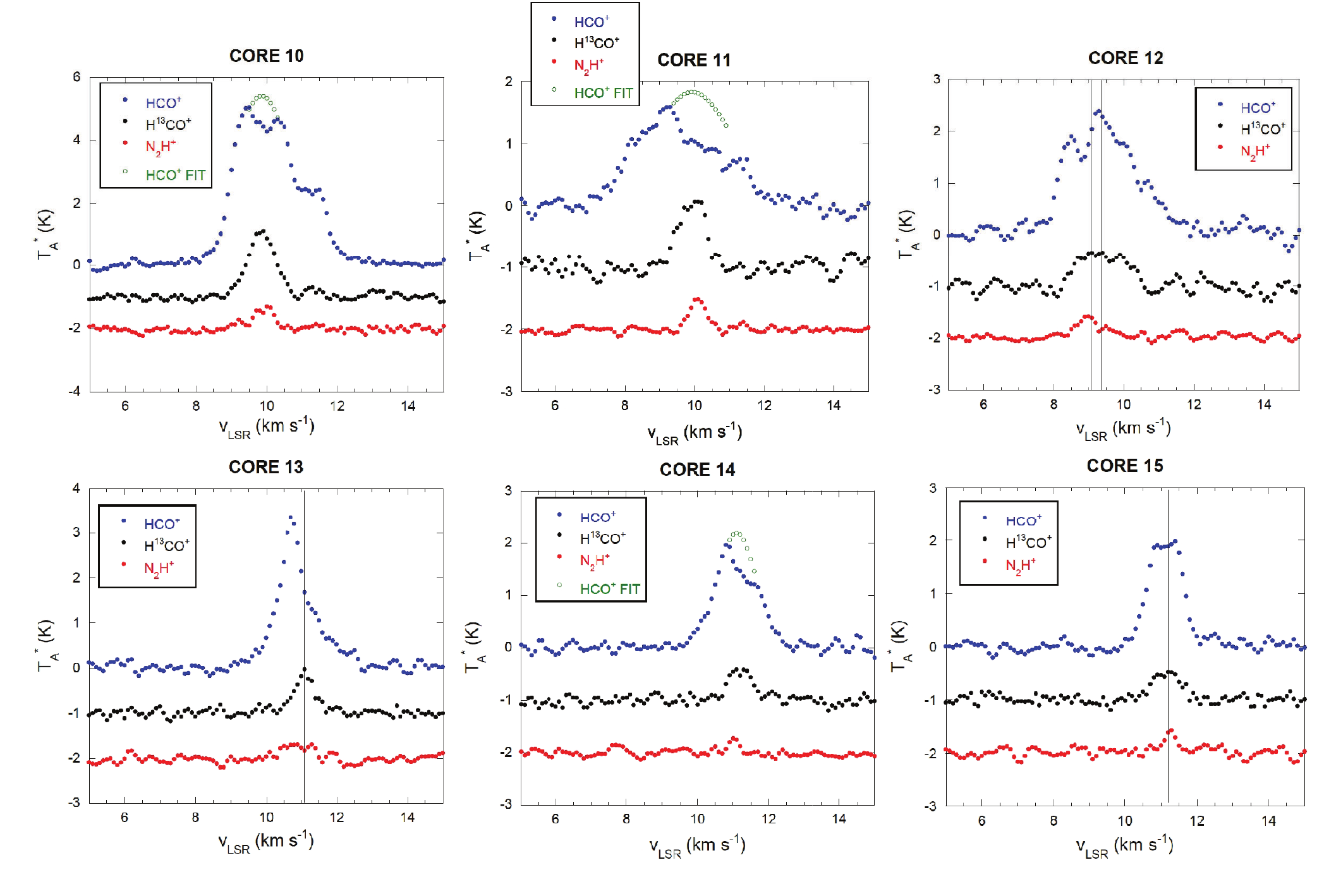}
\caption{The $J$ = 1$\rightarrow$0 line profiles of
HCO$^+$ (filled blue circles), H$^{13}$CO$^+$ (filled black circles), and N$_2$H$^+$ (filled red circles) toward cores 
10, 11, 12, 13, 14 and 15.  
The open green circles show interpolated Gaussian fitting to the HCO$^+$ profile
by fixing the line center velocity to 
the Gaussian fitted peak velocity of the 
H$^{13}$CO$^+$ emission for the blue-skewed profile cores.
The vertical line shows the peak velocity of the 
H$^{13}$CO$^+$ emission for the other cores,
for which Gaussian interpolation to HCO$^+$ was not made.
\label{fig:PP10-15SP}}
\end{figure}

\begin{figure}
\epsscale{0.7}
\figurenum{3}
\includegraphics[bb=0 0 600 400,width=24cm,angle=90]{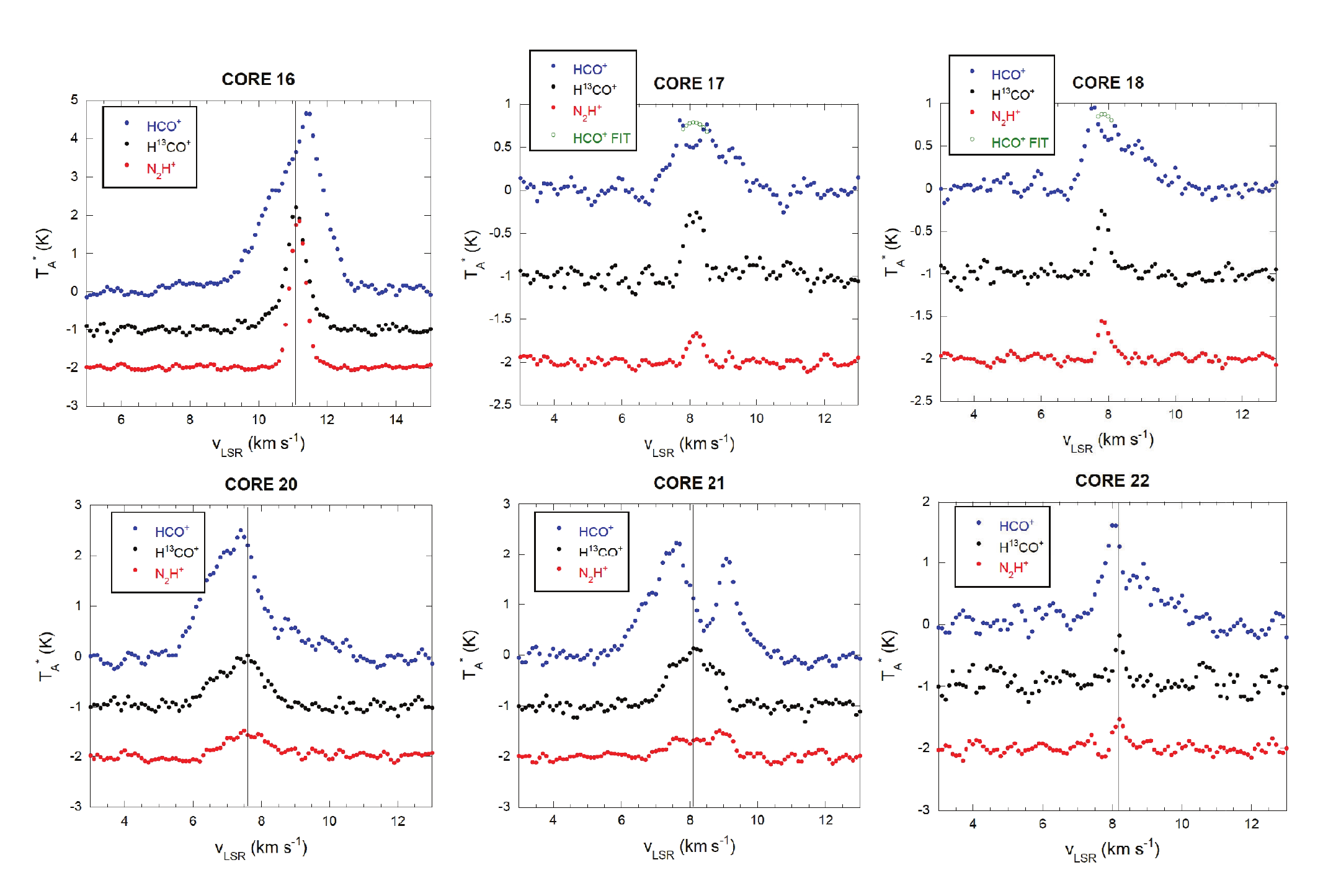}
\caption{
Same as Figure \ref{fig:PP10-15SP} 
except for cores 16, 17, 18, 20, 21, and 22.
\label{fig:PP16-22SP}}
\end{figure}

\begin{figure}
\epsscale{0.7}
\figurenum{4}
\includegraphics[bb=0 0 800 600,width=22cm,angle=90]{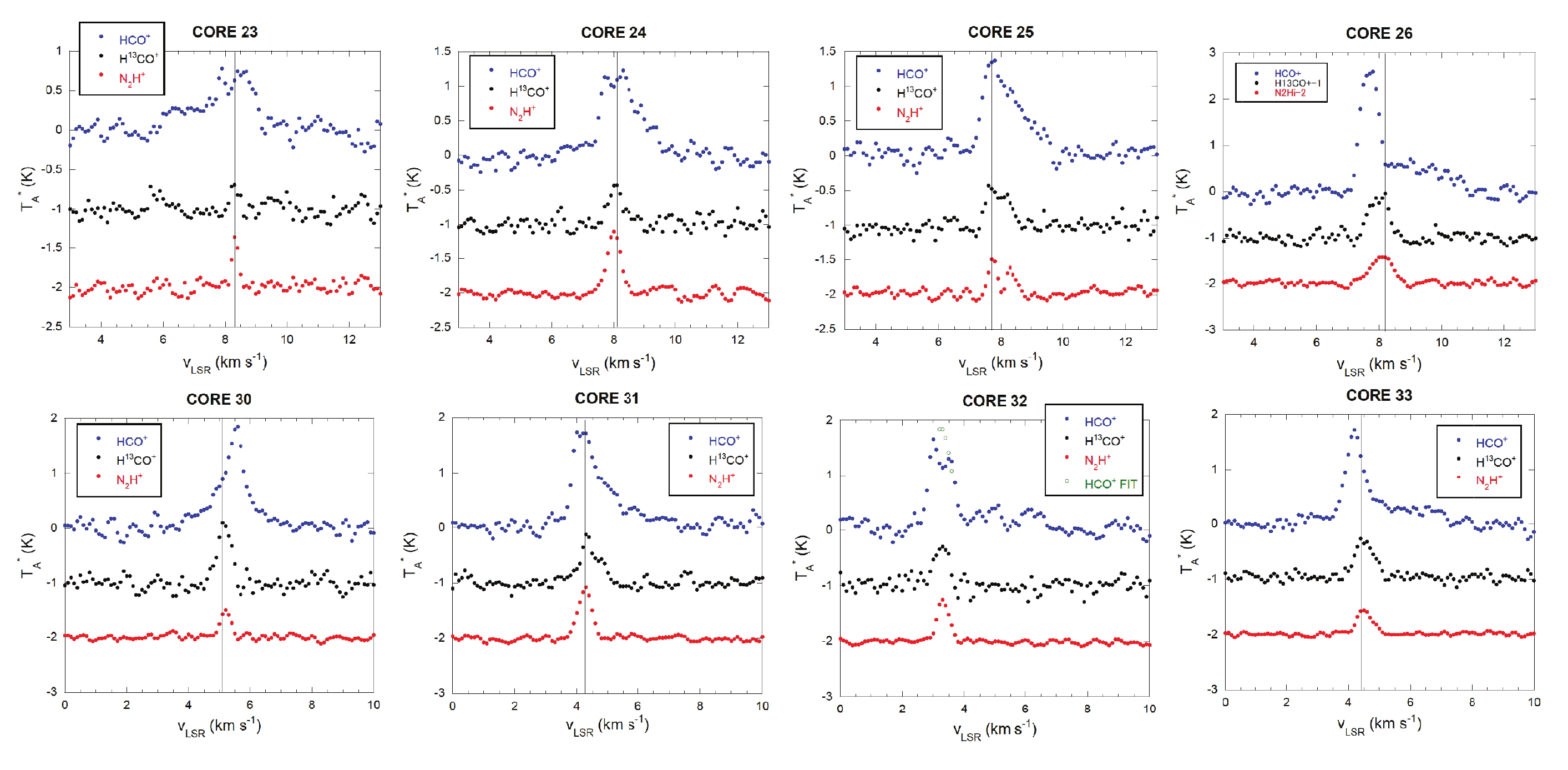}
\caption{
Same as Figure \ref{fig:PP10-15SP}
except for cores 23, 24, 25, 26, 30, 32, and 33.
\label{fig:PP23-33SP}}
\end{figure}

Second, we investigated the HCO$^+$, H$^{13}$CO$^+$, and N$_2$H$^+$ profiles.
To improve the S/N ratio, we collected data within a radius of 10$\arcsec$,
which is slightly larger than the telescope beam $radius$, 
on the Bessel--Gauss stacked pixel data.
If the absorption is not sufficiently strong, absorption may cause line asymmetry without a dip in the profile.
The blue-skewed line profile with a dip provides stronger evidence of possible inward motion than simple line profile asymmetry.
Figures
\ref{fig:PP10-15SP} to \ref{fig:PP23-33SP}
shows the 
HCO$^+$, H$^{13}$CO$^+$, and N$_2$H$^+$ ($F_1, F$ = 0, 1$\rightarrow$1, 2) profiles toward
cores which possibly show either blue-skewed or red-skewed
HCO$^+$ profiles.
Table \ref{tbl:fig} summarizes which figure corresponds to which core.
The relative intensity of the 1st and 2nd intense peaks would be reliable,
and we judged either blue-skewed or red-skewed if
the intensity difference was significant w.r.t. the noise level.
For possibly blue-skewed profiles, we tried Gaussian interpolation
by fixing the Gaussian central velocity to that of optically thin H$^{13}$CO$^+$
emission.
When we obtained successful Gaussian interpolation, we judged that the profile had a dip.
Toward cores 16-18, the H$^{13}$CO$^+$ emission could be optically thick, because their intensities were close to those of HCO$^+$.
Exceptionally, cores 12 and 25 had inconsistent velocity differences of $v$(HCO$^+$)$-v$(H$^{13}$CO$^+$) and $v$(HCO$^+$)$-v$(N$_2$H$^+$).
Core 12 had fairly different line shapes and line widths between H$^{13}$CO$^+$ and N$_2$H$^+$.
Core 25 had two velocity components in  H$^{13}$CO$^+$ and N$_2$H$^+$, and it seems that different optical depths of the lines 
result in different velocity differences.
Among the 12 starless cores mentioned in the previous paragraph, 
we saw hints of the blue-skewed line profile
in cores 10, 11, 14, 18, 21, and 32.

\begin{figure}
\epsscale{0.6}
\figurenum{5}
\includegraphics[bb=0 0 800 600,width=20cm,angle=90]{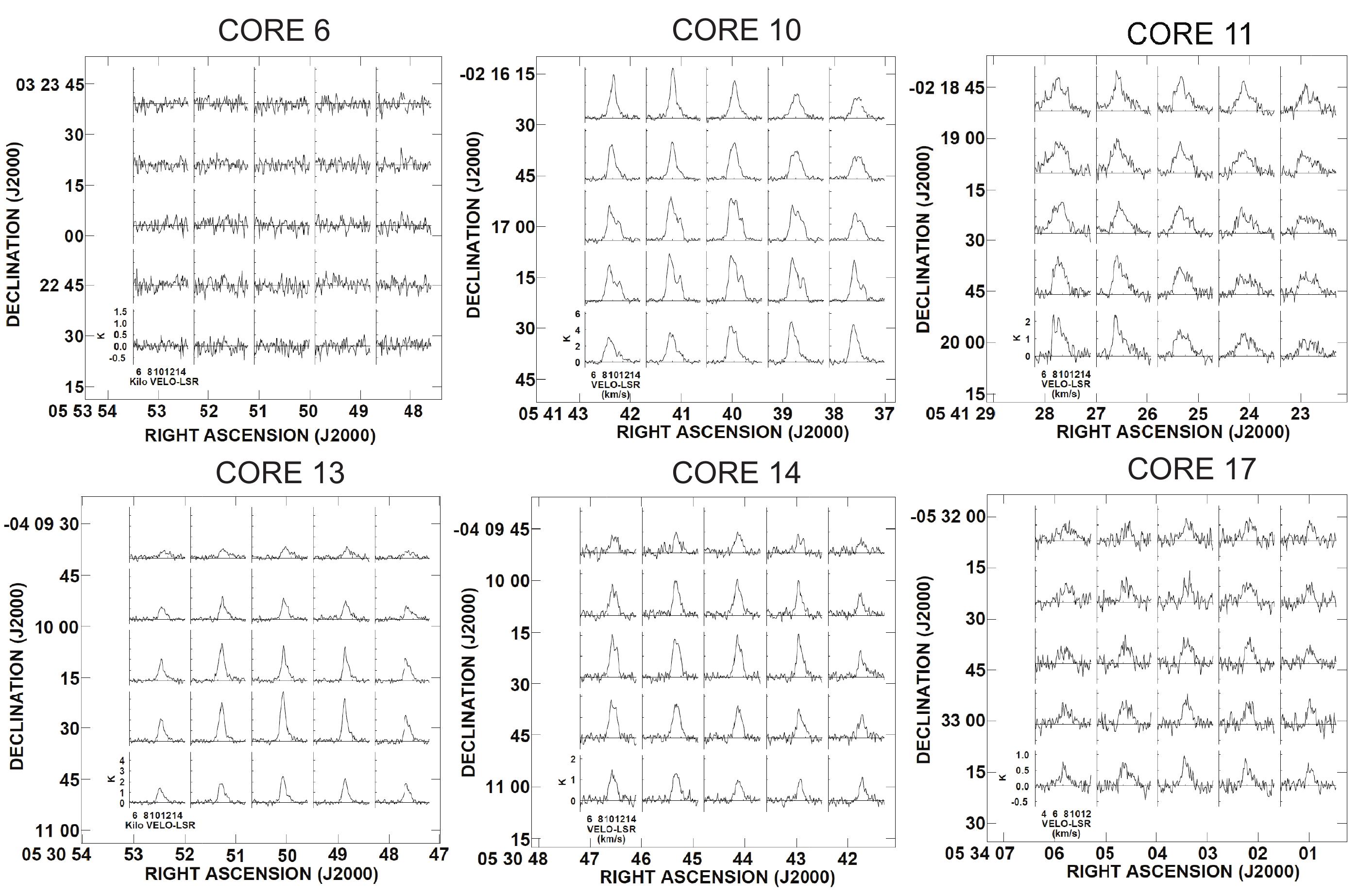}
\caption{
Stamp map of the HCO$^+$ emission toward
cores 6, 10, 11, 13, 14, and 17.
\label{fig:SS6-17-HCO-PLCUB}}
\end{figure}

\begin{figure}
\epsscale{0.6}
\figurenum{6}
\includegraphics[bb=0 0 800 500,width=23cm,angle=90]{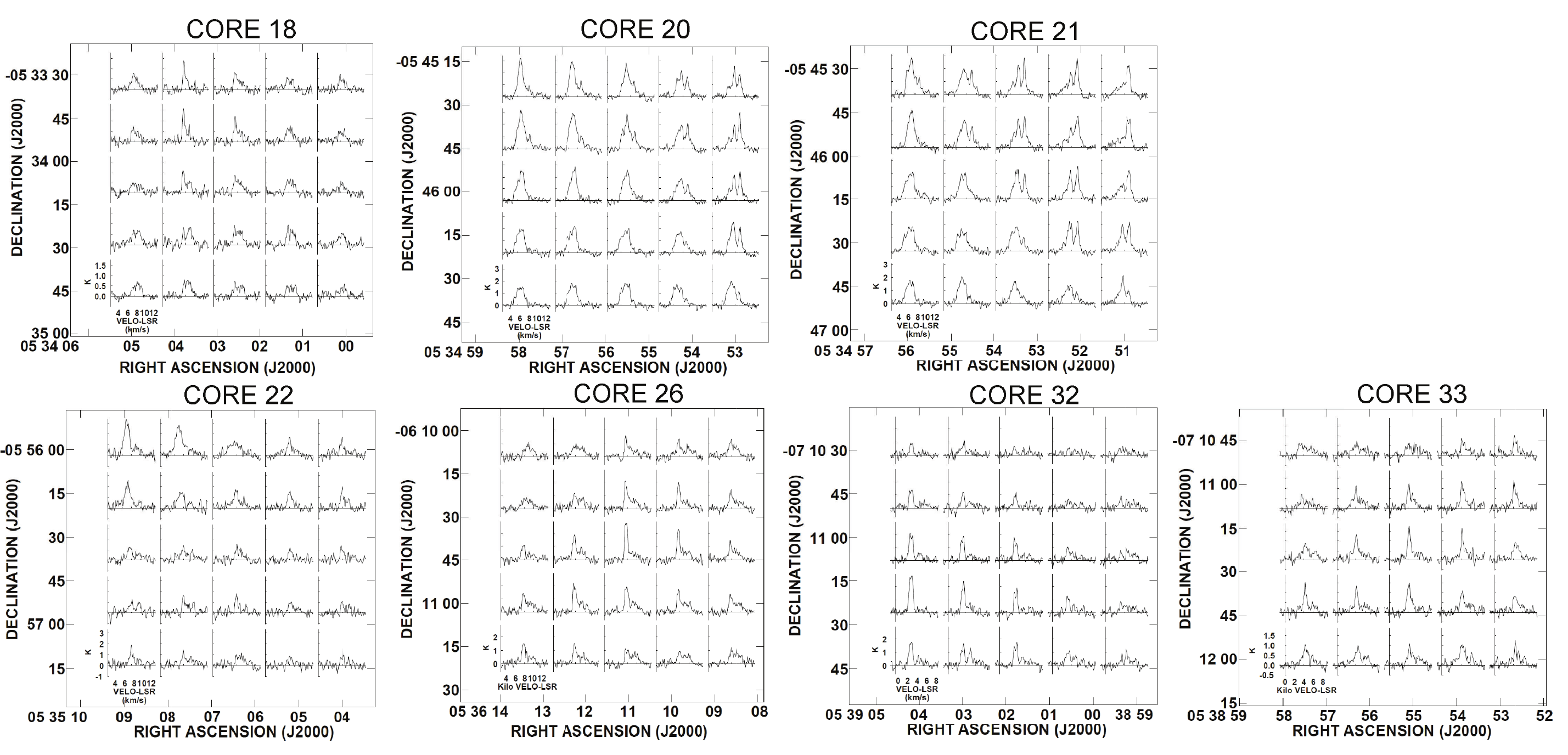}
\caption{
Same as Figure \ref{fig:SS6-17-HCO-PLCUB} except for
cores 18, 20, 21, 22, 26, 32, and 33.
\label{fig:SS18-33-HCO-PLCUB}}
\end{figure}

Third, we investigated stamp maps illustrating the spatial distribution of the HCO$^+$ profiles on a grid of
the sky coordinates.
Figures \ref{fig:SS6-17-HCO-PLCUB} and \ref{fig:SS18-33-HCO-PLCUB}
show 5$\times$5 stamp maps of HCO$^+$ profiles 
on a grid of 18$\arcsec$ (approximately equal to the telescope beam size) 
centered at the core center position,
for cores including the seven cores mentioned above.
The spectra were taken from the Bessel--Gauss stacked pixel data.
The sky coordinates are shown as the intersection point of the profile coordinate axes.
Cores 20 and 21 observed in the same field showed two velocity components.
Core 21 was thought to have two velocity components
rather than a blue-skewed line profile with a dip.
The N$_2$H$^+$ spectrum toward core 21 shown in Figure \ref{fig:PP16-22SP} supported this interpretation.
Subsequently, we had 
five cores (cores 10, 11, 14, 18, and 32) out of the 30 starless cores (or 17\%)
as the HCO$^+$ blue-skewed candidates.
All the five
starless cores showed dips in the HCO$^+$ spectra.
However, for cores 18 and 32, the S/N ratios in the HCO$^+$ spectra were not sufficiently high
or the sense of line asymmetry was not observed consistently around the core center.
Finally, we identified starless cores 10, 11, and 14 to have the HCO$^+$ ``blue-skewed line profiles,''
whereas cores 18 and 32 were identified as ``candidate blue-skewed line profiles.''
When the HCO$^+$ profile showed clear dips,
we interpolated the line profile from the outer part (open green circles in Figures
\ref{fig:PP10-15SP} to \ref{fig:PP23-33SP}).
We here fixed the HCO$^+$ Gaussian peak velocity to the value obtained from the Gaussian fitting
to the optically thinner, H$^{13}$CO$^+$ emission.
We determined the velocity range of absorption by comparing the HCO$^+$ and H$^{13}$CO$^+$
profiles. 
In other cases, we indicate the vertical lines of
the H$^{13}$CO$^+$ peak velocity (velocity corresponding to the intensity maximum),
to illustrate the velocity difference of $v$(HCO$^+) - v($H$^{13}$CO$^+)$.
Core 12 in Figure \ref{fig:PP16-22SP} has two vertical lines because the H$^{13}$CO$^+$ profile
has two maxima.
All our starless cores showing blue-skewed and candidate blue-skewed profiles satisfied the significance criterion for
the asymmetry parameter of $\delta V \leq -0.25$.
Out of the 30 observed starless cores,
three cores (10 \%) were identified as cores having HCO$^+$ (solid) ``blue-skewed'' profiles,
and another 
two cores (7 \%)
were identified as ``candidate blue-skewed.''
For the 
five starless
cores with blue-skewed and candidate blue-skewed profiles,
the mean of the asymmetry parameter $\delta V$ was $-$0.5$\pm$0.3.
The core masses of these 
five
blue-skewed starless cores including candidate identification
had
a mean of 1.77~$M_{\odot}$, a median of 1.25~$M_{\odot}$, a minimum of 0.15~$M_{\odot}$, and a maximum of 5.4~$M_{\odot}$,
whereas all the 30 starless cores had
a mean of 1.81~$M_{\odot}$, a median of 1.18~$M_{\odot}$, a minimum of 0.15~$M_{\odot}$, and a maximum of 7.1~$M_{\odot}$.
There is no preferred tendency in mass for the blue-skewed starless cores.

We then considered the protostellar cores.
Out of the six protostellar cores,
four had significantly negative $\delta V$, one (core 35) had $\delta V$ = 0, 
and the remaining one (core 16) had a significantly positive $\delta V$.
Although the sample number of the protostellar core was small,
we observed more cases with inward motions.
It seems that inward motions dominate outward motions implying outflow
even in the protostellar phase in our observations.
The protostellar cores showing blue-skewed profiles have no dips in their spectra.
Core 26 seemed to have a strong redshifted absorption, but also exhibits a prominent redshifted wing emission,
which probably represents a molecular outflow.

\begin{figure}
\epsscale{1.0}
\figurenum{7}
\includegraphics[bb=0 0 505 800, width=10cm]{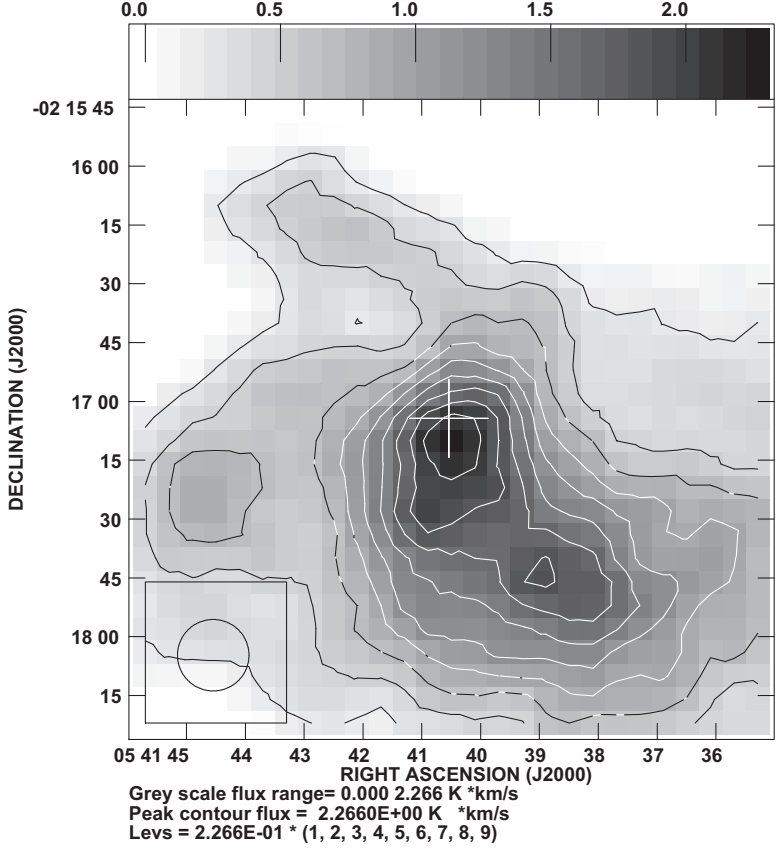}
\caption{
Gray-scale map of the velocity integrated intensity
of the H$^{13}$CO$^+$ emission toward
core 10 (G206.93$-$16.61East1).
The plus sign represents the SCUBA-2 core center position,
and its maximum size (20$\arcsec$) represents the diameter
of the circular area where we collected data to obtain a composite spectrum such as in
Figure \ref{fig:PP10-15SP}.
The circle in the bottom-left corner represents the half-power beam size
(18$\farcs$2 diameter).
\label{fig:S10-H13CO}}
\end{figure}

\begin{figure}
\epsscale{1.0}
\figurenum{8}
\includegraphics[bb=0 0 505 800, width=10cm]{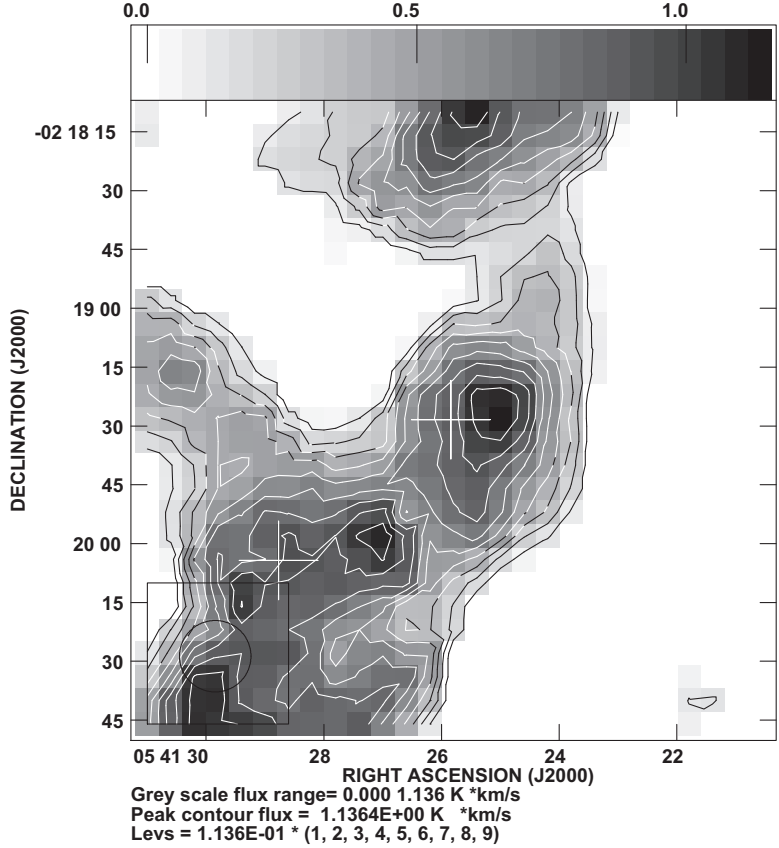}
\caption{
Same as Figure \ref{fig:S10-H13CO}
except for core 11 (G206.93$-$16.61West4)
at the center and
core 12 (G206.93$-$16.61West5)
on its south-eastern side.
\label{fig:S11-H13CO}}
\end{figure}

\begin{figure}
\epsscale{1.0}
\figurenum{9}
\includegraphics[bb=0 0 505 800, width=10cm]{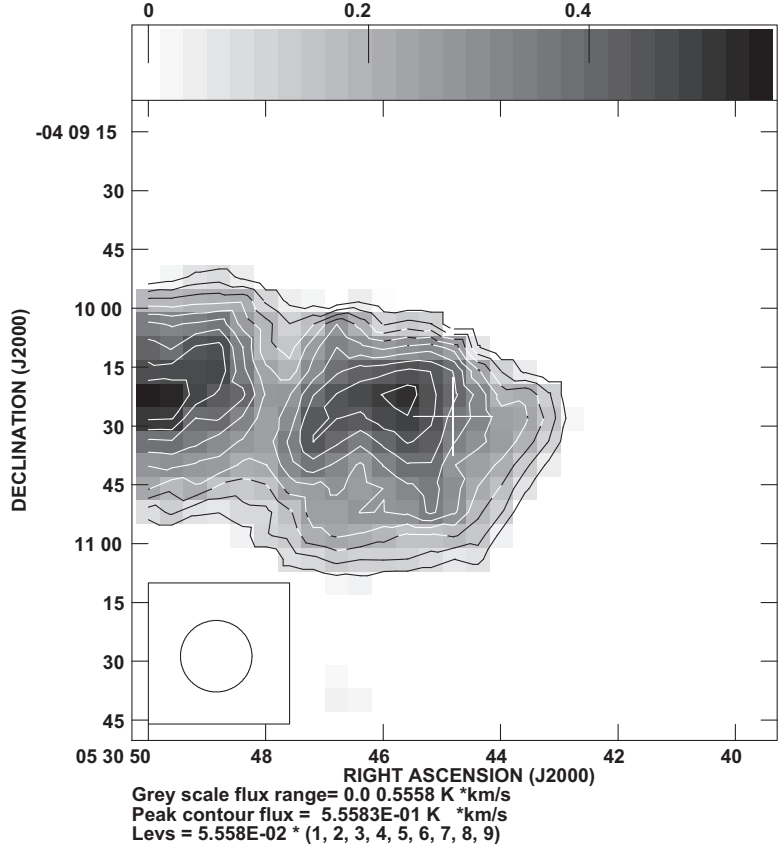}
\caption{
Same as Figure \ref{fig:S10-H13CO}
except for core 14 (G207.36$-$19.82North4).
\label{fig:S14-H13CO}}
\end{figure}

\begin{figure}
\epsscale{1.0}
\figurenum{10}
\includegraphics[bb=0 0 505 800, width=10cm]{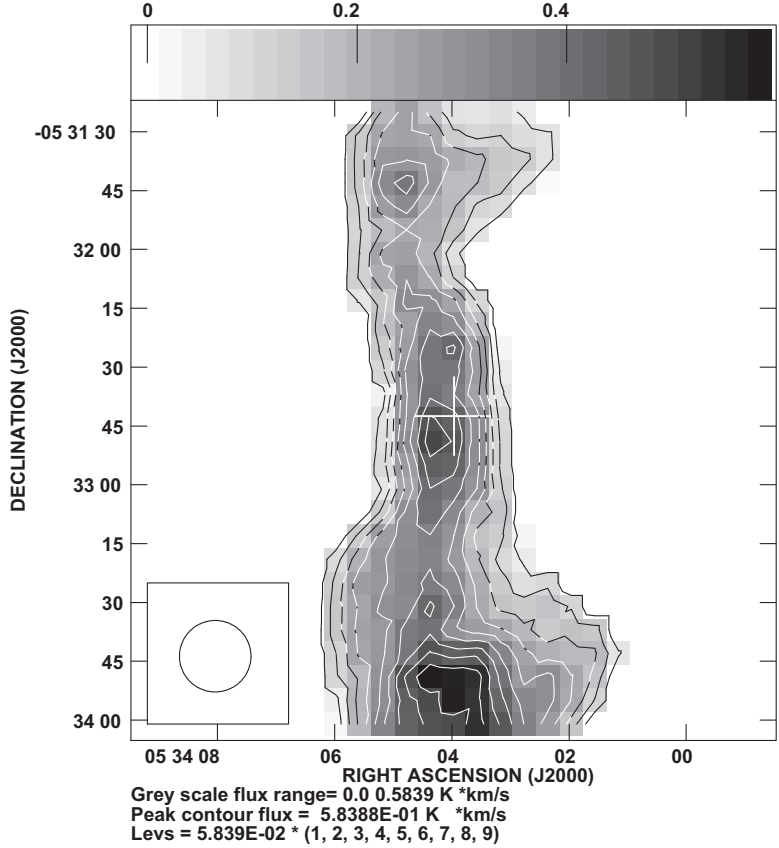}
\caption{
Same as Figure \ref{fig:S10-H13CO}
except for core 17 (G209.05$-$19.73North).
\label{fig:S17-H13CO}}
\end{figure}

\begin{figure}
\epsscale{1.0}
\figurenum{11}
\includegraphics[bb=0 0 505 800, width=10cm]{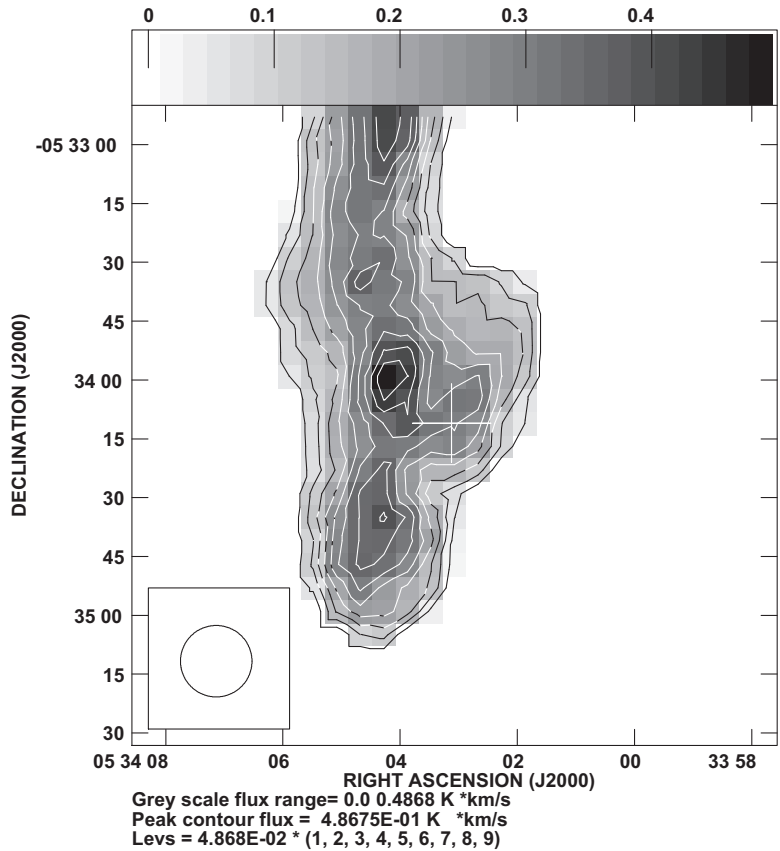}
\caption{
Same as Figure \ref{fig:S10-H13CO}
except for core 18 (G209.05$-$19.73South).
\label{fig:S18-H13CO}}
\end{figure}

\begin{figure}
\epsscale{1.0}
\figurenum{12}
\includegraphics[bb=0 0 505 800, width=10cm]{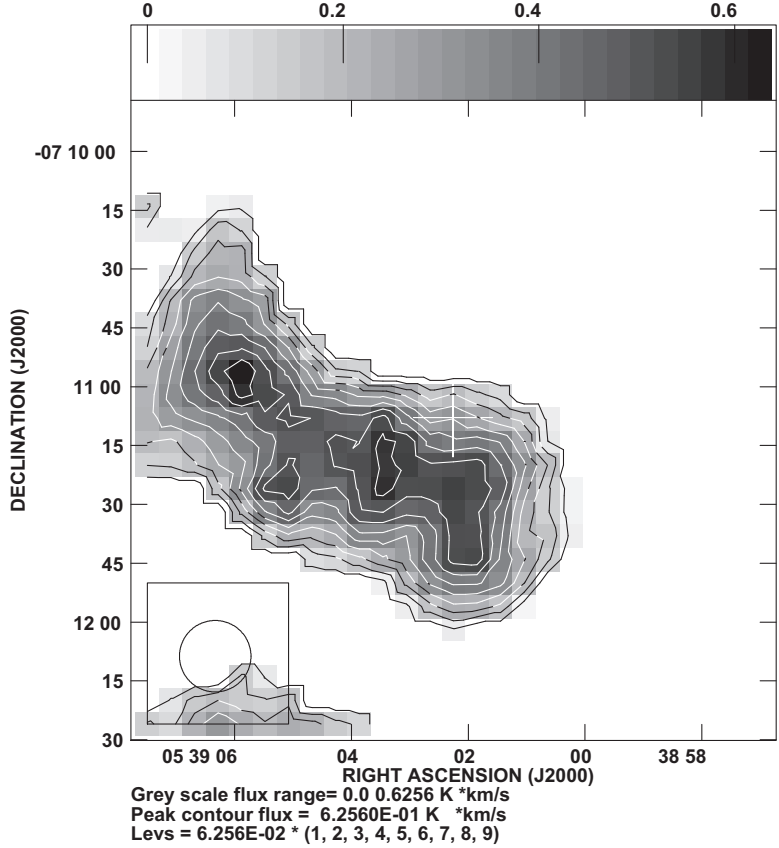}
\caption{
Same as Figure \ref{fig:S10-H13CO}
except for core 32 (G211.16$-$19.33North3).
\label{fig:S32-H13CO}}
\end{figure}

Finally, we investigated the core properties that are assumed to be related to the core evolutionary stages:
the association with the core in H$^{13}$CO$^+$ emission, detection in the dust continuum using the ALMA ACA,
and the  CEF, based on the deuterium fraction (Table \ref{tbl:fil_cef}).
The association of the core in H$^{13}$CO$^+$ emission is indicated if there is a clear H$^{13}$CO$^+$ emission peak
within 30$\arcsec$ from the SCUBA-2 position.
For the association of the core in H$^{13}$CO$^+$ emission,
`1' means that the SCUBA-2 core is associated with the core in H$^{13}$CO$^+$ emission, and `0' means that it is not.
We often observe offsets between the H$^{13}$CO$^+$ emission peak and the SCUBA-2 position.
It is possible that the depletion \citep{2001ApJ...552..639A,2002ApJ...570L.101B} of H$^{13}$CO$^+$ affects its distribution.
More generally, variation in the abundance, including but not restricted to depletion, as well as variation in excitation
affects the distributions of molecular lines. 
\cite{2021ApJS..256...25T} compared the distributions of 
the SCUBA-2 850 $\micron$ dust continuum, 
the $J = 1\rightarrow0$ transitions of 
N$_2$H$^+$ and HC$_3$N, and the 
$J_N = 7_6\rightarrow6_5$ and $J_N = 8_7\rightarrow7_6$
transitions of CCS
in the SCUBA-2 cores in Orion,
and found that molecular emission distributions were often offset from the continuum peak.
We continued to use the SCUBA-2 position as the core center to avoid the effect of spatial variation in abundance and excitation.
We adopted CEF2.0  (CEF version 2.0), which is a measure of the chemical core evolution based on the logarithmic abundance ratio of
DNC/HNC and {\ntd}/{\nth} for starless cores.
The values of CEF2.0 were taken from \cite{2020ApJS..249...33K} and \cite{2021ApJS..256...25T}.
The CEF2.0 values of the starless cores with HCO$^+$ blue-skewed profiles or candidate blue-skewed profiles run from $-$41 to $-$20.
Note that the CEF2.0 values for the overall SCUBA-2 cores in Orion run from $\sim-70$ to $\sim-10$ \citep{2020ApJS..249...33K,2021ApJS..256...25T}.
The association of the core in H$^{13}$CO$^+$ emission,
detection status 
in the 1.3 mm continuum
with the
ALMA/ACA, and
CEF2.0
are listed in Table \ref{tbl:fil_cef}.
The detectability in the dust continuum with the ACA seems particularly effective to identify the core evolutionary stage \citep{2020ApJ...899...10T}.
The detection status in the
1.3 mm
dust continuum using the ALMA ACA
was obtained from \cite{2020ApJS..251...20D} 
(RMS noise level = 0.6$-$2 mJy beam$^{-1}$).
Table \ref{tbl:coord} lists ALMA core names used by \cite{2020ApJS..251...20D} 
as well as 
those obtained from the ALMA Science Portal\footnote{almascience.org} for the sources that have been observed using the ALMA but have not been published yet.
In addition, we adopted the ACA detection status
in the 1.2 mm continuum
for core 32 from \cite{2020ApJ...895..119T} (RMS noise level = 0.24 mJy beam$^{-1}$).
The detection status in the ALMA in the combined image from configuration TM2 and the ACA
in the 1.3 mm continuum
was reported by \cite{2021ApJ...907L..15S}, and is listed as `TM2+ACA Detection.'
Figures
\ref{fig:S10-H13CO}
to
\ref{fig:S32-H13CO}
show
the velocity-integrated intensity map of the H$^{13}$CO$^+$ emission
toward the starless cores with HCO$^+$ blue-skewed profiles or candidate blue-skewed profiles.
All 
five 
starless cores of the blue-skewed or candidate blue-skewed line profiles with dips
accompany H$^{13}$CO$^+$ cores.

The association with the core in H$^{13}$CO$^+$ emission may suggest
that the 
five
starless cores are at a relatively late stage of the starless phase
\citep{2002ApJ...575..950O,2020ApJ...899...10T}, but
using the CEF2.0 values, we cannot conclude that they are predominantly at extremely late stages.
Regarding the ACA detectability \citep{2020ApJ...899...10T},
two cores (cores 14 and 32) out of the five
starless cores with blue-skewed or candidate blue-skewed profiles were observed using the ALMA ACA
by \cite{2020ApJS..251...20D}
in the ALMASOP collaboration and \cite{2020ApJ...895..119T},
and these cores were detected in the dust continuum.
However, core 14 was not detected in the ALMA TM2+ACA 
by \cite{2020ApJS..251...20D}.

To summarize the statistics on blue-skewed profiles, 
five 
starless cores out of 30 showed blue-skewed or candidate blue-skewed profiles, 
whereas four protostellar cores out of six showed blue-skewed profiles.

\floattable
\begin{deluxetable}{lrrrrrrrrrr}
\tablecaption{Peak Intensity and Velocity \label{tbl:intensity}}
\tablecolumns{11}
\tablenum{4}
\tablewidth{0pt}
\tablehead{
\multicolumn{1}{c}{Core Number} &
\multicolumn{2}{c}{HCO$^+$} &
\multicolumn{2}{c}{H$^{13}$CO$^+$} &
\multicolumn{2}{c}{HNC} &
\multicolumn{2}{c}{HN$^{13}$C} &
\multicolumn{2}{c}{N$_2$H$^+$} \\
\cline{2-11}
\colhead{} &
\colhead{$T_A^*$} &
\colhead{$v_{LSR}$} &
\colhead{$T_A^*$} &
\colhead{$v_{LSR}$} &
\colhead{$T_A^*$} &
\colhead{$v_{LSR}$} &
\colhead{$T_A^*$} &
\colhead{$v_{LSR}$} &
\colhead{$T_A^*$} &
\colhead{$v_{LSR}$} \\
\colhead{} &
\colhead{K} &
\colhead{km s$^{-1}$} &
\colhead{K} &
\colhead{km s$^{-1}$} &
\colhead{K} &
\colhead{km s$^{-1}$} &
\colhead{K} &
\colhead{km s$^{-1}$} &
\colhead{K} &
\colhead{km s$^{-1}$}
}
\startdata
1  &  1.2   &  9.8 &    $<$  0.34   &    . . .  &  0.6   &  10.8 &    $<$  0.32   &    . . .  &      . . .  &    . . .  \\
2  &  2.3   &  11.0 &  0.9   &  11.1 &  1.2   &  11.1 &    $<$  0.27   &    . . .  &    $<$  0.23   &    . . .  \\
3  &  1.4   &  10.9 &  0.5   &  10.8 &  0.9   &  10.9 &    $<$  0.31   &    . . .  &    $<$  0.28   &    . . .  \\
4  &  0.5   &  9.7 &  0.7   &  9.4 &  1.2   &  10.0 &  0.5   &  9.4 &  0.3   &  9.4 \\
5  &  0.6   &  9.7 &  0.8   &  10.0 &  1.2   &  9.8 &  0.6   &  10.2 &  0.4   &  10.2 \\
6  &    $<$  0.33   &    . . .  &  0.5   &  9.9 &  1.2   &  9.7 &  0.4   &  10.5 &  0.3   &  10.4 \\
7  &    $<$  0.40   &    . . .  &    $<$  0.40   &    . . .  &    $<$  0.39   &    . . .  &    $<$  0.34   &    . . .  &    $<$  0.23   &    . . .  \\
8  &  3.5   &  9.9 &  1.0   &  9.8 &  2.7   &  9.7 &    $<$  0.30   &    . . .  &  1.1   &  9.8 \\
9  &  0.8   &  9.6 &    $<$  0.27   &    . . .  &  0.4   &  9.2 &    $<$  0.28   &    . . .  &    $<$  0.27   &    . . .  \\
10  &  5.0   &  9.5 &  2.1   &  9.9 &  3.5   &  9.7 &    $<$  0.30   &    . . .  &  0.7   &  10.0 \\
11  &  1.6   &  9.3 &  1.1   &  10.2 &  0.9   &  9.6 &    $<$  0.27   &    . . .  &  0.5   &  10.1 \\
12  &  2.4   &  9.3 &  0.7   &  9.4 &  1.4   &  9.9 &  0.4   &  9.7 &  0.4   &  9.0 \\
13  &  3.4   &  10.7 &  1.0   &  11.1 &  1.6   &  10.9 &    $<$  0.25   &    . . .  &  0.3   &  10.9 \\
14  &  1.9   &  10.9 &  0.6   &  11.1 &  1.0   &  10.9 &    $<$  0.23   &    . . .  &  0.3   &  11.0 \\
15  &  2.0   &  11.4 &  0.5   &  11.2 &  1.3   &  11.4 &    $<$  0.27   &    . . .  &  0.4   &  11.3 \\
16  &  4.7   &  11.4 &  3.2   &  11.1 &  3.7   &  11.4 &  0.9   &  11.1 &  3.9   &  11.2 \\
17  &  0.8   &  7.7 &  0.7   &  8.2 &  1.0   &  8.4 &    $<$  0.25   &    . . .  &  0.3   &  8.2 \\
18  &  1.0   &  7.6 &  0.7   &  7.8 &  0.9   &  7.6 &    $<$  0.22   &    . . .  &  0.4   &  7.8 \\
19  &  1.9   &  8.4 &  0.5   &  8.6 &  1.3   &  8.5 &  0.4   &  8.8 &  0.7   &  8.5 \\
20  &  2.5   &  7.4 &  1.0   &  7.6 &  1.0   &  7.0 &    $<$  0.29   &    . . .  &  0.4   &  7.9 \\
21  &  2.2   &  7.6 &  1.1   &  8.1 &  1.6   &  7.5 &    $<$  0.25   &    . . .  &  0.5   &  8.9 \\
22  &  1.6   &  8.0 &  0.8   &  8.2 &  1.1   &  8.0 &    $<$  0.44   &    . . .  &  0.5   &  8.2 \\
23  &  0.8   &  7.9 &  0.3   &  8.3 &  1.1   &  8.9 &    $<$  0.27   &      &  0.6   &  8.3 \\
24  &  1.2   &  8.3 &  0.6   &  8.1 &  1.3   &  8.6 &  0.5   &  7.9 &  0.9   &  8.0 \\
25  &  1.4   &  7.8 &  0.6   &  7.6 &  1.6   &  7.6 &  0.4   &  8.3 &  0.5   &  7.7 \\
26  &  2.6   &  7.8 &  1.0   &  8.2 &  2.7   &  7.8 &  0.6   &  8.0 &  0.6   &  8.1 \\
27  &  3.3   &  5.5 &  1.1   &  5.9 &  2.4   &  5.5 &  0.4   &  5.7 &  0.4   &  5.8 \\
28  &  1.5   &  7.4 &  1.2   &  8.1 &  2.3   &  7.3 &  0.7   &  8.1 &  0.5   &  8.2 \\
29  &  1.8   &  6.4 &  0.4   &  6.4 &  1.3   &  6.4 &    $<$  0.37   &    . . .  &  0.2   &  6.3 \\
30  &  1.8   &  5.6 &  1.1   &  5.1 &  1.9   &  5.4 &  0.4   &  5.4 &  0.5   &  5.2 \\
31  &  1.7   &  4.0 &  0.9   &  4.3 &  2.1   &  4.0 &  0.4   &  4.5 &  0.9   &  4.3 \\
32  &  1.7   &  3.0 &  0.7   &  3.3 &  1.6   &  3.1 &  0.4   &  3.4 &  0.7   &  3.3 \\
33  &  1.7   &  4.2 &  0.7   &  4.4 &  1.5   &  4.2 &    $<$  0.30   &  4.3 &  0.4   &  4.5 \\
34  &    $<$  0.39   &      &    $<$  0.39   &    . . .  &    $<$  0.39   &    . . .  &    $<$  0.50   &    . . .  &    $<$  0.18   &    . . .  \\
35  &  1.5   &  3.9 &  0.9   &  3.9 &  2.5   &  4.0 &  0.7   &  4.0 &  0.5   &  4.2 \\
36  &  1.1   &  11.2 &    $<$  0.47   &    . . .  &  1.2   &  11.2 &    $<$  0.55   &    . . .  &  0.5   &  11.5 \\
\enddata
\end{deluxetable}

\clearpage

\floattable
\begin{deluxetable}{ccccccc}
\rotate
\tablecaption{Velocity Difference and Core Properties \label{tbl:veldiff}}
\tablecolumns{7}
\tablenum{5}
\tablewidth{0pt}
\tablehead{
\colhead{Core Number} &
\colhead{YSO Ass.} &
\colhead{$v$(HCO$^+)-v($H$^{13}$CO$^+)$} &
\colhead{$v$(HCO$^+)-v($N$_2$H$^+)$} &
\colhead{$v$(HNC)$-v$(HN$^{13}$C)} &
\colhead{$\Delta v($H$^{13}$CO$^+)$} &
\colhead{$\delta V$} \\
\colhead{} &
\colhead{} &
\colhead{km s$^{-1}$} &
\colhead{km s$^{-1}$} &
\colhead{km s$^{-1}$} &
\colhead{km s$^{-1}$} &
\colhead{} 
}
\startdata
1	&	starless	&		. . .	&		. . .	&		. . .	&		. . .	&			. . .	\\
2	&	starless	&	-0.1 		&		. . .	&		. . .	&	0.45 		&		-0.22 		\\
3	&	starless	&	0.1 		&		. . .	&		. . .	&	0.77 		&		0.13 		\\
4	&	starless	&	0.3 		&	0.3 		&	0.6 		&	0.50 		&		0.59 		\\
5	&	starless	&	-0.3 		&	-0.5 		&	-0.4 		&	1.28 		&		-0.23 		\\
6	&	starless	&		. . .	&		. . .	&	-0.8 		&	1.07 		&			. . .	\\
7	&	starless	&		. . .	&		. . .	&	0.0 		&		. . .	&			. . .	\\
8	&	starless	&	0.1 		&	0.1 		&		. . .	&	0.60 		&		0.17 		\\
9	&	starless	&		. . .	&		. . .	&		. . .	&		. . .	&			. . .	\\
10	&	starless	&	-0.4 		&	-0.5 		&		. . .	&	0.92 		&		-0.44 		\\
11	&	starless	&	-0.9 		&	-0.8 		&		. . .	&	0.89 		&		-1.01 		\\
12	&	starless	&	-0.1 		&	0.3 		&	0.2 		&	1.44 		&	(	-0.07 	)	\\
13	&	protostellar	&	-0.4 		&	-0.2 		&		. . .	&	0.70 		&		-0.57 		\\
14	&	starless	&	-0.2 		&	-0.1 		&		. . .	&	0.80 		&	(	-0.25 	)	\\
15	&	starless	&	0.2 		&	0.1 		&		. . .	&	0.89 		&	(	0.22 	)	\\
16	&	protostellar	&	0.3 		&	0.2 		&	0.3 		&	0.75 		&		0.40 		\\
17	&	starless	&	-0.5 		&	-0.5 		&		. . .	&	0.57 		&		-0.88 		\\
18	&	starless	&	-0.2 		&	-0.2 		&		. . .	&	0.43 		&		-0.47 		\\
19	&	starless	&	-0.2 		&	-0.1 		&	-0.3 		&	0.63 		&		-0.32 		\\
20	&	starless	&	-0.2 		&	-0.5 		&		. . .	&	1.69 		&		-0.12 		\\
21	&	starless	&	-0.5 		&	-1.3 		&		. . .	&	1.52 		&	(	-0.33 	)	\\
22	&	starless	&	-0.2 		&	-0.2 		&		. . .	&	0.34 		&		-0.59 		\\
23	&	starless	&	-0.4 		&	-0.4 		&		. . .	&	0.29 		&		-1.38 		\\
24	&	starless	&	0.2 		&	0.3 		&	0.7 		&	0.57 		&		0.35 		\\
25	&	starless	&	0.2 		&	0.1 		&	-0.7 		&	0.82 		&	(	0.24 	)	\\
26	&	protostellar	&	-0.4 		&	-0.3 		&	-0.2 		&	0.72 		&		-0.56 		\\
27	&	starless	&	-0.4 		&	-0.3 		&	-0.2 		&	0.85 		&		-0.47 		\\
28	&	starless	&	-0.7 		&	-0.8 		&	-0.8 		&	1.09 		&	(	-0.64 	)	\\
29	&	starless	&	0.0 		&	0.1 		&		. . .	&	0.37 		&		0.00 		\\
30	&	starless	&	0.5 		&	0.4 		&	0.0 		&	0.50 		&		0.99 		\\
31	&	protostellar	&	-0.3 		&	-0.3 		&	-0.5 		&	0.79 		&		-0.38 		\\
32	&	starless	&	-0.3 		&	-0.3 		&	-0.3 		&	0.54 		&		-0.55 		\\
33	&	protostellar	&	-0.2 		&	-0.3 		&	-0.1 		&	0.73 		&		-0.27 		\\
34	&	starless	&		. . .	&		. . .	&		. . .	&		. . .	&			. . .	\\
35	&	protostellar	&	0.0 		&	-0.3 		&	0.0 		&	0.75 		&		0.00 		\\
36	&	starless	&		. . .	&	-0.3 		&		. . .	&	0.67 		&			. . .	\\
\enddata
\end{deluxetable}

\clearpage

\floattable
\begin{deluxetable}{ccc}
\tablecaption{Number of Velocity Components \label{tbl:velcomp}}
\tablecolumns{3}
\tablenum{6}
\tablewidth{0pt}
\tablehead{
\colhead{Core Number} &
\colhead{H$^{13}$CO$^+$} &
\colhead{N$_2$H$^+$} 
}
\startdata
1  &  . . .  &  . . .  \\
2  &  1  &  . . .  \\
3  &  1  &  . . .  \\
4  &  1  &  1  \\
5  &  1  &  1  \\
6  &  2  &  1  \\
7  &  . . .  &  . . .  \\
8  &  1  &  1  \\
9  &  . . .  &  . . .  \\
10  &  1  &  1 or 2  \\
11  &  1  &  1  \\
12  &  2  &  1  \\
13  &  1  &  1 or 2  \\
14  &  1 or 2  &  1  \\
15  &  1 or 2  &  1  \\
16  &  1  &  1  \\
17  &  1  &  1  \\
18  &  1  &  1  \\
19  &  1  &  1  \\
20  &  1  &  1  \\
21  &  complicated  &  1-3  \\
22  &  1  &  1  \\
23  &  1  &  1  \\
24  &  1  &  1  \\
25  &  2  &  2  \\
26  &  1  &  1  \\
27  &  1  &  1  \\
28  &  3  &  2  \\
29  &  1  &  . . .  \\
30  &  1  &  1  \\
31  &  1  &  1  \\
32  &  1  &  1  \\
33  &  1  &  1  \\
34  &  . . .  &  1  \\
35  &  1  &  1  \\
36  &  1  &  1  \\
\enddata
\end{deluxetable}

\clearpage

\floattable
\begin{deluxetable}{ccccc}
\tablecaption{Figure Correspondence \label{tbl:fig}}
\tablecolumns{5}
\tablenum{7}
\tablewidth{0pt}
\tablehead{
\colhead{Core Number} &
\colhead{starless/protostellar} &
\colhead{Feature} &
\colhead{Figure} &
\colhead{Chapter} 
}
\startdata
6  &  starless      &  no HCO$^+$                & 5    & 3.4  \\
10  &  starless      &  blue-skewed                & 2, 5 & 3.2  \\
11  &  starless      &  blue-skewed                & 2, 5 & 3.2  \\
12  &  starless      &  complicated HCO$^+$    & 2     &   \\
13  &  protpstellar  &  blue-skewed                 & 2, 5 & 3.3  \\
14  &  starless      &  blue-skewed                 & 2, 5 & 3.2  \\
15  &  starless      &  flat-top HCO$^+$           & 2    &   \\
16  &  protostellar  &  red-skewed                  & 3    &   \\
17  &  starless     &  dip, skewness is not clear & 3, 5 & 3.2  \\
18  &  starless     &  candidate blue-skewed    & 3, 6 & 3.2  \\
20  &  starless     & two velocity components & 3, 6 & 3.1  \\
21  &  starless     & two velocity components & 3, 6 & 3.1  \\
22  &  starless      &suspicious but not convincing & 3, 6 &   \\
23  &  starless      &  two peaks without skew  & 4    & 3.4  \\
24  &  starless      &  red-skewed                  & 4    &   \\
25  &  starless      &  wing-like                      & 4    &   \\
26  &  protostellar  &  blue-skewed                 & 4, 6 & 3.3  \\
30  &  starless      &  red-skewed                  & 4     &   \\
31  &  protostellar  &  blue-skewed                 & 4     & 3.3  \\
32  &  starless     &  candidate blue-skewed    & 4, 6 & 3.2  \\
33  &  protostellar  &  blue-skewed                 & 4, 6 & 3.3  \\
\enddata
\end{deluxetable}

\clearpage

\floattable
\begin{deluxetable}{cccccc}
\rotate
\tablecaption{Association with Core in H$^{13}$CO$^+$ Emission, 
Filament Association, ALMA Detection Status, and CEF2.0 \label{tbl:fil_cef}}
\tablecolumns{6}
\tablenum{8}
\tablewidth{0pt}
\tablehead{
\colhead{Core Number} &
\colhead{Core in H$^{13}$CO$^+$ Emission} &
\colhead{SCUBA-2 Filament} &
\colhead{ACA Detection} &
\colhead{TM2+ACA Detection} &
\colhead{CEF2.0} \\
\colhead{} &
\colhead{} &
\colhead{} &
\colhead{} &
\colhead{} &
\colhead{} 
}
\startdata
1  &    . . .  &  0   &  NO (weak?)    &  NO    &    . . .    \\
2  &  1 &  0   &  NO    &  NO    &    . . .    \\
3  &  1 &  0   &  NO (weak?)    &  NO    &    . . .    \\
4  &  1 &  0   &    . . .  &    . . .  &  -43  $\pm$  27  \\
5  &  1 &  1   &  NO (weak?)    &  NO    &    . . .    \\
6  &  0 &  1   &  YES    &  NO    &    . . .    \\
7  &  0 &  1   &  YES    &  YES    &    . . .    \\
8  &  1 &  1   &  YES    &  NO    &    . . .    \\
9  &  0 &  1   &  YES    &  NO    &    . . .    \\
10  &  1 &  0   &    . . .  &    . . .  &  -41  $\pm$  27  \\
11  &  1 &  1   &    . . .  &    . . .  &  -27  $\pm$  14  \\
12  &  1 &  1   &  NO    &  NO    &  -34  $\pm$  3  \\
13  &  1 &  1   &  YES    &  NO    &    . . .    \\
14  &  1 &  1   &  YES    &  NO    &    . . .    \\
15  &  1 &  0   &  NO    &  NO    &  -41  $\pm$  8  \\
16  &  1 &  1   &  YES    &  YES    &    . . .    \\
17  &  1 &  1   &    . . .  &    . . .  &  -19  $\pm$  15  \\
18  &  1 &  1   &    . . .  &    . . .  &  -20  $\pm$  19  \\
19  &  0 &  1   &  YES (weak?)    &  NO    &  -58  $\pm$  17  \\
20  &  1 &  1   &  YES    &  YES    &  -42  $\pm$  5  \\
21  &  1 &  1   &  NO (weak?)    &  NO    &    . . .    \\
22  &  0 &  1   &  YES    &  NO    &  -27  $\pm$  15  \\
23  &  0 &  0   &    . . .  &    . . .  &  -56  $\pm$  27  \\
24  &  0 &  1   &  NO    &  NO    &  -50  $\pm$  27  \\
25  &  0 &  1   &  YES    &  NO    &    . . .    \\
26  &  1 &  1   &  YES    &  YES    &    . . .    \\
27  &  1 &  0   &  NO (weak?)    &  NO    &  -37  $\pm$  14  \\
28  &  1 &  1   &  YES (weak?)    &  NO    &  -22  $\pm$  20  \\
29  &  0 &  0   &  YES    &  NO    &  -7  $\pm$  28  \\
30  &  0 &  1   &  YES    &  NO    &  -24  $\pm$  14  \\
31  &  1 &  1   &  YES    &  NO    &    . . .    \\
32  &  1 &  1   &    YES&    . . .  &  -22  $\pm$  5  \\
33  &  1 &  1   &  YES (weak)    &  NO    &    . . .    \\
34  &    . . .  &  1   &    . . .  &    . . .  &    . . .    \\
35  &  1 &  1   &  YES    &  YES    &    . . .    \\
36  &  0 &  0   &  YES (weak?)    &  NO    &    . . .    \\
\enddata
\end{deluxetable}

\clearpage

\clearpage

\clearpage

\floattable
\begin{deluxetable}{cccc}
\tablecaption{HCO$^+$ Absorption Strength \label{tbl:area}}
\tablecolumns{4}
\tablenum{9}
\tablewidth{0pt}
\tablehead{
\colhead{Core Number} &
\colhead{Dip Integ. Intensity} &
\colhead{Dip Width} &
\colhead{HCO$^+$ Fit Intensity} \\
\colhead{} &
\colhead{K km s$^{-1}$} &
\colhead{km s$^{-1}$} &
\colhead{K} 
}
\startdata
10  &  0.52 &  0.10 &  5.42   \\
11  &  1.09 &  0.60 &  1.83   \\                
14  &  0.52 &  0.23 &  2.28   \\                
17  &  0.11 &  0.14 &  0.79   \\
18  &  0.11 &  0.13 &  0.88   \\                
32  &  0.18 &  0.10 &  1.86   \\
\enddata
\end{deluxetable}

\subsection{Individual Starless Cores with Blue-skewed or Possibly Blue-skewed Profiles}

(a) Core 10  (G206.93$-$16.61East1)

This starless core belongs to NGC 2023 in the Orion B cloud.
Figure \ref{fig:SS6-17-HCO-PLCUB} shows
complicated profiles. It appears that
this region contains another velocity component at $\sim$11.5~km~s$^{-1}$
in HCO$^+$.
However, we observe double-peaked profile separately and clearly
with peaks at 9.5 and 10.3~km~s$^{-1}$ in HCO$^+$,
and the peak velocity of the H$^{13}$CO$^+$ emission well corresponds to
that of the dip. 
Table \ref{tbl:velcomp} lists one or two velocity components in H$^{13}$CO$^+$; however,
the velocity difference between the two velocity components is very small.
This will not affect our interpretation as a core with a HCO$^+$ blue-skewed profile with a dip.

(b) Core 11 (G206.93$-$16.61West4)

This starless core belongs to NGC 2023 in the Orion B cloud and is close to core 10.
This is a clear case of a blue-skewed line profile with a dip in HCO$^+$.
Figure \ref{fig:PP10-15SP} shows that 
the peak velocities of the H$^{13}$CO$^+$ and N$_2$H$^+$ emissions are consistent.
However, that of the HCO$^+$ emission is blue-shifted.

(c) Core 14 (G207.36$-$19.82North4)

This starless core is located in the northern end of the Orion A cloud
and is close to core 13.
This is a clear case of a blue-skewed line profile with a dip.
Figure \ref{fig:SS6-17-HCO-PLCUB} shows double-peaked profile around the core center.
We also see it on the eastern side of the figure.

(d) Core 17 (G209.05$-$19.73North)

This starless core belongs to OMC-4 in the integral-shaped filament in the Orion A cloud.
Figure \ref{fig:SS6-17-HCO-PLCUB} shows that
the dip in the observed spectrum is detected with an S/N ratio of 3.6.
Figure \ref{fig:PP16-22SP} shows that the blue peak
of the double-peaked HCO$^+$ profile is only slightly brighter than the
red peak,
and the intensity difference between the two peaks
is smaller than RMS noise level (an S/N ratio of 0.6).
The existence of a dip is clear, but the skewness is not clear.
We exclude this core from the candidate blue-skewed profile.
Figure \ref{fig:PP16-22SP} shows that 
the peak velocity of the H$^{13}$CO$^+$ emission well corresponds to
that of the dip. 
The peak velocities of the H$^{13}$CO$^+$ and N$_2$H$^+$ emission are consistent.

(e) Core 18 (G209.05$-$19.73South)

This starless core belongs to OMC-4 in the integral-shaped filament in the Orion A cloud and is close to core 17.
Figure \ref{fig:SS6-17-HCO-PLCUB} shows that
the S/N ratio of 
the dip in the observed spectrum is 3.8.
The intensity difference of the two peaks is marginally detected with an S/N ratio of 3.0.
This source is classified a candidate blue-skewed profile.
Figure \ref{fig:PP16-22SP} shows that 
the peak velocities of the H$^{13}$CO$^+$ and N$_2$H$^+$ emissions are consistent whereas
that of the HCO$^+$ emission is blue-shifted, suggesting a blue-skewed profile with a dip.

(f) Core 32 (G211.16$-$19.33North3)

This starless core belongs to L1641 in the Orion A cloud.
Figure \ref{fig:SS18-33-HCO-PLCUB} shows that
we observe double-peaked profiles at the core center and on the eastern side of the figure.
The sense of asymmetry is blue-skewed toward the core center; however it
is opposite at the neighboring positions.
This core was previously observed using the ALMA ACA \citep{2020ApJ...895..119T},
and a blue-skewed line profile with a dip was detected in DCO$^+$ $J = 3\rightarrow2$.

\subsection{Individual Protostellar Cores with Blue-skewed Profiles}

(a) Core 13 (G207.36$-$19.82North2)

This protostellar core is located in the northern end of the Orion A cloud.
This is a clear case of a blue-skewed line profile without a dip.

(b) Core 26 (G209.94$-$19.52North)

This protostellar core belongs to L1641 in the Orion A cloud.
It is a clear case of a blue-skewed line profile without a dip.
We saw red wing emission in HCO$^+$ as pointed out earlier.

(c) Core 31 (G211.16$-$19.33North5)

This protostellar core belongs to L1641 in the Orion A cloud.
It is a clear case of a blue-skewed line profile without a dip.

(d) Core 33 (G211.16$-$19.33North4)

This protostellar core belongs to L1641 in the Orion A cloud.
It is a clear case of a blue-skewed line profile without a dip.

\subsection{Notes on Other Specific Cores}

Toward starless core 6 (G203.21$-$11.20East1), we did not detect HCO$^+$ emission, 
although we detected H$^{13}$CO$^+$ emission.
It appears that the HCO$^+$ intensity sharply drops toward the east around R.A. = 5$^h$53$^m$50$^s$.
We observed an absorption dip in HCO$^+$ 
18$\arcsec$ west of the center of
core 6
with an S/N ratio of 4.5
(Figure \ref{fig:SS6-17-HCO-PLCUB}),
suggesting contamination of the line emission at the reference off position.
\cite{2021ApJ...912L...7L} observed a similar situation in their observation using the ALMA that the cold core M1 in NGC6334S
showed strong H$^{13}$CO$^+$ emission, but did not exhibit HCO+ emission.

Starless core 9 (G206.21$-$16.17South)
does not show H$^{13}$CO$^+$ emission.  The intensities of the HCO$^+$ and N$_2$H$^+$ emission are weak (0.8 K and 0.2 K, respectively).
Instead, a local peak in the 850 $\mu$m continuum and N$_2$H$^+$ emissions $\sim$30$\arcsec$ is located south-east of the SCUBA-2 core coordinates
of \cite{2018ApJS..236...51Y}
\citep[Figure 19.3 in][]{2021ApJS..256...25T}. 
\cite{2020ApJS..251...20D} identified the continuum emission using the ALMA ACA 
toward this south-eastern position as G206.21$-$16.17S.
We do not see a hint of the blue/red-skewed profile in HCO$^+$ even around this south-eastern position (figures are omitted).

Starless core 23 (G209.77$-$19.40West)
has a large velocity difference between HCO$^+$ and H$^{13}$CO$^+$.
It has two velocity peaks in HCO$^+$.
However, the profile does not show a blue-skewed profile.
We do not identify a core in H$^{13}$CO$^+$ emission here.

\section{DISCUSSION \label{sec:dis}}

\subsection{HCO$^+$ Absorption Strength and Likely Epochs of Infall Motions}

We attempted to determine if inward motions occur at specific epochs in the starless core phase.
With regards to the core evolutionary stages, we used the core density and CEF2.0.
Both these values are considered to increase with evolution.
We adopted the integrated intensity of the absorption dip in the HCO$^+$ spectrum as the strength of absorption.
We measured the integrated intensity enclosed between the HCO$^+$ profile (filled blue circles) and the interpolated
profiles (open green circles) in Figures \ref{fig:PP10-15SP}, \ref{fig:PP16-22SP}, and \ref{fig:PP23-33SP}.
The integration velocity range for the dip was chosen to be the portion sandwiched by the two velocity ranges used for Gaussian interpolation,
which were chosen by eye.
We also calculated the dip width, which
was the dip integrated intensity divided by the HCO$^+$ peak intensity estimated from the interpolated Gaussian fitting.
The dip width is a kind of the ``equivalent width'' expressed in velocity units.
Table \ref{tbl:area} lists the dip integrated intensity and dip width in the HCO$^+$ emission.

Figures \ref{fig:R_HCOa} and \ref{fig:n_HCOa} plot
the core radius and density from the SCUBA-2 observations
against the HCO$^+$ dip integrated intensity.
Larger symbols represent blue-skewed cores, whereas smaller symbols represent candidate blue-skewed cores.
It was observed that the core radius decreases and the core density increases
with increasing HCO$^+$ dip integrated intensity.
If we assume that the increase in the core density represents the core evolution,
the HCO$^+$ dip integrated intensity increases with core evolution, and
the core radius decreases with core evolution.
However, if we considere only the blue-skewed cores,
the aforementioned trend is not very clear.
Figures \ref{fig:R_NHCOa} and \ref{fig:n_NHCOa} are the same as Figures \ref{fig:R_HCOa} and \ref{fig:n_HCOa} but
for the dip width as an abscissa axis; however,
the trend is not very clear.
We conclude that we do not observe strong evidence of evolutionary change in HCO$^+$
absorption strength.

Next, we investigated the speed of inward motions.
Three-dimensional fields of density and velocity of inward motions and their relationship with the line of sight
as well as radiative transfer will determine the actual line profile.
For simplicity, we take the velocity difference, 
$v$(HCO$^+$)$-v$(H$^{13}$CO$^+$), as a
velocity extent of inward motions.
Apparently, this is not always true, because
gas with inward motions is not necessarily observed as {\it absorption}.
If the inward-motion velocity increases toward the center of the core, the maximum 
line-of-sight velocity may appear as a maximum velocity offset from the systemic velocity in the {\it emission}.
We adopted the above simple measure of 
the velocity of the {\it identifiable} inward motion,
although we might overlook even faster components. 
The absolute values of $v$(HCO$^+$)$-v$(H$^{13}$CO$^+$) of the 
five
starless cores 
with the blue-skewed lines or candidate blue-skewed line profiles
range from 0.2 to 0.9~km~s$^{-1}$.
Here, $v$(HCO$^+$) adopted corresponds to more intense peaks of the double peaks.
Moreover, the lower limit of 0.2~km~s$^{-1}$ comes from the significance criterion,
and we cannot detect subsonic motions well.
The sound speed
$C_s$
of the gas at temperatures of 10$-$17 K \citep{2019MNRAS.485.2895E}
of a mean molecular weight per particle of 2.33 $m_{\rm H}$ and the corresponding FWHM velocity width
$\Delta v = \sqrt{8~{\rm ln}~2}~C_s$
are 0.19$-$0.25~km~s$^{-1}$ and 0.44$-$0.58~km~s$^{-1}$, respectively.
Then, the absolute value of the $v$(HCO$^+$)$-v$(H$^{13}$CO$^+$) suggested
that inward motions may include supersonic values.

Figures \ref{fig:CEF2.0_HCOa} and \ref{fig:CEF2.0_NHCOa} plot
CEF2.0 against the dip integrated intensity and dip width, respectively.
Note that CEF2.0 is defined as an increasing function of time evolution 
such that CEF2.0 in starless cores ranges from $\sim -100$ to $\sim$0.
We did not note any clear correlation.
Given that CEF2.0 represents core evolution,
the dip integrated intensity or dip width does not appear at specific epochs during the evolution of the starless core phases.
Therefore, we have no strong evidence to prove that inward motion phenomena are
observed predominantly at the
last stage in the starless phase.

For starless dark cloud cores, \cite{2005ApJ...632..982S} suggested an evolutionary sequence 
of L1521B, L1498, and L1544 in this order on the basis of chemical evolution signatures \citep{2003ApJ...583..789L},
central density \citep{2001ApJ...557..193E}, and density structure.
\cite{2006ApJ...646..258H} showed that the deuterium fraction increased in this order
for these objects, and suggested the same evolutionary sequence for them.
CEF2.0 is defined as an increasing function of the deuterium fraction.
However, Figure \ref{fig:n-CEF2.0} does not show that the density increases with CEF2.0 
which traces the deuterium fraction.
We then attempted to determine if starless Orion cores evolved like starless dark cloud cores.
Figures 
\ref{fig:vdiff_Dv_CEF2.0} and
\ref{fig:vdiff_Dv_n}
show the relationship between CEF2.0, core density, 
the velocity difference between HCO$^+$ and H$^{13}$CO$^+$, and the H$^{13}$CO$^+$ line width.
No outstanding trends are observed except for the H$^{13}$CO$^+$ line width against density;
the H$^{13}$CO$^+$ line width appears to increase with increasing density.
Least-squares fitting led to 
$\Delta v$(H$^{13}$CO$^+$)/km~s$^{-1}$
= 0.43 log ($n$(H$_2$)/10$^5$~cm$^{-3}$) +0.57 with a correlation coefficient of 0.55.
If starless cores dissipate turbulence toward star formation,
the line width may decrease with time.
If the density increases during starless core evolution, Figure {\ref{fig:vdiff_Dv_n} suggests that the line width also increases.
We may need to investigate the best evolutionary tracer for the evolution of starless cores in Orion.
The density range suggested by \cite{2005ApJ...632..982S} approximately runs from $< 10^4$~cm$^{-3}$ to $10^6$~cm$^{-3}$.
By contrast, the density range studied by \cite{2018ApJS..236...51Y} runs only from $10^5$~cm$^{-3}$ to $10^6$~cm$^{-3}$.
\cite{2005ApJ...619..379C} compared the deuterium fraction based on N$_2$D$^+$ from observations and chemical simulations
and it appears that the observed values match simulations from $10^2$~cm$^{-3}$ to $10^4$~cm$^{-3}$.
Density may depend on 
core evolution stages as well as local environments for cores.
Thus, it is
likely 
that a variation in the core density within a narrower range may have
limited reliability as an evolutionary stage tracer.

\begin{figure}
\epsscale{0.7}
\figurenum{13}
\includegraphics[bb=0 0 505 800, width=10cm]{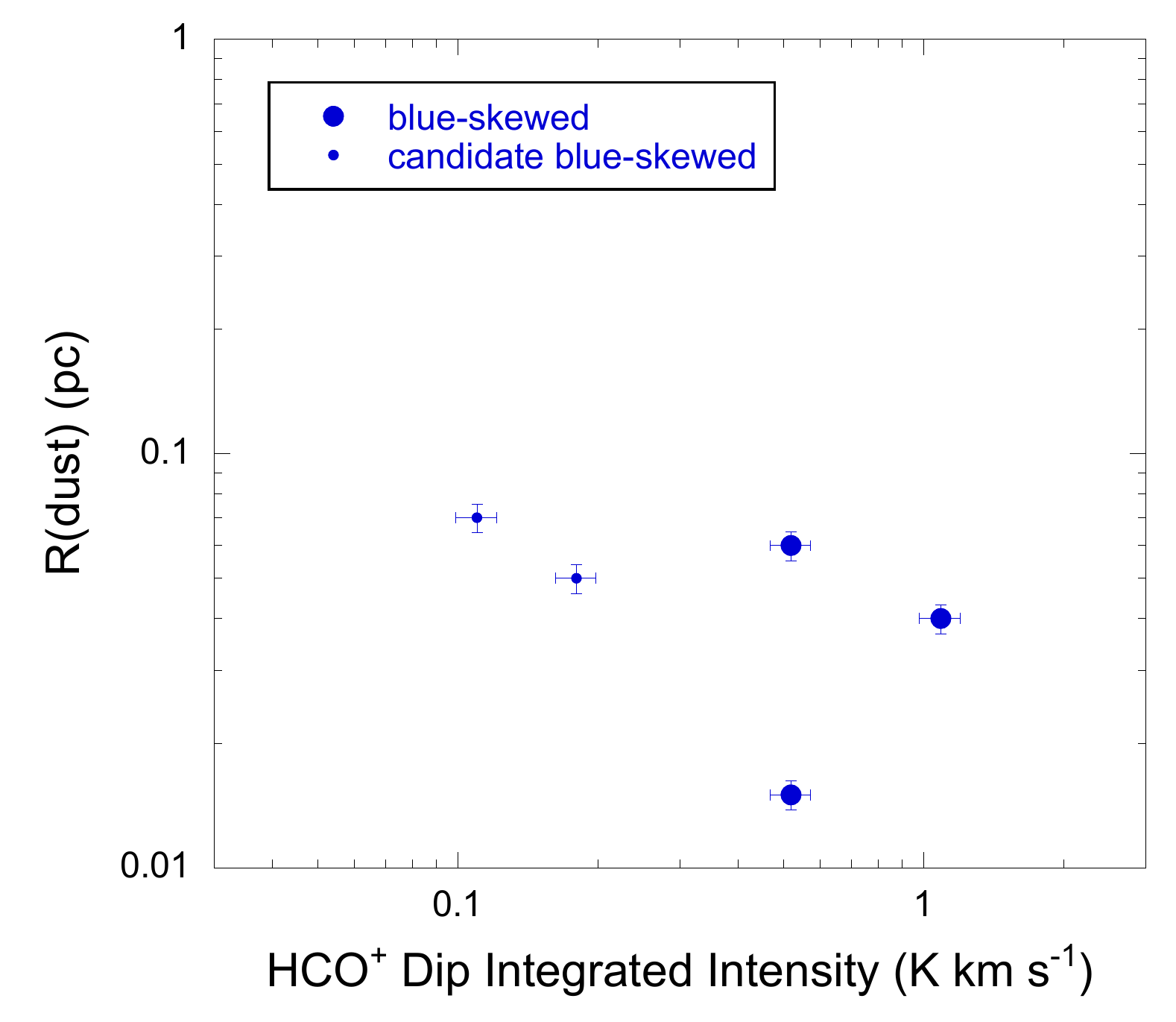}
\caption{Core radius is plotted
against the HCO$^+$ dip integrated intensity
for the starless cores with (candidate) blue-skewed profiles. 
The larger symbols represent the blue-skewed cores,
whereas the smaller symbols represent the candidate blue-skewed cores.
\cite{2018ApJS..236...51Y} did not provide the uncertainty in the core radius (or diameter) explicitly, but
we adopted the uncertainty in radius to be 8\% from the uncertainty in their adopted distance,
as \cite{2021ApJS..256...25T} did.
The uncertainty in the dip integrated intensity is assumed to be 10\% from the typical absolute intensity calibration accuracy.
\label{fig:R_HCOa}}
\end{figure}

\begin{figure}
\epsscale{0.7}
\figurenum{14}
\includegraphics[bb=0 0 505 800, width=10cm]{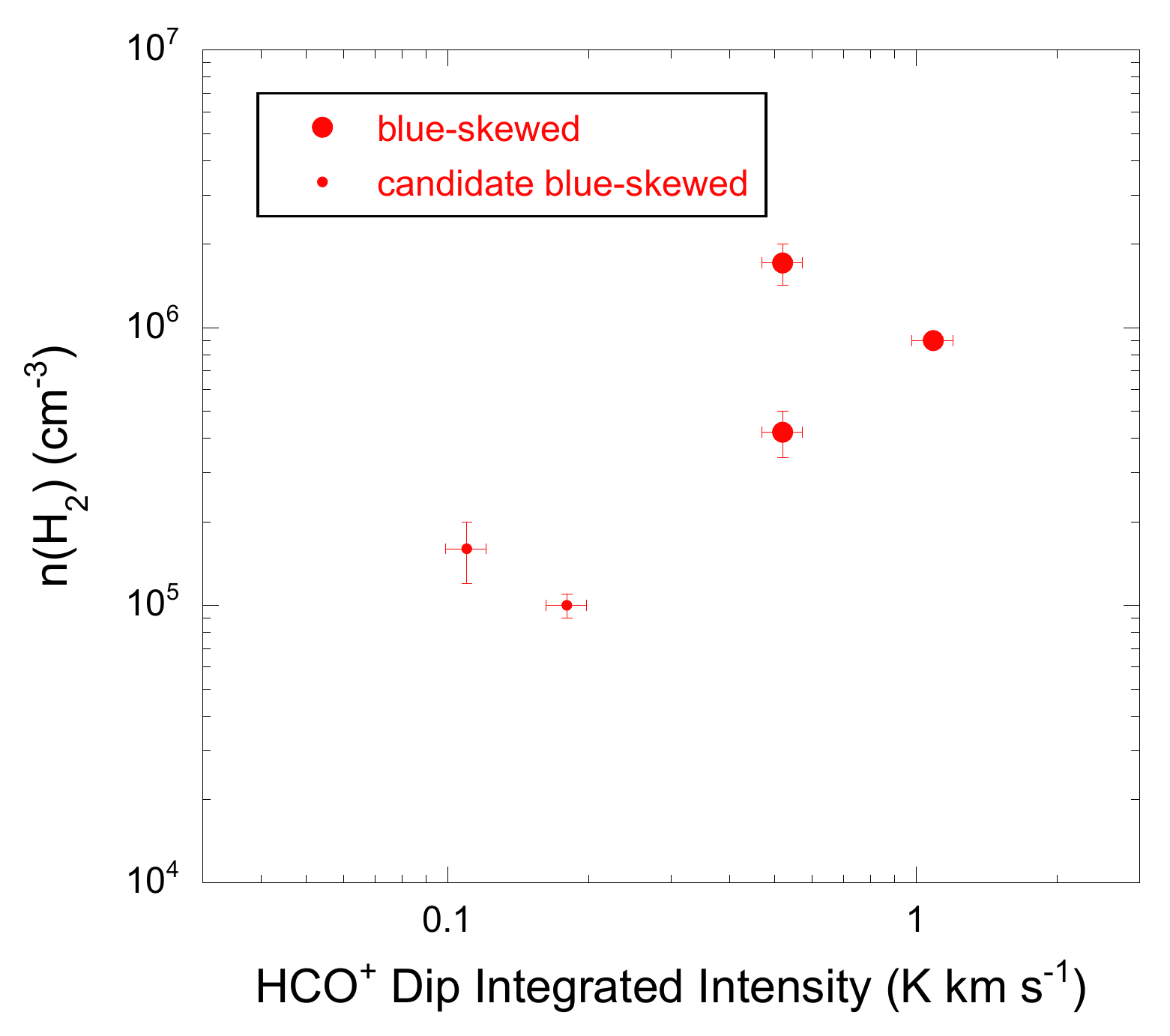}
\caption{Core density is plotted
against the HCO$^+$ dip integrated intensity
for the starless cores with (candidate) blue-skewed profiles. 
The larger symbols represent the blue-skewed cores,
whereas the smaller symbols represent the candidate blue-skewed cores.
\label{fig:n_HCOa}}
\end{figure}

\begin{figure}
\epsscale{0.7}
\figurenum{15}
\includegraphics[bb=0 0 505 800, width=10cm]{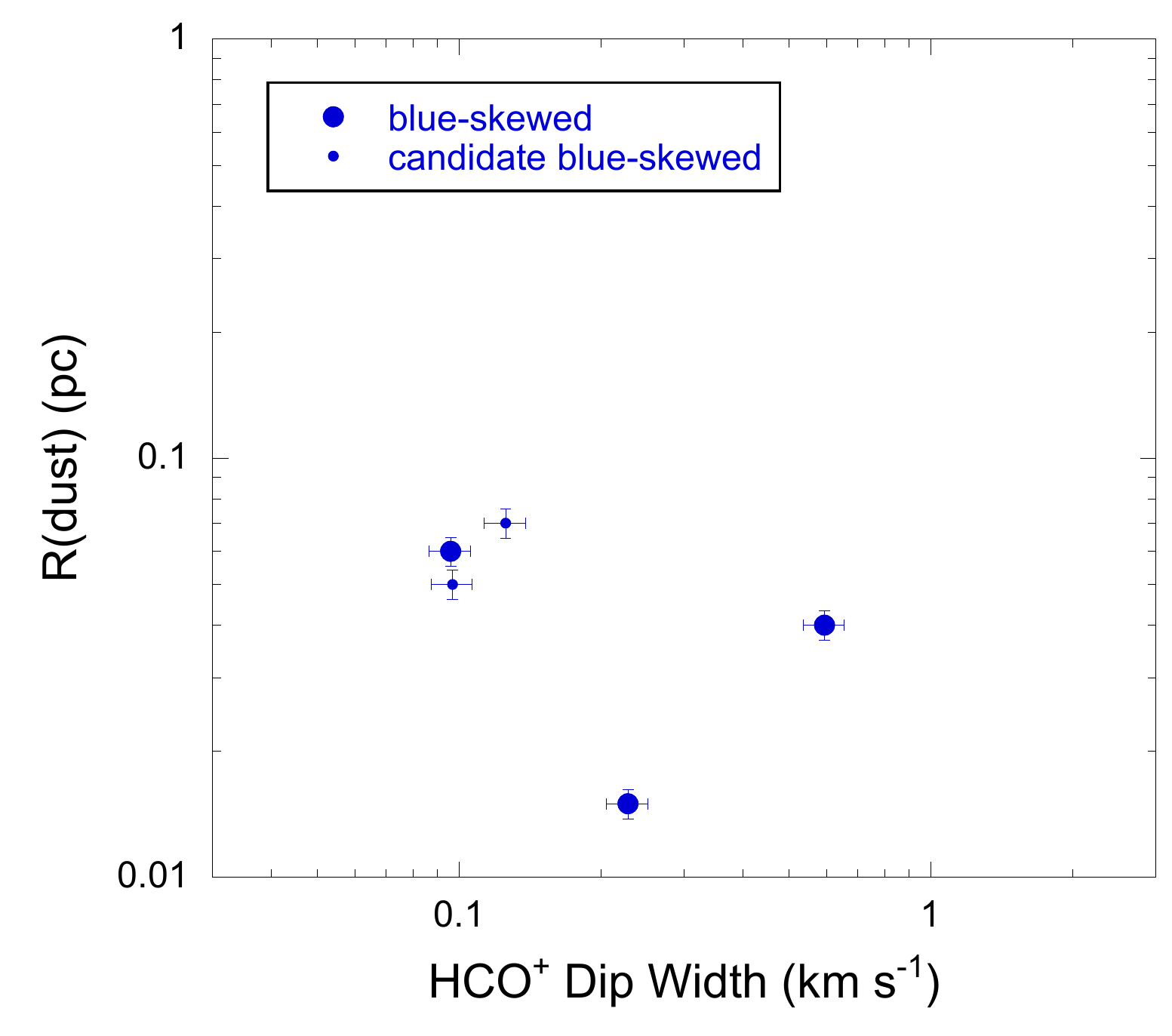}
\caption{Core radius is
plotted against the HCO$^+$ dip width
for the starless cores with (candidate) blue-skewed profiles. 
The larger symbols represent the blue-skewed cores,
whereas the smaller symbols represent the candidate blue-skewed cores.
The uncertainty in the dip width is assumed to be 10\%.
\label{fig:R_NHCOa}}
\end{figure}

\begin{figure}
\epsscale{0.7}
\figurenum{16}
\includegraphics[bb=0 0 505 800, width=10cm]{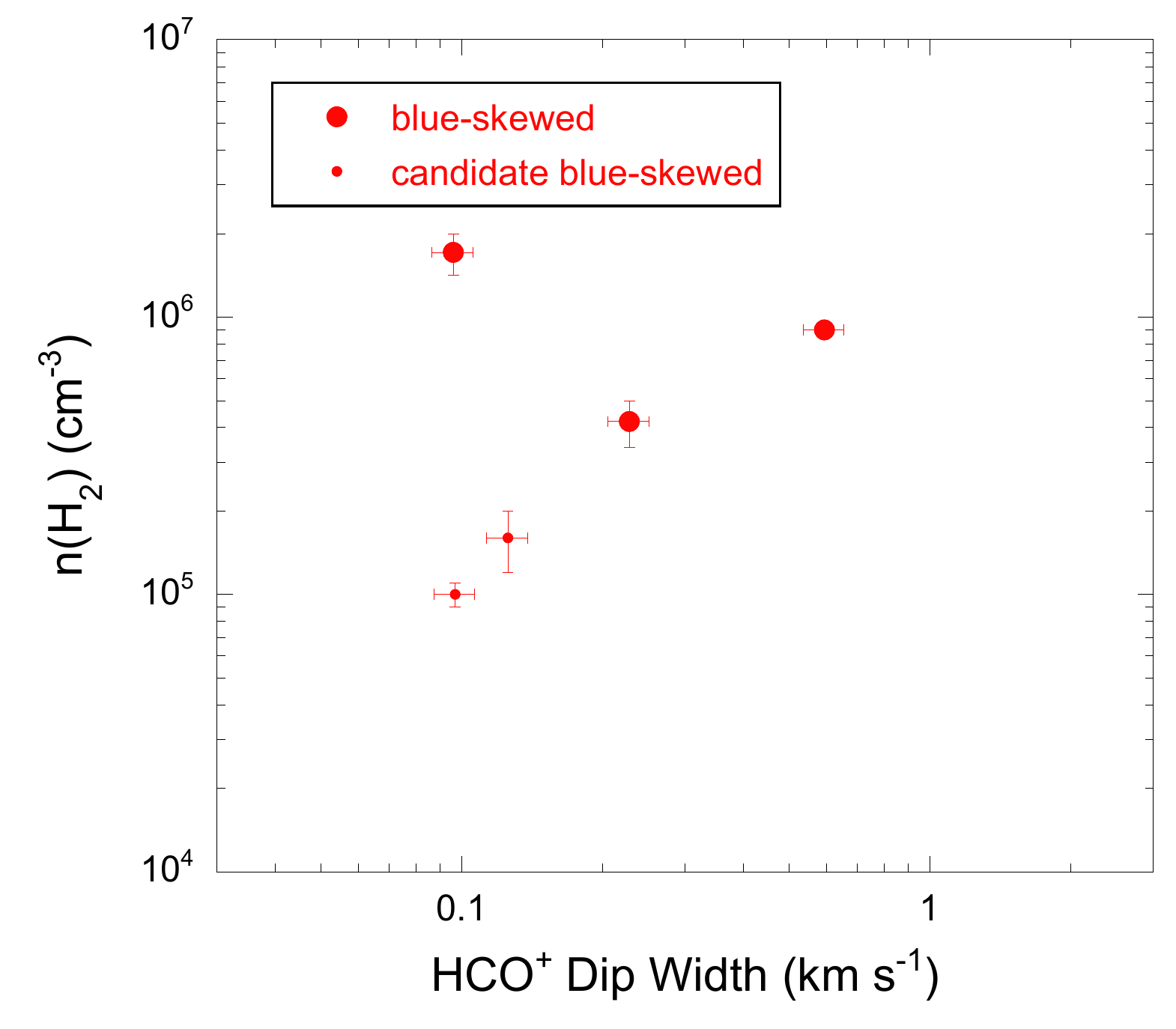}
\caption{Core radius is
plotted against the HCO$^+$ dip width
for the starless cores with (candidate) blue-skewed profiles. 
The larger symbols represent the blue-skewed cores,
whereas the smaller symbols represent the candidate blue-skewed cores.
\label{fig:n_NHCOa}}
\end{figure}

\begin{figure}
\epsscale{0.7}
\figurenum{17}
\includegraphics[bb=0 0 505 800, width=10cm]{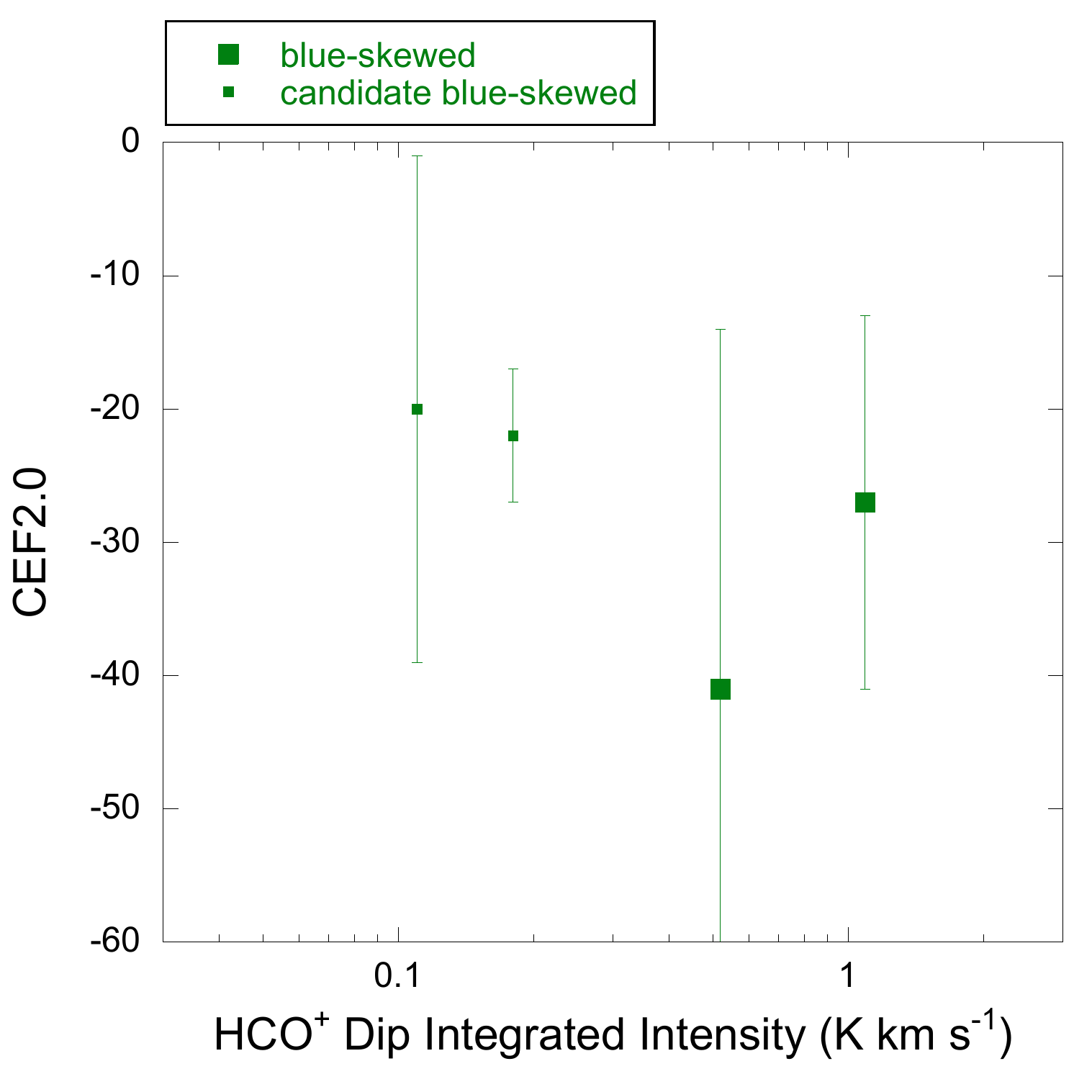}
\caption{CEF2.0 is plotted against the HCO$^+$ dip integrated intensity
for the starless cores with (candidate) blue-skewed profiles.
 \label{fig:CEF2.0_HCOa}}
\end{figure}

\begin{figure}
\epsscale{0.7}
\figurenum{18}
\includegraphics[bb=0 0 505 800, width=10cm]{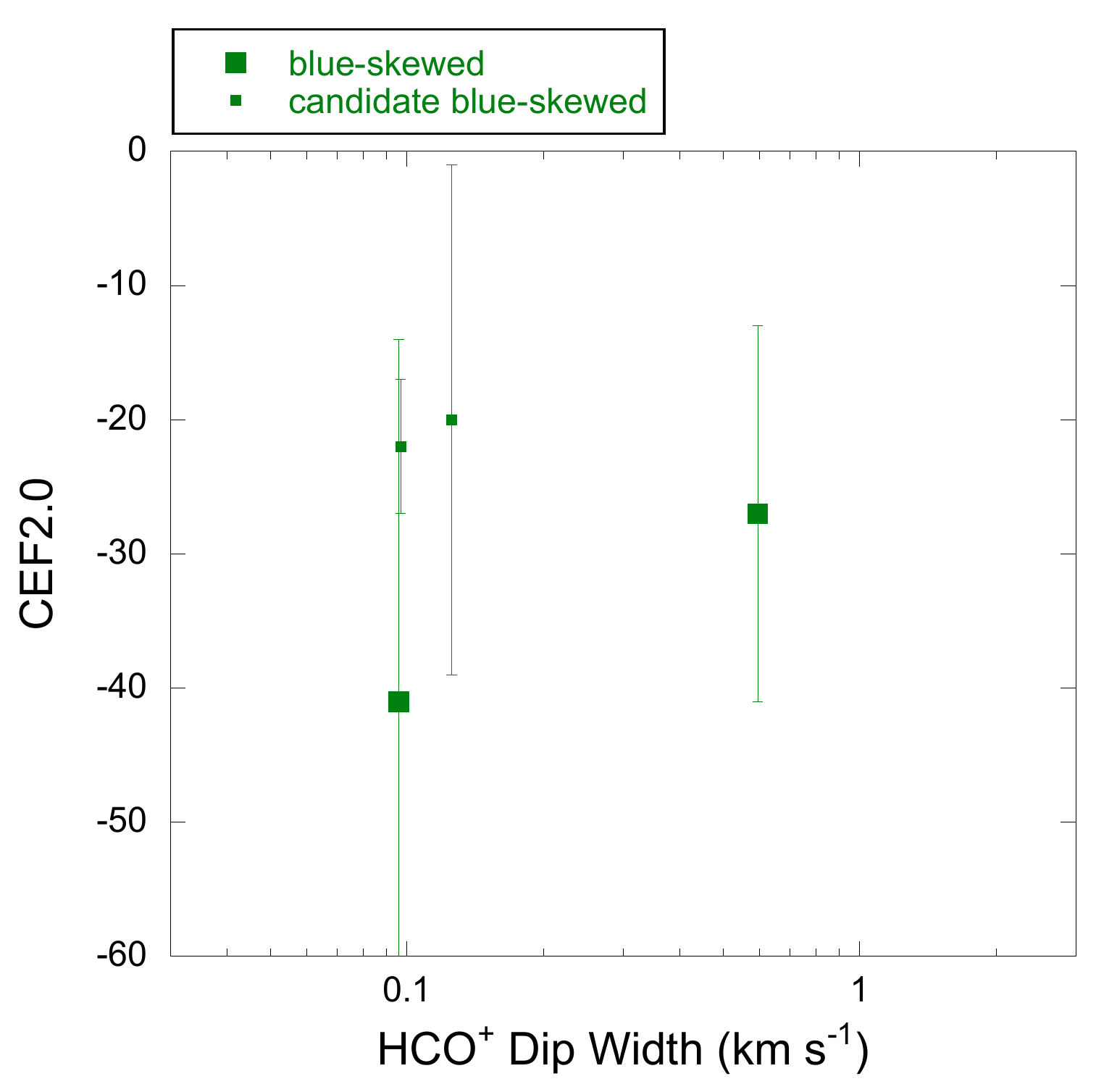}
\caption{CEF2.0 is plotted against the HCO$^+$ dip width
for the starless cores with (candidate) blue-skewed profiles.
 \label{fig:CEF2.0_NHCOa}}
\end{figure}

\begin{figure}
\epsscale{0.7}
\figurenum{19}
\includegraphics[bb=0 0 505 800, width=10cm]{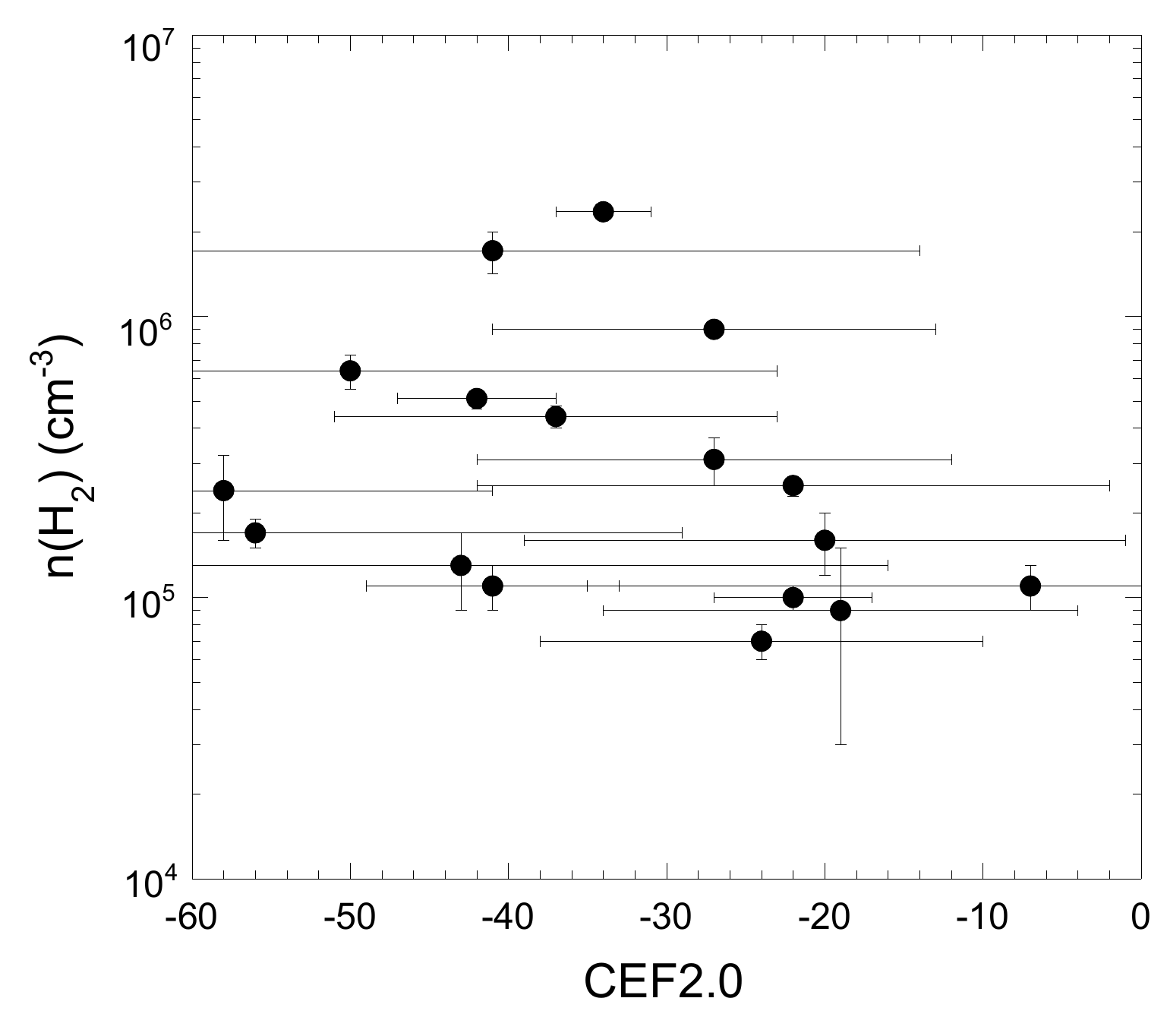}
\caption{Core density is plotted against CEF2.0
for the starless cores regardless of (candidate) blue-skewed profiles.
 \label{fig:n-CEF2.0}}
\end{figure}

\begin{figure}
\epsscale{0.7}
\figurenum{20}
\includegraphics[bb=0 0 505 800, width=10cm]{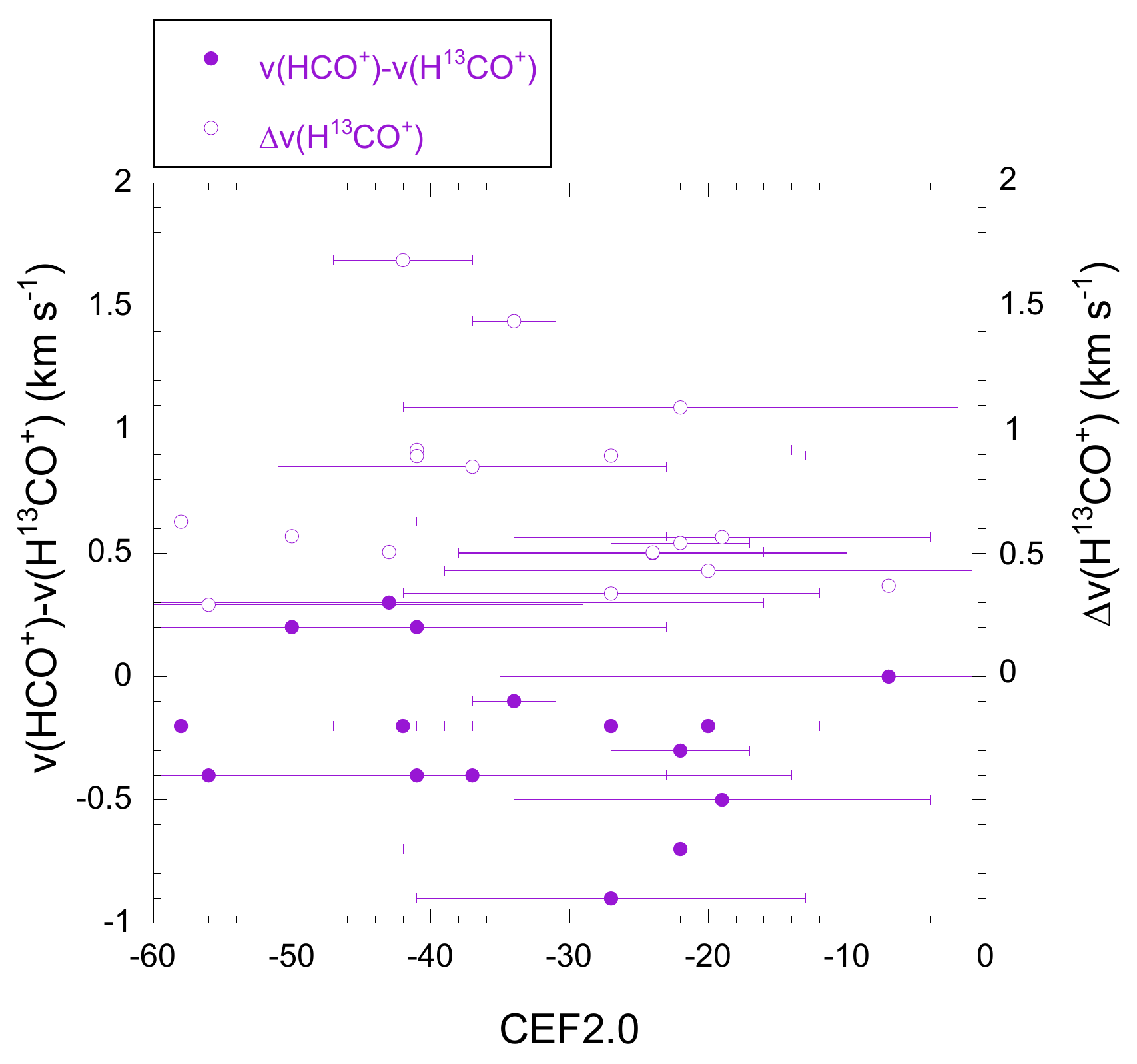}
\caption{Velocity difference between HCO$^+$ and H$^{13}$CO$^+$, and the H$^{13}$CO$^+$ line width 
are plotted against CEF2.0
for the starless cores regardless of (candidate) blue-skewed profiles.
 \label{fig:vdiff_Dv_CEF2.0}}
\end{figure}

\begin{figure}
\epsscale{0.7}
\figurenum{21}
\includegraphics[bb=0 0 505 800, width=10cm]{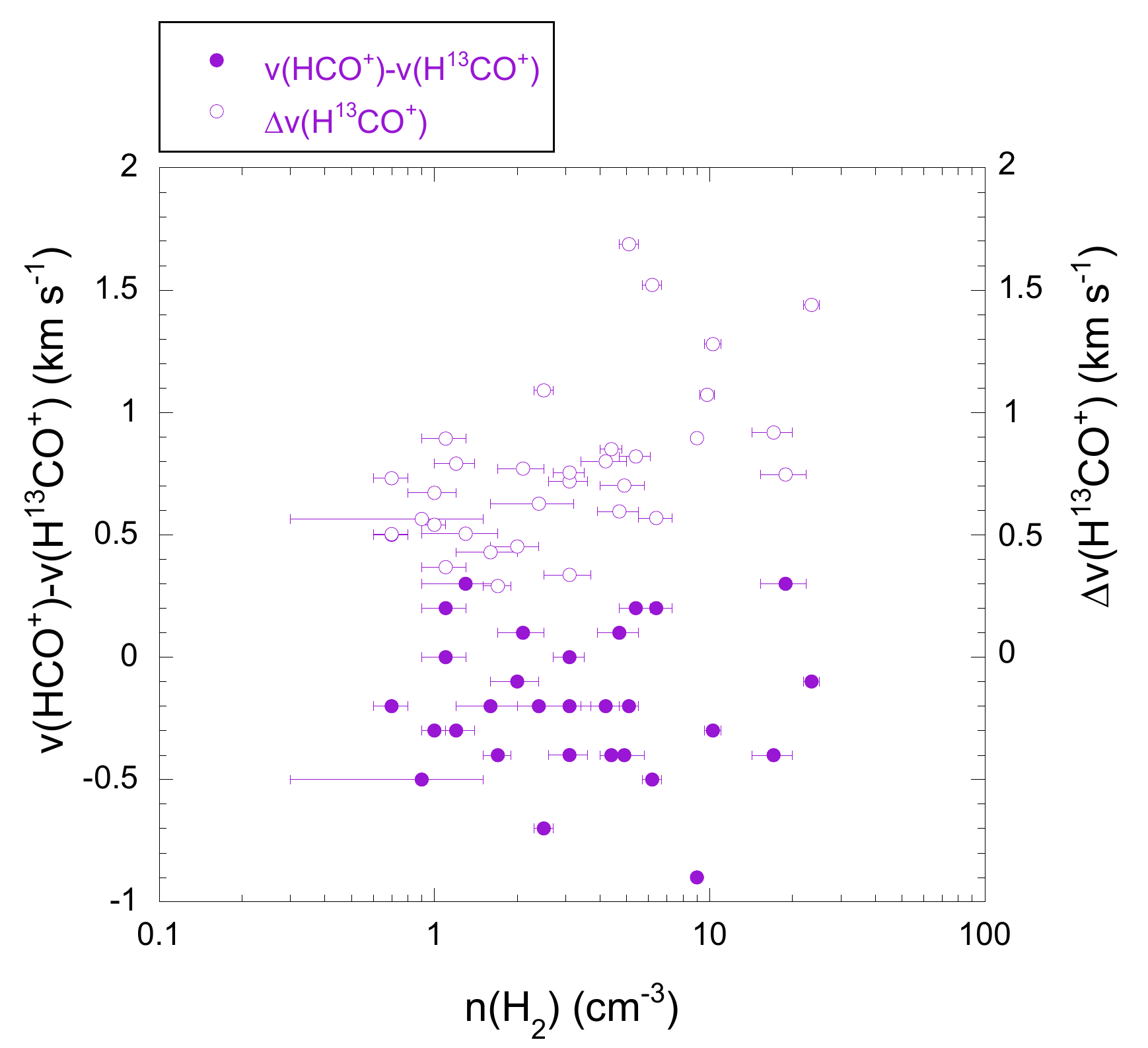}
\caption{Velocity difference between HCO$^+$ and H$^{13}$CO$^+$, and the H$^{13}$CO$^+$ line width 
are plotted against density
for both the starless and protostellar cores regardless of (candidate) blue-skewed profiles.
 \label{fig:vdiff_Dv_n}}
\end{figure}

\subsection{Reliability 
and Implication
of the Blue-skewed Line Profile with Dip}

Signatures of 
the candidate 
blue-skewed line profiles with dips in the present study are not as convincing.
Independent observations of relevant molecules and in appropriate transitions
are desirable.
\cite{2014MNRAS.444..874C} suggested using optically thicker high-J transitions, such as 
$J = 3\rightarrow2$, $J = 4\rightarrow3$, and $J = 5\rightarrow4$,
of HCN and HCO$^+$.
\cite{2018ApJ...861...14C} suggested that HCO$^+$ may be a better tracer of inward motions among other lines. 

We have $J = 3\rightarrow2$ data for core 32.
\cite{2020ApJ...895..119T} pointed out the possibility that G211.16$-$19.33North3 (core 32) 
showed a hint of the red-shifted dip 
in the DCO$^+$ $J = 3\rightarrow2$ emission using the ALMA ACA 7 m Array.
The profile is the inverse P Cygni-like, possibly because the ALMA ACA did not include the Total Power Array.
Probably, the core is mostly resolved out with the 7 m Array,
whereas a part of the emission from the core (center) and foreground absorption of gas with inward motion
are resolved using it.
In the current observations, we found that this core showed a candidate blue-skewed line profile 
with a dip in the HCO$^+$ $J = 1\rightarrow0$ emission
(Fig. \ref{fig:ACA_Nobeyama}).
The peak velocity of DCO$^+$ $J = 3\rightarrow2$ coincides with
that of the H$^{13}$CO$^+$ $J = 1\rightarrow0$ ($v_{LSR} \sim 3.3$~km~s$^{-1}$).
The velocity of the absorption dip in DCO$^+$ $J = 3\rightarrow2$ was $v_{LSR} = 3.65$~km~s$^{-1}$.
Those of the first and secondary peaks of HCO$^+$ $J = 1\rightarrow0$ were 
$v_{LSR} \sim 3.0$~km~s$^{-1}$ and $\sim 3.5$~km~s$^{-1}$, respectively.
It seems that inward motions traced by DCO$^+$ $J = 3\rightarrow2$
observed with a telescope beam $radius$ of 2$\farcs$5 (1050 au),
have a larger velocity offset ($0.3-0.4$~km~s$^{-1}$)
from the systemic velocity than that traced by HCO$^+$ $J = 1\rightarrow0$ ($\sim 0.2$~km~s$^{-1}$)
observed with a telescope beam $radius$ of 9$\arcsec$ (3800 au).
If we take the maximum velocity offset the DCO$^+$ $J = 3\rightarrow2$ dip at $v_{LSR} = 3.9$~km~s$^{-1}$
the offset from the DCO$^+$ $J = 3\rightarrow2$ peak emission is 0.6~km~s$^{-1}$.
Using two independent observations with the ACA and Nobeyama 45 m telescope, core 32 (G211.16$-$19.33North3)
might show better evidence of inward motions.
We decided to study the implications of these different velocity offsets.
Different inward velocities at different radii may represent acceleration due to gravity.
The ``inside-out'' collapse of \cite{1977ApJ...214..488S} may have 
inward velocities of 0.5 and 0.2~km~s$^{-1}$
at radii of $1.5\times 10^{16}$ and $6\times 10^{16}$~cm (or 1000 and 4000 au), respectively
\citep[Figure 2 of ][]{1992ApJ...394..204Z}.
The absorption features of core 32 might represent the motions of 
the inside-out infall of \cite{1977ApJ...214..488S}.
However, the Shu infall will form a detectable protostar quickly ($< 3\times10^5$ yr from the start of infall)
\citep{1998ApJ...504..900T}.
Core 32 is starless, and might not be consistent with combination of expected velocities and infall elapse time from the Shu infall.
If we employ the initial density distribution obtained
by multiplying the equilibrium density distribution of the critical Bonnor-Ebert sphere
by a constant factor, 
we can delay the formation of a protostar, but instead it is hard to see 
decreasing infall velocity with increasing radius for $r = 1000-3000$ au because
mass has not been accumulated enough at the center through infall
\citep{2005ApJ...620..330A,2015MNRAS.446.3731K,1993ApJ...416..303F}.
One possibility is that the core has a static central compact ``kernel'' with enough mass to accelerate the surrounding gas inward
but has not formed a protostar
\citep{1998ApJ...496L.109M,2019ApJ...874...89C}.
The SCUBA-2 mass of core 32 is $0.41\pm0.12~M_{\odot}$.
CEF2.0 of core 32 is $-22\pm5$, which corresponds to the late starless core stage close to the onset of star formation.
If CEF2.0 reflects the core evolution well, the core density distribution may have become centrally peaked.

\begin{figure}
\epsscale{1.0}
\figurenum{22}
\includegraphics[bb=0 0 505 800, width=10cm]{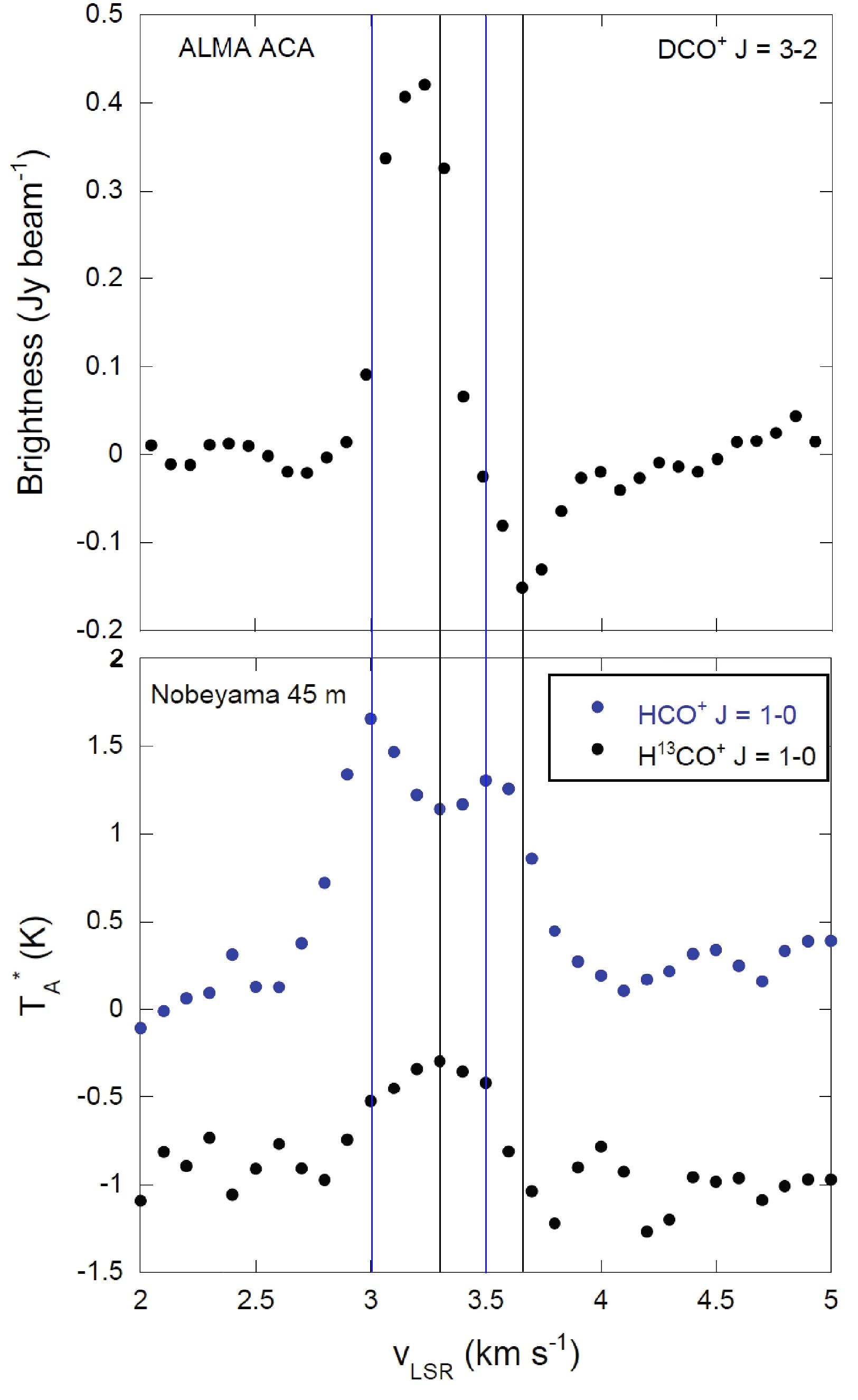}
\caption{
Comparison of the spectra in
DCO$^+$ $J$ = 3$\rightarrow$2 observed using the ALMA ACA,
HCO$^+$ $J$ = 1$\rightarrow$0 and
H$^{13}$CO$^+$ $J$ = 1$\rightarrow$0 observed with the Nobeyama 45 m telescope,
toward
core 32 (G211.16$-$19.33North3).
The vertical lines are drawn at peak and dip velocities to guide the reader's eye.
We assume that
the peak velocity of H$^{13}$CO$^+$ $J = 1\rightarrow0$ ($v_{LSR} \sim 3.3$~km~s$^{-1}$)
represents the systemic velocity.
\label{fig:ACA_Nobeyama}}
\end{figure}

\subsection{Comparison with \cite{2021ApJS..254...14Y}}

We compared $\delta V$ of \cite{2021ApJS..254...14Y} with a spatial resolution of
30$\arcsec$ with that in the present study with a resolution of 18$\farcs$2.
There are 22 overlapping cores between their cores with the $\delta V$ estimate and 
those in this study.
Their $\delta V$ values of the 22 cores had a mean of $-0.13\pm0.34$ and a median of -0.04,
whose absolute values are smaller than those in the present study
(mean = $-0.22\pm0.49$, median = $-$0.26).
Their H$^{13}$CO$^+$ line widths 
(mean = 0.86$\pm$0.35~km~s$^{-1}$, median = 0.88~km~s$^{-1}$ for the overlapping 22 cores)
are 
slightly larger than ours
(mean = 0.78$\pm$0.34~km~s$^{-1}$, median = 0.74~km~s$^{-1}$).
It is possible that different spatial resolutions are one of the main reasons for the slightly different line widths,
because it is known that the line width increases with increasing radius for molecular clouds \citep{1981MNRAS.194..809L}
Figure \ref{fig:dVcomp} compares the $\delta V$ values obtained by them and those obtained in the present study. 
Correlation is observed to be poor.

As shown earlier, 
the velocity differences between 
$v$(HCO$^+$)$-v$(H$^{13}$CO$^+$), 
$v$(HCO$^+$)$-v$(N$_2$H$^+$), and 
$v$(HNC)$-v$(HN$^{13}$C) in the Nobeyama observations (Table \ref{tbl:veldiff})
are rather consistent, and seems reliable.
To test the above interpretation, 
we collected data within a radius of 15$\arcsec$,
which corresponds to the telescope beam $radius$ employed by \cite{2021ApJS..254...14Y}. 
Table \ref{tbl:15arcsec} lists the peak velocities, $^{13}$CO$^+$ line widths, and asymmetry parameters.
The ratio of the asymmetry parameter of the 15$\arcsec$ radius sampling to that of the original 10$\arcsec$ radius sampling
had a mean of 0.6 and a median of 0.9.
The ratio of the asymmetry parameter of that of \cite{2021ApJS..254...14Y} to that of our original 10$\arcsec$ radius sampling
had a mean of 0.6 and a median of 0.4.
These values are not far different.
Figure \ref{fig:dVcomp15-10} compares the $\delta V$ values obtained with 15$\arcsec$ radius sampling and those with 10$\arcsec$ radius sampling. 
The correlation is rather good.
An outlier at $\delta V$ (this study, 10$\arcsec$ radius) = 1.38 corresponds to core 24, which
has two HCO$^+$ peaks with similar line strengths (Figure \ref{fig:PP23-33SP}).
Adoption of a different sampling radius led to a different choice from the two peaks.
Figure \ref{fig:dVcompYi-15} compares the $\delta V$ values by \cite{2021ApJS..254...14Y} and those with 15$\arcsec$ radius sampling. 
The correlation is poor.
We concluded that differences between beam sizes could not fully explain the poor correlation between the asymmetry parameters
of \cite{2021ApJS..254...14Y} and ours.

The observations in \cite{2021ApJS..254...14Y} were made in single pointing toward the SCUBA-2 cores,
whereas this study mapped the areas around the cores.
Thus, the identification of blue-skewed cores in this study is thought to be more reliable because of spatial information.

\begin{figure}
\epsscale{0.6}
\figurenum{23}
\includegraphics[bb=0 0 505 575, width=15cm]{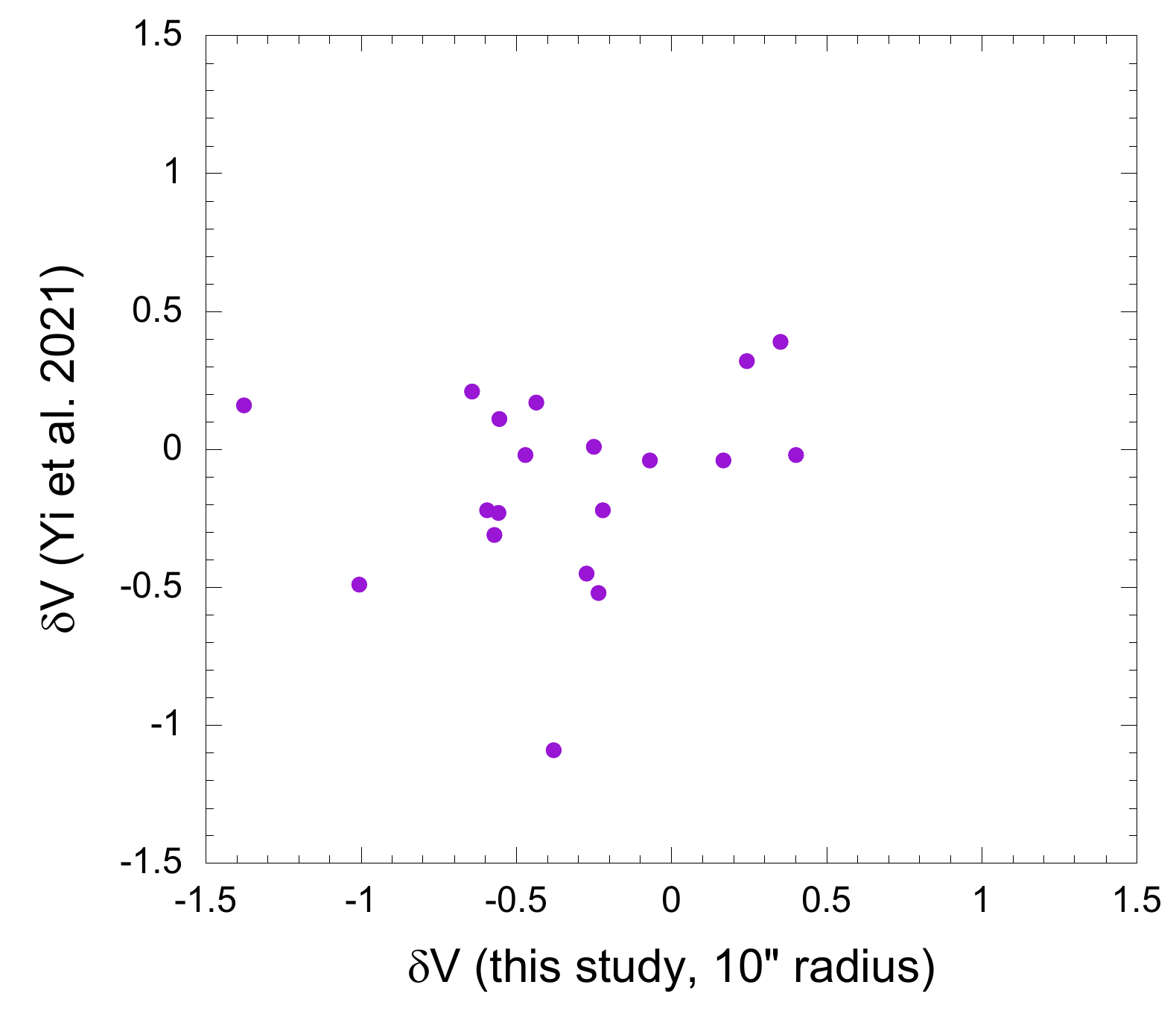}
\caption{
Asymmetry parameter by \cite{2021ApJS..254...14Y}
is compared with that of the present study.
\label{fig:dVcomp}}
\end{figure}

\begin{figure}
\epsscale{0.6}
\figurenum{24}
\includegraphics[bb=0 0 505 575, width=15cm]{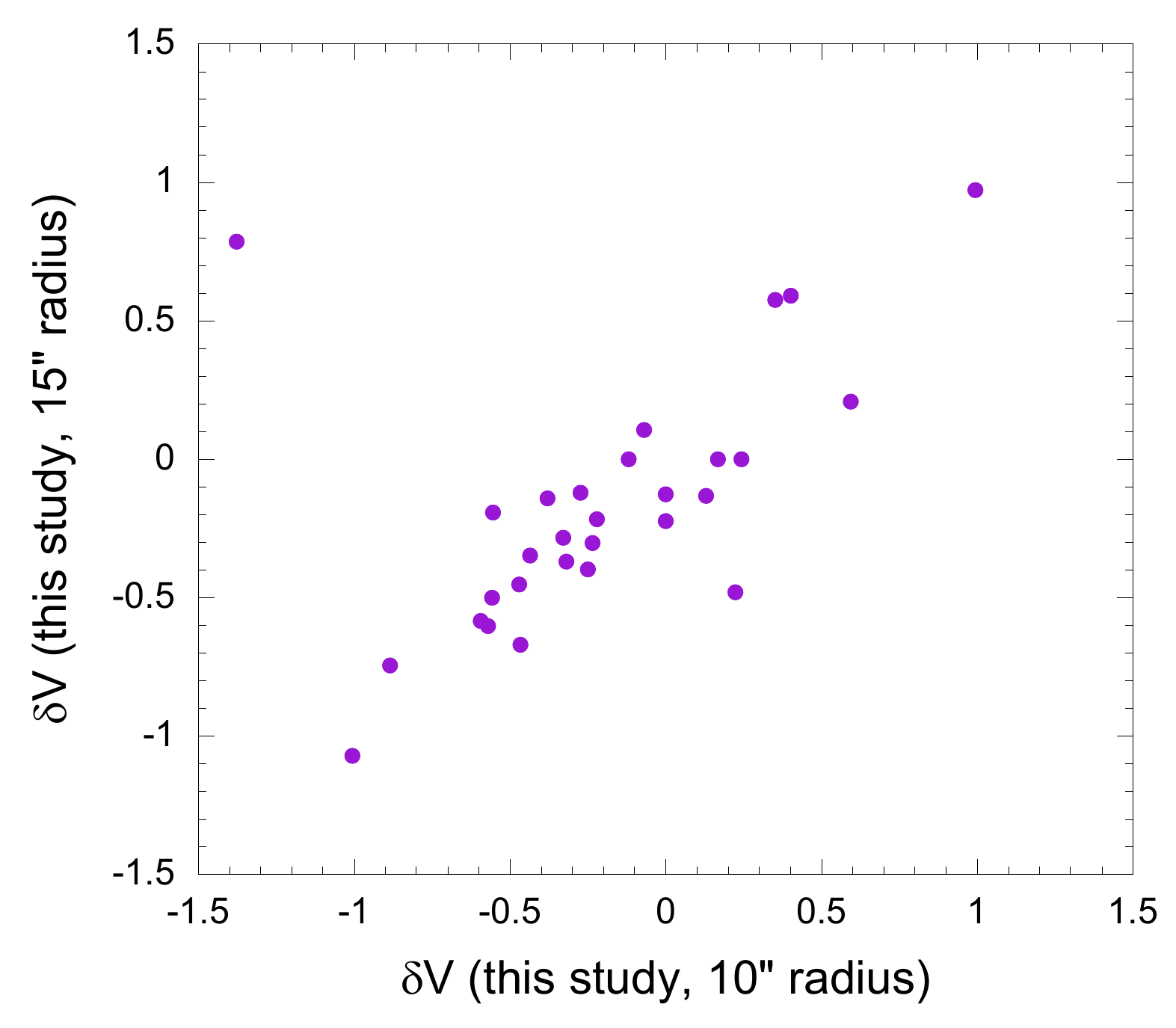}
\caption{
Asymmetry parameter with 15$\arcsec$ radius sampling
is compared with that with 10$\arcsec$ radius sampling.
\label{fig:dVcomp15-10}}
\end{figure}

\begin{figure}
\epsscale{0.6}
\figurenum{25}
\includegraphics[bb=0 0 505 575, width=15cm]{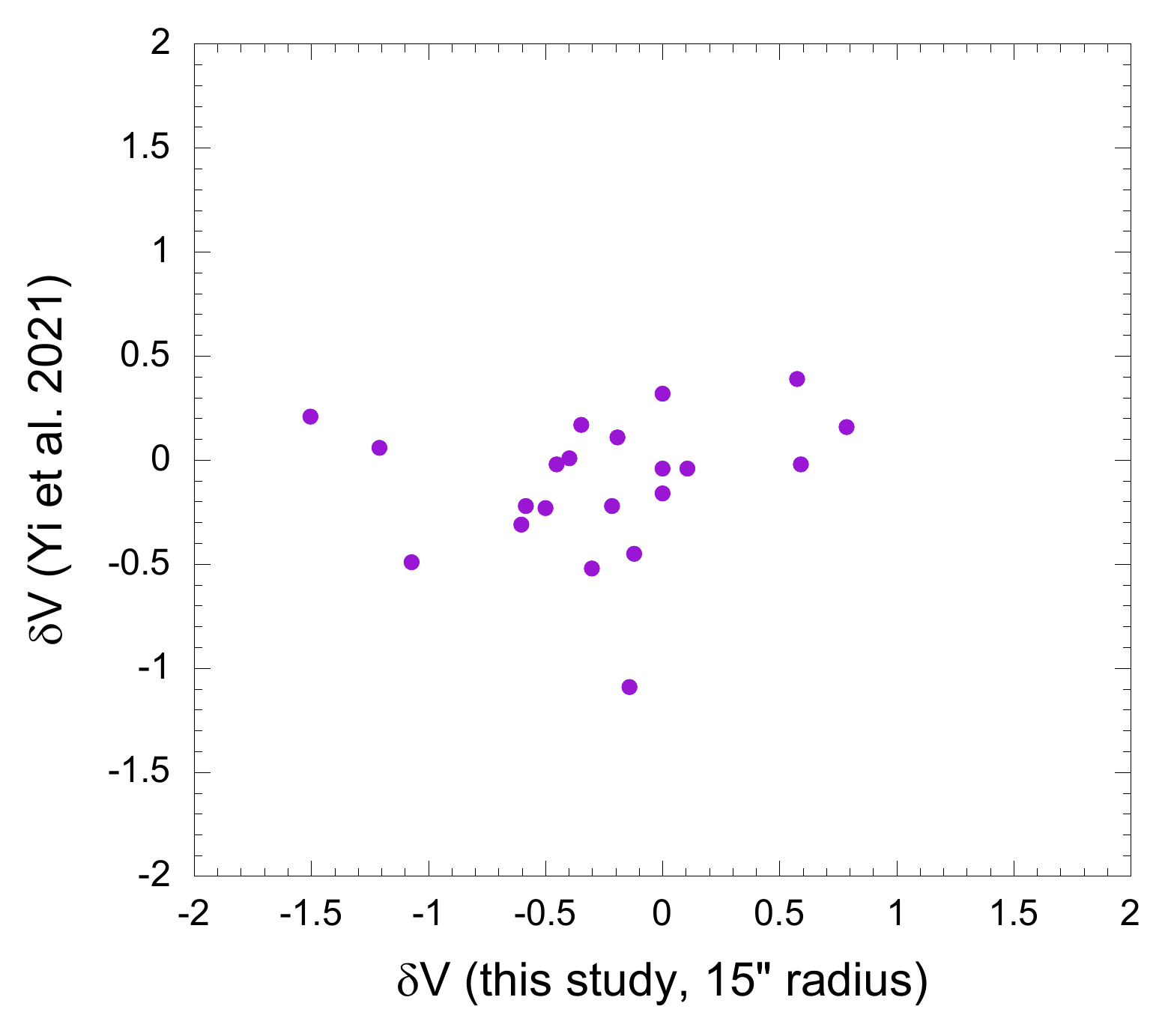}
\caption{
Asymmetry parameter by \cite{2021ApJS..254...14Y}
is compared with that with 15$\arcsec$ radius sampling.
\label{fig:dVcompYi-15}}
\end{figure}

\floattable
\begin{deluxetable}{crrcc}
\tablecaption{Asymmetry Parameter with 15$\arcsec$ Radius Sampling \label{tbl:15arcsec}}
\tablecolumns{5}
\tablenum{10}
\tablewidth{0pt}
\tablehead{
\colhead{Core Number} &
\colhead{$v$(HCO$^+)$} &
\colhead{$v($H$^{13}$CO$^+)$} &
\colhead{$\Delta v($H$^{13}$CO$^+)$} &
\colhead{$\delta V$} \\
\colhead{} &
\colhead{km s$^{-1}$} &
\colhead{km s$^{-1}$} &
\colhead{km s$^{-1}$} &
\colhead{} 
}
\startdata
2  &  11.0 &  11.1 &  0.46 &  -0.2 \\
3  &  10.7 &  10.8 &  0.76 &  -0.1 \\
4  &  9.7 &  9.6 &  0.48 &  0.2 \\
5  &  9.6 &  10.0 &  1.32 &  -0.3 \\                                              
8  &  9.8 &  9.8 &  0.61 &  0.0 \\
9  &  9.6 &  9.8 &  0.17 &  -1.2 \\
10  &  9.5 &  9.8 &  0.86 &  -0.3 \\
11  &  9.1 &  10.0 &  0.84 &  -1.1 \\
12  &  9.3 &  9.1 &  1.88 &  0.1 \\
13  &  10.7 &  11.1 &  0.66 &  -0.6 \\
14  &  10.8 &  11.1 &  0.75 &  -0.4 \\
15  &  10.9 &  11.3 &  0.83 &  -0.5 \\
16  &  11.5 &  11.1 &  0.68 &  0.6 \\
17  &  7.7 &  8.2 &  0.67 &  -0.7 \\
18  &  7.6 &  7.9 &  0.45 &  -0.7 \\
19  &  8.4 &  8.6 &  0.54 &  -0.4 \\
20  &  7.4 &  7.4 &  1.70 &  0.0 \\
21  &  7.6 &  8.1 &  1.76 &  -0.3 \\
22  &  8.0 &  8.2 &  0.34 &  -0.6 \\
23  &  8.6 &  8.3 &  0.38 &  0.8 \\
24  &  8.3 &  8.0 &  0.52 &  0.6 \\
25  &  7.7 &  7.7 &  0.82 &  0.0 \\
26  &  7.8 &  8.2 &  0.80 &  -0.5 \\
27  &  5.5 &  5.9 &  0.88 &  -0.5 \\
28  &  7.4 &  8.1 &  0.47 &  -1.5 \\
29  &  6.4 &  6.5 &  0.45 &  -0.2 \\
30  &  5.6 &  5.1 &  0.51 &  1.0 \\
31  &  4.2 &  4.3 &  0.71 &  -0.1 \\
32  &  3.1 &  3.2 &  0.52 &  -0.2 \\
33  &  4.3 &  4.4 &  0.83 &  -0.1 \\
34  &  4.8 &  3.7 &  0.34 &  3.2 \\
35  &  3.9 &  4.0 &  0.79 &  -0.1 \\
36  &  11.2 &  11.2 &  0.64 &  0.0 \\
\enddata
\end{deluxetable}

\subsection{Association with Filaments}

\cite{2021ApJS..256...25T} investigated whether core properties such as radius, velocity dispersion, core mass, and virial parameter
depended on the association with filaments; no clear differences between the cores associated with filaments and those not associated with filaments were found.
We judged the association with filaments with the SCUBA-2 maps
of \cite{2018ApJS..236...51Y} by eye.
Table \ref{tbl:fil_cef} summarizes association with filaments.
20 out of the 30 starless cores (66\%) are associated with filaments.
We found that four (except for core 10) out of the 
five starless cores 
(80\%)
showing blue-skewed or candidate blue-skewed profiles
were associated with filaments.
The current result does not suggest any outstanding statistics regarding
the percentage of filament association in the cores showing blue-skewed or candidate blue-skewed line profiles with dips.

It is not easy to relate the filament dynamics and the cores showing blue-skewed or candidate blue-skewed line profiles with dips.
Figure \ref{fig:S17-H13COMOM1} to \ref{fig:S32-H13COMOM1}
show moment 1 (intensity-weighted velocity) maps toward the three cores
with a hint of the velocity gradient along the filament.

One outstanding example is the filament connecting cores 17 and 18 
(Figures \ref{fig:S17-H13COMOM1} and \ref{fig:S18-H13COMOM1}, respectively).
A velocity gradient along the north--south filament is clearly seen.
If we can assume that this velocity gradient represents inflows, these cores showing the candidate blue-skewed line profiles with dips may accrete the material
along the filament.
Another example is core 32 (Figure \ref{fig:S32-H13COMOM1}). 
For comparison, Figure 18 of \cite{2021ApJS..256...25T} illustrated the N$_2$H$^+$ moment 1 map toward this core.
From these figures, we see
an east--west velocity gradient consistently; however the velocity field shown in the latter for a wider area seems more complicated.

\begin{figure}
\epsscale{1.0}
\figurenum{26}
\includegraphics[bb=0 0 505 575, width=10cm]{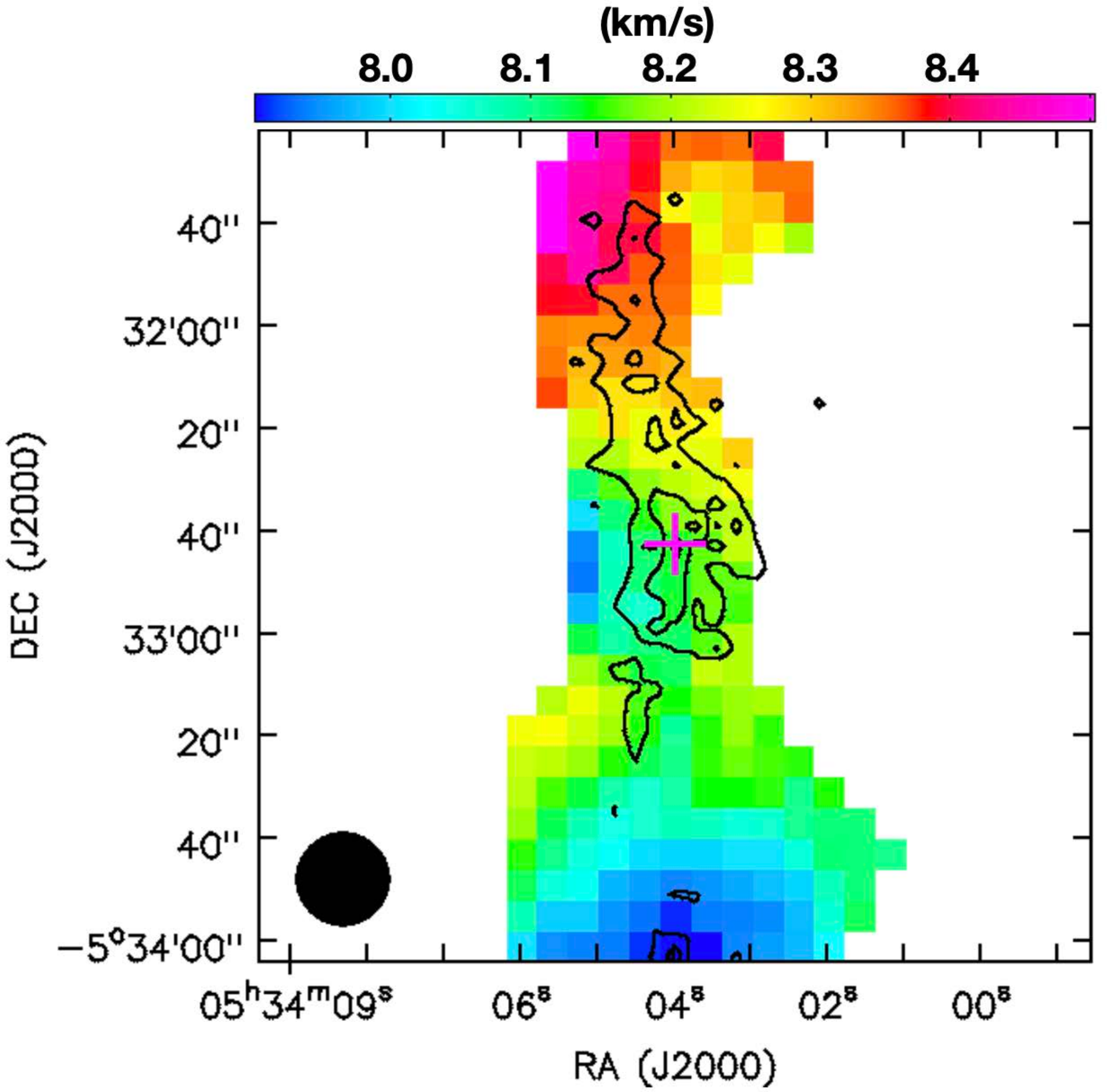}
\caption{
Gray-scale map of the moment 1 (intensity-weighted velocity) map 
of the H$^{13}$CO$^+$ emission p
toward
core 17 (G209.05$-$19.73North).
The plus sign represents the SCUBA-2 core center position.
The circle in the bottom-left corner represents the half-power beam size
(18$\farcs$2 diameter).
Contours of the SCUBA-2 850~$\micron$ continuum emission are drawn at levels of
50\%, 70\%, and 90\% of the maximum intensity,
which is 259.3~mJy~beam$^{-1}$.  
This core is associated with a SCUBA-2 filament elongated in the north--south direction.
\label{fig:S17-H13COMOM1}}
\end{figure}

\begin{figure}
\epsscale{1.0}
\figurenum{27}
\includegraphics[bb=0 0 505 575, width=10cm]{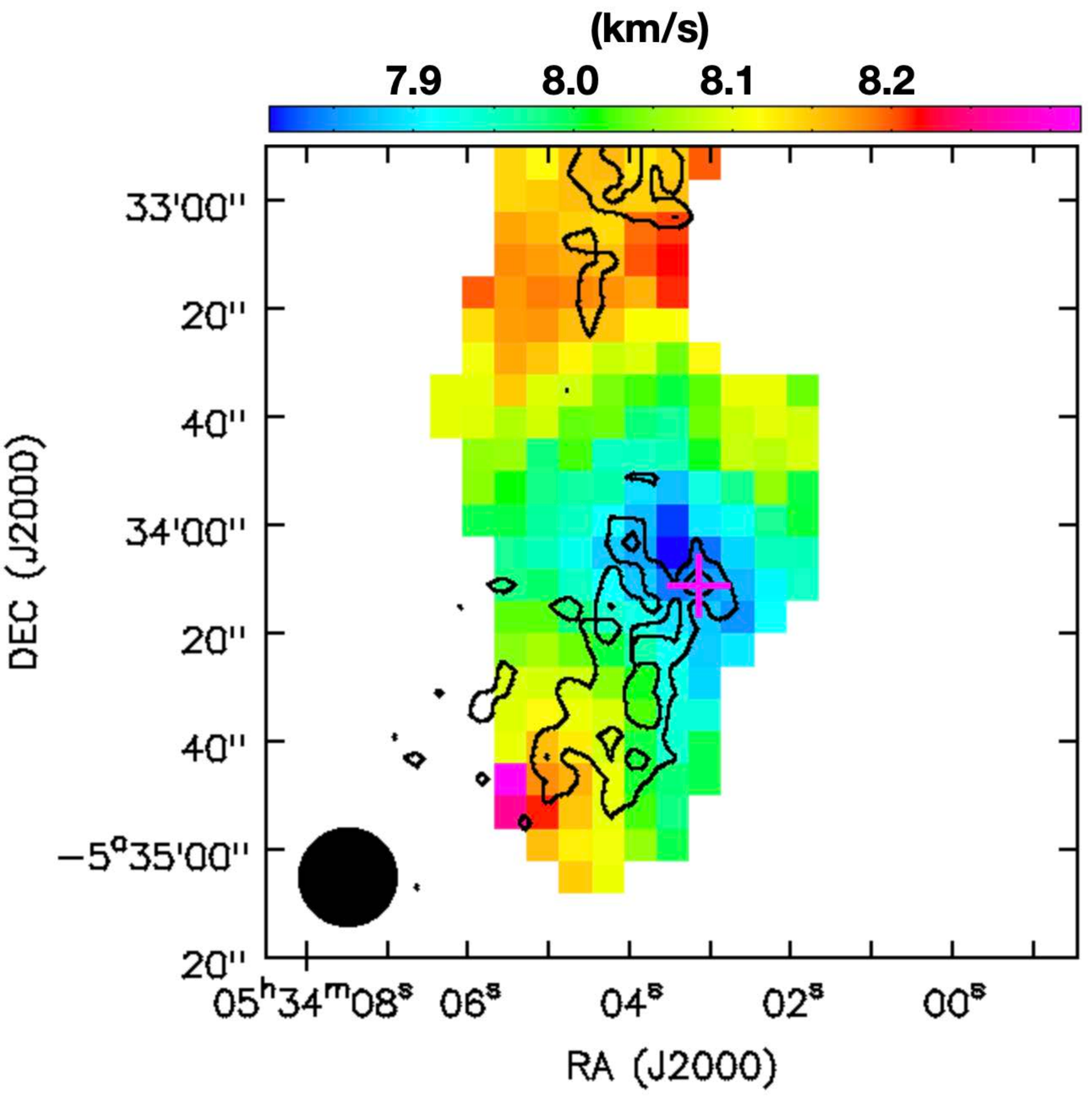}
\caption{
Same as Figure \ref{fig:S17-H13COMOM1} but for
core 18 (G209.05$-$19.73South).
This core is associated with a SCUBA-2 filament oriented in the north--south direction.
Note that this Figure is adjacent to Figure \ref{fig:S17-H13COMOM1},
and cores 17 and 18 are connected in the same north--south filament.
The maximum intensity of the SCUBA-2 850~$\micron$ emission is
260.6~mJy~beam$^{-1}$.  
\label{fig:S18-H13COMOM1}}
\end{figure}

\begin{figure}
\epsscale{1.0}
\figurenum{28}
\includegraphics[bb=0 0 505 575, width=10cm]{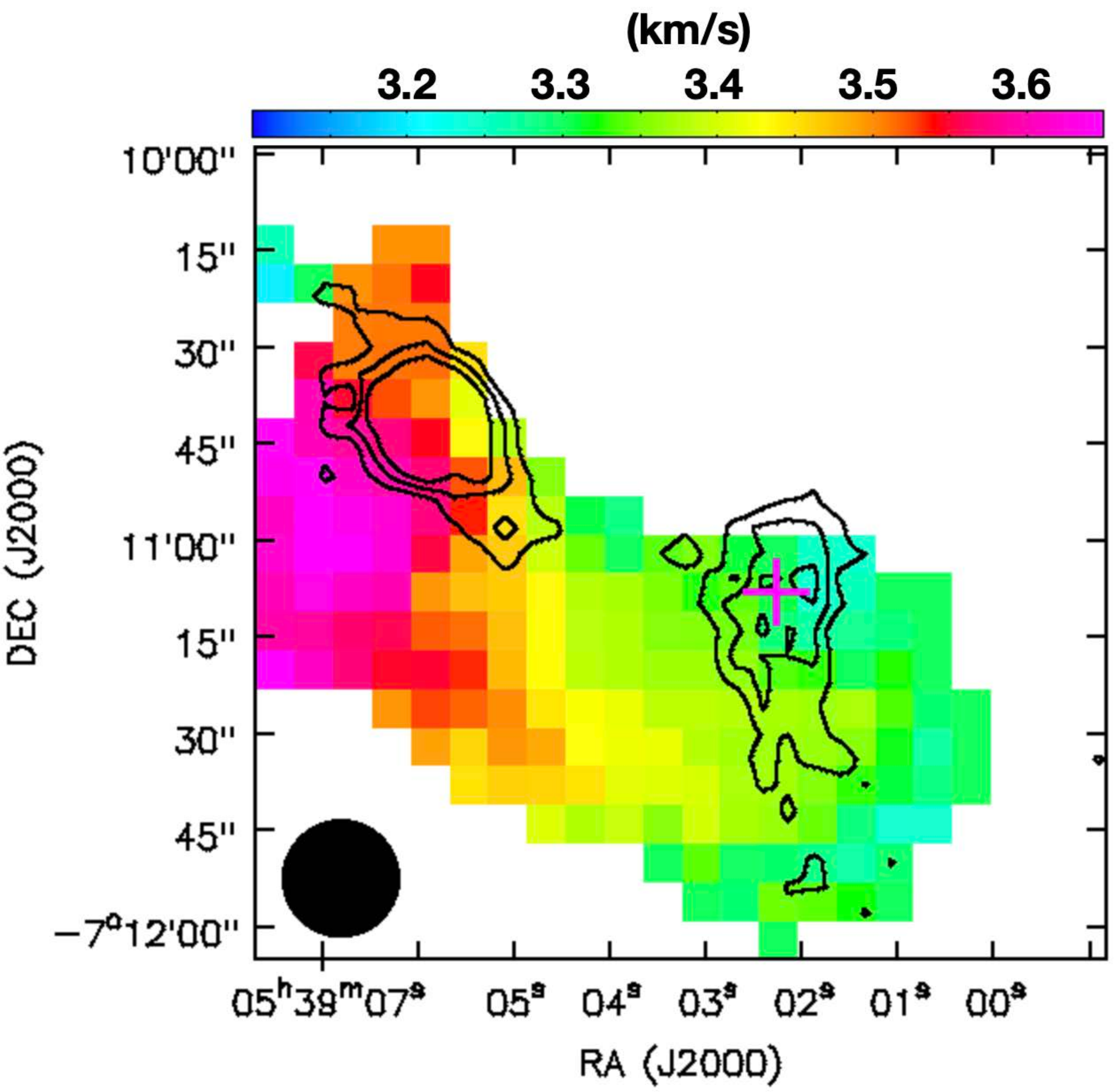}
\caption{
Same as Figure \ref{fig:S17-H13COMOM1} but for
core 32 (G211.16$-$19.33North3).
This core is associated with a SCUBA-2 filament oriented in the east--west direction.
The maximum intensity of the SCUBA-2 850~$\micron$ emission is
83.9 mJy beam$^{-1}$.  
\label{fig:S32-H13COMOM1}}
\end{figure}

\subsection{Mass Rates of Inward Motions}

Without detailed models of each source, we offered only two complementary estimates for the mass rates of inward motions.
(1) We assumed that the cores were in or near a state of equilibrium. This is plausible if they are isolated dense cores just at the beginning of infall, but
the model is more general.
(2) We assumed that all the mass of the core was flowing freely at constant velocity and used the crossing time. This will correspond to the inertial inflow model of 
\cite{2020ApJ...900...82P}

(1) The mass inflow rate for any cloud that is initially in approximate
hydrostatic equilibrium with 
magnetic fields and turbulence as well as thermal pressure
is proportional to $a^3/G$ 
\citep{1977ApJ...218..834H,1980ApJ...241..637S,2007ARAA..45..565M},
where $G$ is the gravitational constant and $a$ is the effective sound speed.
The coefficient for $a^3/G$ is known to be of order unity, and we simply assumed unity.
To estimate $a$, we adopted the total line width 
$\Delta v_{NT}$ \citep{1992ApJ...384..523F},
which is a sum of the nonthermal line width $\Delta v_{NT}$ and 
the thermal contribution corresponding to the mean molecular weight
in quadrature,
to obtain the effective speed of sound $a$
from the line width of the H$^{13}$CO$^+$ line, which is optically thinner,
by assuming a kinetic temperature of $T_k$ = 15 K;

\begin{equation}
\Delta v_{NT}^2 = \Delta v_{obs}^2 - (8~{\rm ln}~2) kT_k/m_{obs},
\end{equation}

\begin{equation}
\Delta v_{TOT}^2 = \Delta v_{NT}^2 + (8~{\rm ln}~2) kT_k/\mu,
\end{equation}

and 

\begin{equation}
a = \sigma_{TOT} = \Delta v_{TOT}/\sqrt{8~{\rm ln}~2},
\end{equation}

\noindent where 
$\Delta v_{obs}$, 
$m_{obs}$, and $\mu$ are the observed FWHM line width, the mass of the molecule H$^{13}$CO$^+$ and the mean molecular mass per particle (2.33~$m_{\rm H}$),
respectively. 
The mass rate $\dot{M}$~(equilibrium) in this method for the 
five (candidate) blue-skewed starless
cores ranges
from $4.0 \times 10^{-6}~M_{\odot}$~yr$^{-1}$ to $1.9 \times 10^{-5}~M_{\odot}$~yr$^{-1}$.
For all 36 cores, it was $(1.6\pm2.1) \times 10^{-5}~M_{\odot}$~yr$^{-1}$
(the mean value and standard deviation in the sample).

(2) Second, we estimated another mass inflow rate using the crossing time.
We estimated the crossing time by dividing the SCUBA-2 core radius by the effective sound speed above, 
and then estimated the mass rate
by dividing mass by the crossing time.
We obtained the mass rate
$\dot{M}$~(v$_{cross}$) in this method for the five (candidate) blue-skewed starless cores to be $(2.5-39) \times 10^{-6}~M_{\odot}$~yr$^{-1}$.
For all 36 cores, it range from $4.7 \times 10^{-7}~M_{\odot}$~yr$^{-1}$ to $8.4 \times 10^{-5}~M_{\odot}$~yr$^{-1}$.
The mean of $\dot{M}$~(v$_{cross}$) was $(1.3\pm1.7) \times 10^{-5}~M_{\odot}$~yr$^{-1}$
(the mean value and standard deviation in the sample).
The standard deviation is larger than the mean value, because of the very large range of values.

For comparison, the ratio of the mass rate in these two methods, $\dot{M}$~(v$_{cross}$)/$\dot{M}$~(equilibrium) for all 36 cores was 1.0$\pm$0.8.
The estimates from these two different methods are similar. 
If the mean value of the ratio is larger than unity, it may be because the second method
assumes freely flowing gas with no pressure support. 
Alternatively, if the ratio is smaller than unity, 
freely flowing matter does not deliver mass to a core a great deal faster 
than infall onto a star could process it.
Approximately, it seems that freely flowing matter delivers mass to a core so that infall onto a star could process it.

For the filament connecting cores 17 (with a dip but no skewness) and 18 (candidate blue-skewed), 
we estimated the possible inflow rate along the filament assuming that the velocity gradient of the filament 
represents inflow.
We estimated the local thermodynamic equilibrium (LTE) mass of the filament to be 4~$M_{\odot}$
from the H$^{13}$CO$^+$ data 
with
an excitation temperature of $\sim$5 K and an H$^{13}$CO$^+$ abundance of $4.8\times 10^{-11}$ 
\citep[][and references therein]{2007ApJ...665.1194I}.
The velocity gradient along the filament was $\sim$1~km~s$^{-1}$ pc$^{-1}$.
We estimated an inflow timescale of 10$^6$ yr from the velocity gradient,
and then estimated the rate by dividing mass by the inflow timescale.
The inflow rate along the filament, which may be fed into the core, could be of the order of $4~\times 10^{-6}~M_{\odot}$~yr$^{-1}$.

These values may provide us the order of magnitude estimate of the rate for
inflows along filaments and infall toward the core center, if they exist.

\section{SUMMARY \label{sec:sum}}

We searched for inward motions toward the 36 SCUBA-2 cores (30 starless cores and 6 protostellar cores) in Orion.
We used the Nobeyama 45 m radio telescope, and mapped
them in the $J = 1\rightarrow0$ transitions of HCO$^+$, H$^{13}$CO$^+$, N$_2$H$^+$, HNC, and HN$^{13}$C.
The asymmetry parameter (normalized velocity difference) $\delta V$, which is the ratio of the velocity difference
between HCO$^+$ and H$^{13}$CO$^+$, to the H$^{13}$CO$^+$ line width
is biased toward negative values, suggesting that
inward motions are more dominant than outward motions.
Three starless cores (10\%) were found to have HCO$^+$ blue-skewed line profiles
and 
two more starless cores (7\%)
had candidate blue-skewed line profiles.
The peak velocity difference between HCO$^+$ and H$^{13}$CO$^+$ reaches 0.9~km~s$^{-1}$, suggesting that inward motions are at least partially supersonic.
The mean of $\delta V$ of the aforementioned 
five starless
cores
was derived to be 
$-$0.5$\pm$0.3.
All these cores were found to be associated with cores with H$^{13}$CO$^+$ emission.
G211.16$-$19.33North3 also observed using the ALMA ACA in DCO$^+$ $J = 3\rightarrow2$
showed blue-skewed features.
The velocity offset of gas with inward motions in the ALMA ACA observations was larger than that in the Nobeyama 45 m telescope observations,
which can be explained in terms of gravitational acceleration of inward motions.
Although there was no clear indication that inward motions were observed predominantly at a specific phase
in general, it seems that this core is at the last stage in the starless phase.

\vspace{5mm}



\acknowledgments
K.T. was supported by JSPS KAKENHI (Grant Number 20H05645). 
P.S. was partially supported by a Grant-in-Aid for Scientific Research (KAKENHI Number 18H01259) from the JSPS.

\vspace{5mm}
\facilities{No:45m}

\software{AIPS \citep{1996ASPC..101...37V}, NOSTAR \citep{2008PASJ...60..445S}}

\end{document}